\documentclass[12pt]{article}

\usepackage{fancyhdr}
\usepackage{isomath}
\usepackage{amsmath}
\usepackage{amsbsy}
\usepackage{amssymb}
\usepackage{amscd}
\usepackage{amsfonts}
\usepackage{graphics}
\usepackage{verbatim}
\usepackage{subfigure}
\usepackage{xspace}
\usepackage{euscript}
\usepackage{alltt}
\usepackage{boxedminipage}
\usepackage{float}
\usepackage{color}
\usepackage[,colorlinks,urlcolor=blue,citecolor=blue]{hyperref}
\usepackage[all]{xy}
\usepackage{t1enc}
\usepackage{times,exscale}
\usepackage{graphicx,calc}
\usepackage{mdwlist}
\usepackage{stmaryrd}
\usepackage{units}
\usepackage{setspace}
\usepackage{pdfpages}
\usepackage[usenames,dvipsnames]{xcolor}
\usepackage{enumitem}
\usepackage{placeins}
\usepackage{isomath}
\usepackage{amsmath}
\usepackage{amssymb}
\usepackage{amscd}
\usepackage{amsfonts}

\newcommand{\bfa}{{\mathbold a}}
\newcommand{\bfb}{{\mathbold b}}

\newcommand{\bfd}{{\mathbold d}}
\newcommand{\bfe}{{\mathbold e}}

\newcommand{\bfm}{{\mathbold m}}
\newcommand{\bfn}{{\mathbold n}}

\newcommand{\bft}{{\mathbold t}}
\newcommand{\bfu}{{\mathbold u}}
\newcommand{\bfv}{{\mathbold v}}

\newcommand{\bfx}{{\mathbold x}}

\newcommand{\bfC}{{\mathbold C}}

\newcommand{\bfE}{{\mathbold E}}
\newcommand{\bfF}{{\mathbold F}}

\newcommand{\bfI}{{\mathbold I}}

\newcommand{\bfL}{{\mathbold L}}

\newcommand{\bfN}{{\mathbold N}}

\newcommand{\bfS}{{\mathbold S}}

\newcommand{\bfU}{{\mathbold U}}
\newcommand{\bfV}{{\mathbold V}}

\newcommand{\bfX}{{\mathbold X}}

\newcommand{\eps}{{\varepsilon}}

\newcommand{\beq}{\begin{equation}}
\newcommand{\eeq}{\end{equation}}
\newcommand{\beqs}{\begin{eqnarray}}
\newcommand{\eeqs}{\end{eqnarray}}
\newcommand{\half}{\frac{1}{2}}

\newcommand{\calR}{{\cal R}}

\newcommand{\bfsigma}{\mathbold {\sigma}}
\newcommand{\bfepsilon}{\mathbold {\epsilon}}

\newcommand{\bfalpha}{\mathbold {\alpha}}

\newcommand{\grad}{\mathop{\rm grad}\nolimits}
\newcommand{\divergence}{\mathop{\rm div}\nolimits}
\newcommand{\curl}{\mathop{\rm curl}\nolimits}

\newcommand{\const}{\mathrm{const.}}

\newcommand{\parderiv}[2]{\frac{\partial #1}{\partial #2}}

\newcommand{\deriv}[2]{\frac{d #1}{d #2}}
\newcommand{\jump}[1]{\left\llbracket #1 \right\rrbracket}

\newenvironment{packedenum}{
\begin{enumerate}
  \setlength{\itemsep}{1pt}
  \setlength{\parskip}{0pt}
  \setlength{\parsep}{0pt}
}{\end{enumerate}}

\graphicspath{ {./imagedir/} }

\usepackage[margin=20mm]{geometry}
\numberwithin{equation}{section}

\newcommand{\Wcirc}{\mathring{W}}

\title{A Dynamic Multiscale Phase-field Model for Structural Transformations and Twinning:  Regularized Interfaces with Transparent Prescription of Complex Kinetics and Nucleation}
\author{Vaibhav Agrawal\footnote{\url{vaibhava@andrew.cmu.edu}} \  and Kaushik Dayal\footnote{\url{kaushik@cmu.edu}} \\ \mbox{\small Carnegie Mellon University}}
\date{\today}

\begin{document}
\pagestyle{fancyplain}
\lhead{\fancyplain
{\scriptsize A Dynamic Multiscale Phase-field Model for Structural Transformations and Twinning
%(submitted to {\em J. Mech. Phys. Solids})
}
{\scriptsize A Dynamic Multiscale Phase-field Model for Structural Transformations and Twinning
%(submitted to {\em J. Mech. Phys. Solids})
}
}
\rhead{\fancyplain{\scriptsize Vaibhav Agrawal and Kaushik Dayal}{\scriptsize Vaibhav Agrawal and Kaushik Dayal}}
\maketitle
\begin{abstract}
The motion of microstructural interfaces is important in modeling materials that undergo twinning and structural phase transformations.
Continuum models fall into two classes: sharp-interface models, where interfaces are singular surfaces; and regularized-interface models, such as phase-field models, where interfaces are smeared out.
The former are challenging for numerical solutions because the interfaces need to be explicitly tracked, but have the advantage that the kinetics of existing interfaces and the nucleation of new interfaces can be transparently and precisely prescribed.
In contrast, phase-field models do not require explicit tracking of interfaces, thereby enabling relatively simple numerical calculations, but the specification of kinetics and nucleation is both restrictive and extremely opaque.
This prevents straightforward calibration of phase-field models to experiment and/or molecular simulations, and breaks the multiscale hierarchy of passing information from atomic to continuum.
Consequently, phase-field models cannot be confidently used in dynamic settings.

This shortcoming of existing phase-field models motivates our work.
We present the formulation of a phase-field model -- i.e., a model with regularized interfaces that do not require explicit numerical tracking -- that allows for easy and transparent prescription of complex interface kinetics and nucleation.
The key ingredients are a re-parametrization of the energy density to clearly separate nucleation from kinetics; and an evolution law that comes from a conservation statement for interfaces.
This enables clear prescription of nucleation through the source term of the conservation law and of kinetics through an interfacial velocity field.
A formal limit of the kinetic driving force recovers the classical continuum sharp-interface driving force, providing confidence in both the re-parametrized energy  and the evolution statement.
%Traveling wave solutions to our model also display interesting features.

We present a number of numerical calculations that characterize our formulation in one and two dimensions.
These calculations illustrate: 
(i) stick-slip, linear, and quadratic kinetics; 
(ii) highly-sensitive rate-dependent nucleation; 
(iii) independent prescription of the forward and backward nucleation stresses without changing the energy landscape; 
(iv) the competition between nucleation and kinetics in determining the final microstructural state; 
(v) the transition from subsonic to supersonic, where kinetic relations should and should not be imposed respectively; and 
(vi) the effect of anisotropic and non-monotone kinetics.
These calculations demonstrate the ability of this formulation to precisely prescribe complex nucleation and kinetics in a simple and transparent manner.
We also extend our conservation statement to describe the kinetics of the junction lines between microstructural interfaces and boundaries.
This enables us to prescribe an additional kinetic relation for the boundary, and we examine the interplay between the bulk kinetics and junction kinetics.

%Dissemination: 
%John Clayton + Jarek Knap, Phoebus Rosakis, Kaushik Bhattacharya + Ravichandran, Rohan Abeyaratne, John Bassani, Turab Lookman + Marcel Porta + Groger + Saxena, Johannes Zimmer + Michael Herrmann, Richard James, Michael Ortiz, Mort Gurtin, David Owen + Luca Deseri, Qiang Du, Sergio Turteltaub, Pradeep Sharma, Yi-Chung + Jiangyu, Tadmor, Fosdick, Oscar Lopez, Truskinovsky, Suquet, Walkington, Alain Molinari, Isaac Chenchiah, Harish Tippur, Dennis Kochmann, Stewart Silling, Lipton, Rohrer, Allan Bower, Plaza, Weckner, Steigmann, Shaw+Daly, Ball, Arias+Arroyo, Landis, Beyerlein + Koslowski + Abigail Hunter + Carlos Tome + Ricardo Lebensohn, Yash Kulkarni, KT Ramesh + Tim Wright, Ravi-Chander, Rollett, Ravishankar (Pitt), Eliot Fried, Clifton, McDowell, Sehitoglu, Purohit, Mark Tschopp, Dondl, Sekerka, Franz Fischer + Thomas Antretter + Tom Waitz, Ackland

\end{abstract}

%%%%%%%%%%%%%%%%%%%%%
%%%%%%%%%%%%%%%%%%%%%
%%%%%%%%%%%%%%%%%%%%%
%%%%%%%%%%%%%%%%%%%%%

\section{Introduction}

Twinning and structural phase transformations are important in areas as diverse as superelasticity and shape-memory in functional materials \cite{bhatta-book}, forming of structural metals \cite{bozzolo2010misorientations}, nanostructured metals with exceptional properties such as high strength and high ductility \cite{hunter2014predictions, kulkarni2009nanotwinned,waitz2009phase}, and the dynamic response of metals under extreme conditions \cite{ravishankar-nanotwinned}.
The typical microstructure in these settings consists of homogeneously deformed regions separated by interfaces across which the deformation varies extremely rapidly.
Many important aspects of these phenomena are governed by the nucleation, motion, and response of the interfaces.

In the continuum setting, twinning and structural transformations are modeled using nonconvex strain energy density functions $W(\epsilon)$, an approach introduced in the seminal paper of Ericksen \cite{ericksen-1975} in 1D.
The nonconvexity allows for the coexistence of different phases or twins for a given stress value $\sigma = \deriv{W}{\epsilon}$.
The different phases are separated by interfaces across which the strain is discontinuous.
Since the standard continuum theory contains no lengthscale, these interfaces are ``sharp'', i.e. singularly localized.
Ericksen observed that the continuum balance of linear momentum is insufficient to identify a unique spatial location of the interfaces, even assuming the existence of a single interface.
In the static setting without inertia, he used energy minimization as a selection criterion to obtain a unique solution.

Abeyaratne and Knowles \cite{abeyaratne1990driving,abeyaratne1991kinetic} examined nonconvex models in the dynamic setting with inertia.
Again, balance of linear momentum does not provide a unique solution even in the simplest case of a single interface in 1D.
Further, energy minimization is not applicable in dynamic problems and cannot be used to resolve this.
Invoking thermodynamics, viz. positive dissipation, provides some weak restrictions on the motion of the interface, but still leaves a massively nonunique problem with essentially a 1-parameter family of solutions that is parametrized by the location of the interface.
\cite{abeyaratne1990driving,abeyaratne1991kinetic}, and related work in \cite{truskinovskii1982equilibrium}, find that imposing additional closure relations makes the problem unique.
Namely, the closure relations are (i) the kinetic relation that relates the velocity of the interface to the thermodynamic work conjugate driving force, and (ii) the nucleation criterion that provides for the formation of new interfaces.
Physically, the closure relations can be thought of as a macroscopic remnant of the lattice-level atomic motion from one energy well to another that is lost in the continuum theory. 
However, a systematic derivation from a microscopic theory as well as experimental confirmation remain a topic of active research.

The closure relations -- nucleation criterion and kinetic relation -- have the advantage of clear and direct physical interpretations.
In particular, they fit naturally into a multiscale modeling framework by allowing for precisely-defined constitutive input on the behavior of interfaces from either experiment or modeling (e.g., molecular dynamics).
However, numerical computations with this approach are extremely challenging, because the sharp interfaces require complex and expensive tracking algorithms in a numerical discretization.
This is an unfeasible challenge when one expects numerous interfaces that are evolving, interacting, and nucleating.
Therefore, this {\em sharp-interface approach} has not been widely applied to larger problems.

In contrast to this, there is a large body of work on methods that regularize or smooth the interface by adding strain gradients, viscous dissipation, and similar effects to the stress response, e.g. \cite{AK-straingrad, rosakis-straingrad, turteltaub-straingrad,truskinovsky1993kinks,fuaciu2006longitudinal}.
In 1D, the stress in these models is typically given by:
\begin{equation}
	 \sigma = \deriv{W}{\epsilon} + \nu \frac{d\epsilon}{dt} - \kappa \frac{d^2\epsilon}{dx^2}
\end{equation}
In these {\em regularized-interface approaches}, the evolution of interfaces is obtained simply by solving momentum balance $\divergence \sigma = \rho \ddot{u}$.
The solutions are typically unique but depend strongly on the regularization parameters, viz. the viscosity $\nu$ and the capillarity $\kappa$, in addition to the nonconvex energy density $W$.
While the nonconvexity in $W$ favors the formation of interfaces, the gradient regularization $\kappa \frac{d^2\epsilon}{dx^2}$ penalizes the sharpness of interfaces and thereby prevents them from being singular.
Because interfaces are not singular and hence do not need to be explicitly tracked, these approaches are relatively easy to apply to large problems.
Further, nucleation of new interfaces and topology transitions occur naturally without additional computational effort or constitutive input.

In the closely-related phase-field approaches, the situation is similar.
The phase are distinguished by a scalar field $\phi$, and the energy is nonconvex in $\phi$ with a coupling to elasticity.
A typical phase-field energy density \cite{chen2002phase,su2007continuum,zhang2005computational,abdollahi2012phase,yang-dayal-apl2010,yichung-jiangyu-1,yichung-jiangyu-2} has the form
\begin{equation}
	w(\phi) + \half \left( \bfepsilon - \bfepsilon_0 (\phi) \right) : \bfC : \left( \bfepsilon - \bfepsilon_0 (\phi) \right) + \kappa |\nabla\phi|^2
\end{equation}
$w(\phi)$ is a nonconvex energy and favors the formation of interfaces, while the gradient term $\kappa |\nabla\phi|^2$ regularizes them.
These terms are coupled to linear elasticity through the elastic strain $ \left( \bfepsilon - \bfepsilon_0 (\phi) \right)$, that is the difference between the total strain $\bfepsilon\equiv\grad\bfu$ and the stress-free strain $\bfepsilon_0 (\phi)$ that depends on the phase.
The total energy $E$ is obtained by integrating the energy density over the body and accounting for the boundary working.
The evolution is governed by a gradient descent in $\phi$, i.e. $\mu \dot{\phi} = \frac{\delta E}{\delta \phi}$, and linear momentum balance for the evolution of the displacement / strain field.
Phase-field models share the key features of the strain-gradient models: 
\begin{packedenum}
	\item the evolution of interfaces is unique, and governed by the parameters $\mu$ and $\kappa$ in addition to the energy density; 
	\item nucleation and topology transitions occur naturally without additional input, and like kinetics, are governed by $\mu$, $\kappa$ and the energy density; 
	\item nucleation and kinetics are modeled together a single equation; and
	\item they are relatively easy to apply to large problems because interfaces are not singular.
\end{packedenum}

Feature 4 of phase-field models is an important advantage of these approaches.
However, Features 1, 2, and 3 are {\em not} advantages, but instead important shortcomings of these models.
While it is certainly important to obtain unique solutions, the fact that the nucleation and kinetics of interfaces are governed by a small set of parameters implies that the range of behavior that can be modeled is highly restricted.
In addition, the nucleation and kinetics of interfaces are physically distinct processes from the atomic perspective, but in these models are governed through the same evolution equation.

For instance, Feature 1 greatly limits the ability to formulate a model that produces a desired kinetic response.
For example, the kinetics of interfaces in strain-gradient models is analyzed by \cite{AK-straingrad}, and they find that the range of kinetic responses that can be obtained by varying $\nu$ and $\kappa$ is extremely constrained\footnote{In generalized versions, e.g. \cite{fried1994dynamic,rosakis-straingrad}, a larger class of kinetics is possible, but the relation between the model parameters and the induced kinetics is not transparent even for 1D.}.
For instance, an important feature that is widely observed is stick-slip behavior of interfaces, e.g. \cite{faran2011kinetic}, but this cannot be modeled in strain-gradient or phase-field models.
In general, it is difficult to prescribe a given kinetic response directly; making the parameter $\mu$ a function of various quantities {\em may} allow this, but the dependence on these quantities to obtain a desired kinetic response is not transparent.
Therefore, calibrating a desired kinetic response using this route can require much trial-and-error that is tedious, unsystematic, and very expensive.

The situation in Feature 2 is similar to that in Feature 1, except that it is {\em much} more difficult!
Existing phase-field models are completely opaque, even in 1D, about the precise critical condition at which nucleation occurs \cite{dayal-bhatta-jmps2006}.
The nucleation behavior in a three-dimensional setting, with a complex energy landscape and numerous local extrema, combined with both inertial and gradient descent dynamics, is even more complex.
Consequently, the inverse problem is extremely hard: namely, how do we set up the energy and evolution to obtain a desired nucleation response?
That is, given some critical conditions under which nucleation takes place -- perhaps from experimental observation or molecular calculations -- how do we tailor the various functions and parameters in the model to achieve this behavior?
Modifying the energy barriers is an obvious starting point, but is difficult to do systematically; for instance, changing an energy barrier affects the nucleation behavior of both the forward and reverse transformations.
In a situation with numerous possible transformations, modifying the barrier can have unintended effects on all the transformations.
Another strategy is to use spatially-localized defects (or ``soft spots'', e.g. \cite{zhang2005computational}), where the energy landscape is locally modified to have shallow barriers to aid nucleation. 
This approach is difficult to use in situations that have not already been well-characterized by other techniques.
For instance, to obtain a desired nucleation stress, how should the soft spots be spatially arranged? should we have more soft spots with higher barriers, or a few soft spots with lower
barriers? what shape should they be? and so on. 
The current approach is typically ad-hoc, and involves trying a given configuration of soft spots, and doing full-field calculations to test if the given configuration provides the desired nucleation behavior.
Other strategies to induce nucleation, such as adding external driving noise, differ in the details, but have the same basic problem that modeling a desired nucleation behavior essentially requires solving a nasty inverse problem posed in a very large space in an unsystematic and expensive way, when the forward problem itself is not well-understood.
In addition, this inverse problem can be highly geometry- and problem-specific; calibration for a specific geometry will likely not be transferable to other geometries.
This difficult situation is vastly compounded when one begins to consider the realistic case that there is not simply one type of twinning interface, but rather various different ones for orientations, each with a different propensity to nucleate and move, with the entire problem posed in 3D.
Further, it is likely that critical conditions for nucleation in real systems is not simply related to the energy conjugate driving force; rather, there is likely rate-dependence, possibly dependence on hydrostatic stress even in volume-preserving twinning transformations, and so on.

Feature 3 complicates the process of calibrating a model to an observed nucleation and kinetics because the separation between these processes in the model is almost absent.

The failings of existing phase-field models make them impossible to use in a hierarchical multiscale setting.
Hierarchical multiscale approaches rely on the passage of information from fundamental models to larger-scale models.
In the setting of structural transformations, atomic calculations (e.g., \cite{daphalapurkar2014kinetics,ojha2014twin,wang2010atomic,beyerlein2010probabilistic,barrett2012breakdown}) as well as experiments (e.g., \cite{niemczura2006dynamics,escobar1993pressure}) have provided important information on twin kinetics and nucleation.
But almost none of this information can be used in the existing phase-field models, beyond some minor calibrations.
This wastes the wealth of insights that have been gathered from atomistics and experiment, and breaks the multiscale link between the atomic level and the continuum.

The advantages and failings of the different approaches described above provide the motivation for our work.
We present a regularized-interface model that has the advantage that computations are easy and efficient because we do not need to track interface evolution, nucleation, and topology transitions.
However, our formulation is also designed to obtain the key advantage of the sharp interface formulation, namely that we can transparently, precisely, and readily specify complex nucleation and kinetics behavior.
The technical strategy consists of 2 elements: (i) parametrization of the energy in a specific way, and (ii) evolution of $\phi$ through a geometric conservation law.

The first element is to re-parametrize the energy density so that it continues to reproduce the elastic response of each phase away from energy barriers, but leads to a clear separation between kinetics and nucleation.
Briefly, the re-parametrized energy density $\Wcirc(\bfF,\phi)$ is independent of $\phi$ except when $\phi$ is in a narrow range that can be considered to be the transition between the phases; an example is shown in Fig. \ref{fig:energy-landscape} and explained in detail below.
Therefore, the work-conjugate driving force for $\phi$ vanishes when $\phi$ is away from the transition range, and consequently $\phi$ cannot evolve  irrespective of the value of stress and other mechanical quantities.
Hence, when a region of the body is in a single phase, i.e. $\phi$ is uniform in a region, then $\phi$ cannot evolve and a new phase cannot nucleate.
The only region where $\phi$ can evolve is when it is in the transition range, which occurs near an interface.
Therefore, the energy allows a material region to undergo a transformation only when an interface sweeps over it, and nucleation of a new phase away from an interface is completely prohibited.
We will re-introduce nucleation through the balance law in a separate term from the kinetics; the advantage of this approach is that nucleation cannot occur through the kinetic law.
Thereby, our approach makes a clear distinction between kinetics and nucleation as mechanisms for the evolution of $\phi$: the kinetic law cannot cause nucleation, and the nucleation term does not affect kinetics.
This is in sharp contrast to standard phase-field models where a uniform phase may nucleate a new phase if the driving force {\em for kinetics} is sufficiently large, even if the desired critical conditions for nucleation have not been met.
There, the variational derivative of the energy with respect to $\phi$ governs both the kinetics of existing interfaces as well as the nucleation of new phases.
Therefore, the process of nucleation is intimately and opaquely mixed in with the prescribed kinetics in these models, making it hard to prescribe precise nucleation criteria.

The second element is to use a geometrically-motivated conservation law to govern the evolution of $\phi$.
Briefly, we interpret $\nabla\phi$ as a geometric object that provides us with the linear density of interfaces.
Then, for a material line element, we count the number of interfaces that are entering and leaving at each end of the element.
 The statement of the conservation law is that the increase in the number of interfaces threaded by the line element is equal to the net number of interfaces that are entering, plus the creation of interfaces through a source term.
 The motion of interfaces is described by an interface velocity field $v_n^\phi$, distinct from the material velocity field.
 The value of $v_n^\phi$ at each point can have a complex functional dependence on {\em any} mechanical field, e.g. stress, stress rate, nonlocal quantities, and so on, and this provides a route to transparently specify extremely complex kinetic response.
Similarly, the source term in the balance law provides transparent and precise control on the nucleation of new interfaces, by activating the source only when the critical conditions for nucleation are realized.
An important element is that the kinetic term is multiplied by $|\nabla\phi|$; therefore a uniform phase will not show any evolution of $\phi$ due to the kinetic term regardless of the stress level, and the only possible mechanism for the evolution of $\phi$ from a uniform state is by nucleation.

%%%%%%%%%%%%%%%%%%%%%
%%%%%%%%%%%%%%%%%%%%%
%%%%%%%%%%%%%%%%%%%%%
%%%%%%%%%%%%%%%%%%%%%

\subsection{Organization}
\label{sec:organization}
The paper is organized as follows.

\begin{itemize}

	\item In Section \ref{sec:formulation}, we describe the re-parametrization of the energy, the formulation of the interface balance principle, and the driving forces on interfaces obtained by enforcing positive dissipation.  We also examine formally the sharp-interface limit of the dissipation in our model.
	
	\item In Section \ref{sec:trav-waves}, we examine in 1D the behavior of steadily-moving interfaces in our model using a traveling-wave approach to show the relation between the prescribed kinetic response and the effective kinetics in terms of interface velocity and classical driving force.

	\item In Section \ref{sec:1d-dynamics}, we perform 1D dynamic calculations to understand the evolution of interfaces.  As in the section on traveling waves, we aim to find the effective kinetic relation induced by our model.
	
	\item In Section \ref{sec:small-parameter}, we examine the effect of a small parameter that has been introduced in the re-parametrization of the energy.
	
	\item In Section \ref{sec:2D-energy}, we briefly outline the formulation of a 2D energy density that we use in many of the subsequent 2D calculations.
	
	\item In Section \ref{sec:non-monotone}, we examine the effect of non-monotone kinetic response in 1D and 2D.
	
	\item In Section \ref{sec:aniso-kinetics}, we demonstrate the formulation of anisotropic kinetic laws in 2D.
	
	\item In Section \ref{sec:2d-stick-slip}, we examine twinning interfaces with stick-slip kinetics in 2D.
	
	\item In Section \ref{sec:rate-asym-nucleation}, we demonstrate -- in 1D and 2D -- the formulation of complex nucleation criteria that includes rate-dependent critical nucleation stresses, as well as independent prescription of forward and reverse nucleation stresses, without modifying the energy barriers.
	
	\item In Section \ref{sec:2d-hydro-nucleation}, we demonstrate imposing hydrostatic stress-dependence of twin nucleation, in addition to usual shear-stress dependence.

	\item In Section \ref{sec:kin-vs-nucleation}, we examine the competition between nucleation and kinetics in a twinning transformation.
	
	\item In Section \ref{sec:thermo-vs-momentum}, we examine the competition between thermodynamics and momentum balance in setting the kinetics of an interface.  We point out the difficulty our model has in dealing with certain phase interfaces whose evolution is uniquely described by momentum balance and that therefore does not require an additional kinetic relation.
	
	\item In Section \ref{sec:boundary-kinetics}, we examine the prescription of a kinetics associated with the junctions between interfaces and specimen boundaries.
	
	\item In Section \ref{sec:discussion}, we review our work.
	
	\item In Appendix \ref{sec:noether}, we examine briefly the connection to Noether's principle.
	
	\item In Appendix \ref{sec:phase-field-supersonic}, we examine the possibility of supersonic phase interfaces in standard phase-field models.

\end{itemize}

%%%%%%%%%%%%%%%%%%%%%
%%%%%%%%%%%%%%%%%%%%%
%%%%%%%%%%%%%%%%%%%%%
%%%%%%%%%%%%%%%%%%%%%

\subsection{Notation, Definitions and Values of Model Parameters}
\label{sec:notation}

Boldface denotes vectors and tensors.
We have used Einstein convention, i.e. repeated indices imply summation over those indices, except when noted.

\begin{tabular}{ll}
	$\phi$  	&  	phase field \\
	$\bfalpha \equiv \nabla\phi$	 &  	gradient of phase field interpreted as the linear density of interfaces \\
	$\hat{\bfn} \equiv \frac{\bfalpha}{|\bfalpha|}$ 	&  	unit normal vector to the interface between phases \\
	$\hat{\bft}$ 	&  	unit tangent to a curve in space \\
	 & 	 \\
	$\bfx_0$	&	material particle in the reference configuration\\
	$\bfx$ 	&  	material particle in the deformed configuration\\
	$\nabla\equiv\nabla_\bfx$ and $\nabla_{\bfx_0}$ 	&	 gradient with respect to $\bfx$ and $\bfx_0$ respectively; $\nabla_{\bfx} = \bfF^{-T} \nabla_{\bfx_0}$ \\ 
	$\Omega$ and $\Omega_0$ 	&  	the body in the current and reference configuration respectively \\
	$\partial \Omega$ and $\partial \Omega_0$ 	&  	the boundary of $\Omega$ and $\Omega_0$ respectively \\
	$\bfN$ and $\bfN_0$		&	the outward normals to $\partial \Omega$ and $\partial \Omega_0$ respectively \\
	$\bfF\equiv\nabla_{\bfx_0}\bfx$ 	&  	deformation gradient  \\
	$\bfC\equiv\bfF^T\bfF$ 	&  	Right Cauchy-Green deformation tensor \\
	$\bfE\equiv\half(\bfC-\bfI)$ 	&  	Green-Lagrangian strain tensor \\
	$\bfepsilon \equiv \half \left(\bfF + \bfF^T \right) - \bfI$ 	&  	linearized strain tensor\\
	 &  \\
	$W(\bfF)$ 	&  	classical elastic energy density \\
	$\Wcirc(\bfF,\phi)$ 	&  	modified elastic energy density  \\
	$\bfsigma = \frac{\partial W}{\partial \bfF} \text{ or }  \frac{\partial \Wcirc}{\partial \bfF}$ 	&  	First Piola-Kirchhoff stress \\
	$\bfv$	&	material velocity \\
	$v^\phi_n$ 	&  	normal velocity field for interface motion \\
	$\hat{v}_n^\phi$ 	&  	kinetic response function for interface normal velocity\\
	$G$ 	&  	nucleation/source term \\
	& \\
	$f$ 	&  	driving force \\
	$f_{bulk}$ 	&  	bulk driving force \\
	$f_{edge}$ 	&  	edge driving force \\
	 &  \\
	 $\llbracket \cdot \rrbracket$	&	The jump in a quantity across an interface
\end{tabular}

 For simplicity, we abuse notation and use interchangeably $W(\bfF)$ and $W(\bfE)$, and $\Wcirc(\bfF,\phi)$ and $\Wcirc(\bfE,\phi)$.
 For these quantities and $\bfsigma$, we use the same symbol both for the field and for the material response function.

$H_l(x)$ represents a function that resembles the Heaviside step function.
It transitions rapidly but smoothly from $0$ to $1$ and is symmetric about $x=0$, and $l$ represents the scale over which the function transitions.  
It is assumed to be sufficiently smooth for all derivatives in the paper to be well-defined.
The particular choice in this paper is $H_l(x) = \half \left( 1+ \tanh(x/l) \right)$.
The derivative of $H_l(x)$ is written $\delta_l(x)$, and is a smooth function that formally approximates the Dirac mass.

%%%%%%%%%%%%%%%%%%%%%
%%%%%%%%%%%%%%%%%%%%%
%%%%%%%%%%%%%%%%%%%%%
%%%%%%%%%%%%%%%%%%%%%

\section{Formulation}
\label{sec:formulation}

Similar to the standard phase-field models, we use two primary fields, $\bfx$ to describe the deformation, and $\phi$ to track the phase of the material.
The evolution of $\bfx$ is governed by balance of linear momentum, i.e. $\divergence \left( J^{-1} \bfF  \frac{\partial \Wcirc(\bfF,\phi)}{\partial \bfF} \right) = \rho \dot{\bfv}$.
We assume that the elastic energy density $W(\bfF)$ and the kinetic and nucleation relations for interfaces have been well-characterized and are available.
We aim to formulate a regularized-interface model that has the same elastic response and kinetic and nucleation behavior for interfaces.
We describe below how to set up $\Wcirc(\bfF,\phi)$ given $W(\bfF)$, the evolution equation for the kinetics and nucleation of $\phi$, and the thermodynamics associated with our model.

%%%%%%%%%%%%%%%%%%%%%
%%%%%%%%%%%%%%%%%%%%%
%%%%%%%%%%%%%%%%%%%%%
%%%%%%%%%%%%%%%%%%%%%

\subsection{Energetics}

We start by assuming that the classical strain energy density $W(\bfF)$ of the material is available, perhaps by calibrating to experiment or from lower-scale calculations.
We wish to obtain the modified energy density $\Wcirc(\bfF, \phi)$, which will have certain features that provide critical advantages for nucleation, yet stays largely faithful to $W(\bfF)$. 
Therefore, we require the following of $\Wcirc$: 
(i) away from energy barriers, $W(\bfF) = \Wcirc(\bfF, \phi)$ for the value of $\phi$ that corresponds to the appropriate phase; and 
(ii) $\Wcirc$ should be convex in the (linear or nonlinear) strain for a given value of $\phi$, preventing transformations purely through the evolution of $\bfF$, and hence enables the dynamics of $\phi$ to govern the phase transformation. 

We begin by considering two phases with characteristic strains given by $\bfE_1$ and $\bfE_2$.
We emphasize that these strains need {\em not} correspond to stress-free states, but that the tangent modulus at those points is positive-definite.
Then, define the functions for $A=1,2$:
\begin{equation}
	\psi_A(\bfE) = W(\bfE_A) + \underbrace{\bfsigma_T}_{\equiv \left. \frac{\partial W}{\partial\bfE} \right|_{\bfE_A}} : (\bfE - \bfE_A) +  \half  (\bfE - \bfE_A) : \underbrace{\bfC_T}_{\equiv \left. \frac{\partial^2 W}{\partial\bfE\partial\bfE} \right|_{\bfE_A}} : (\bfE - \bfE_A)
\end{equation}
$\psi_A$ approximates the behavior of $W$ near the states $\bfE_1$ and $\bfE_2$, and $\psi_A$ are convex in the arguments.

We now define the re-parametrized energy density $\Wcirc(\bfF,\phi)$:
\begin{equation}
	\Wcirc(\bfE,\phi) = \left(1 - H_l (\phi - 0.5) \right) \psi_1(\bfE) +  \left(H_l (\phi - 0.5) \right) \psi_2(\bfE)
\end{equation}
where $H_l$ is a smooth function that resembles the Heaviside (described in Section \ref{sec:notation}).
Therefore, $\Wcirc(\bfE,\phi\simeq 0) = \psi_1(\bfE)$ and $\Wcirc(\bfE,\phi\simeq 1) = \psi_2(\bfE)$; further, $\Wcirc(\bfE_1,0) = W(\bfE_1)$ and $\Wcirc(\bfE_2,1) = W(\bfE_2)$.

\begin{figure}[htb!]
	\begin{center}
	%\vspace{-3ex}
	\fbox{
	\begin{minipage}{155mm}
		\begin{center}
		\includegraphics[width=75mm]{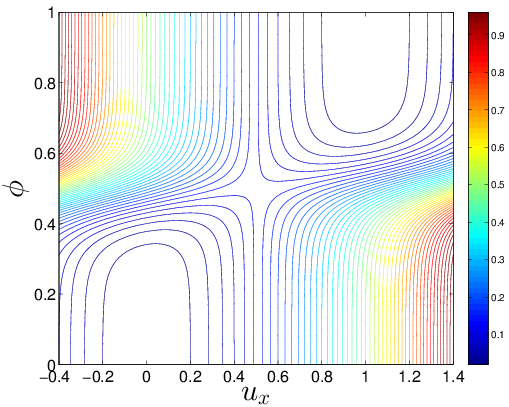}
		\includegraphics[width=75mm]{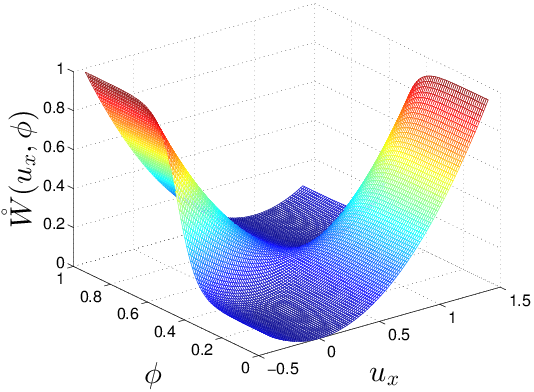}
		\\
		\includegraphics[width=75mm]{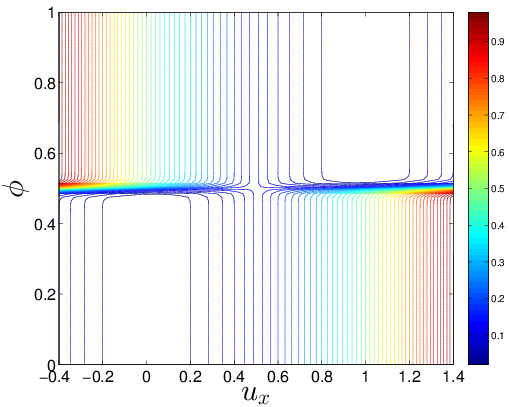}
		\includegraphics[width=75mm]{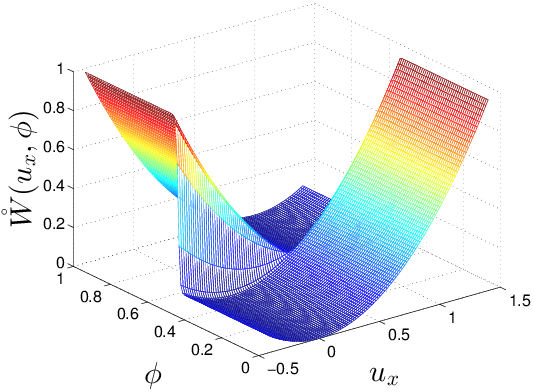}
	\caption{Contour and surface plots of the energy $\Wcirc(\bfE,\phi)$ assuming 1D with only a single strain component, using $l=0.1$ (above) and $l=0.01$ in the function $H_l$ used in the definition of the energy.}
	\label{fig:energy-landscape}
		\end{center}
	\end{minipage}
	}
%	\vspace{-5ex}
	\end{center}
\end{figure}

Fig. \ref{fig:energy-landscape} plots an example of $\Wcirc$ with a scalar strain measure to enable representation on paper.
The low-strain phase corresponds roughly to $0.0 < \phi< 0.3$, and the high-strain phase corresponds roughly to $0.7 < \phi< 1.0$.
The transition range is roughly $0.3 - 0.7$.
In general, $\phi$ is in the transition range only in the vicinity of an interface.
In a uniform phase region, $\phi$ will take on a value appropriate to that phase.

The key reason to re-formulate the energy is to achieve a clear separation between nucleation and kinetics.
In standard phase-field models, the form of the energetic coupling between $\phi$ and strain can lead to the nucleation of a new phase in a single-phase region purely through the kinetic equation, making the separation between nucleation and kinetics impossible.  
Here, $\Wcirc$ is independent of $\phi$ if it is outside the transition range; consequently, there can be no driving force for kinetic evolution when $\phi$ is outside this range, irrespective of the level of stress or other fields.
Consequently, away from an interface, $\phi$ will not evolve through the kinetic response irrespective of the local mechanical state.
Hence, the kinetic response cannot cause nucleation of a new phase in  a single-phase region.
The kinetic equation can  play a role only when $\phi$ is in the transition set in the vicinity of an interface, i.e., it can affect the behavior of an interface but not a uniform phase.
%Therefore, $\phi$ can evolve if it is in the transition range, as in the vicinity of an interface, i.e. by the kinetic motion of an interface.

While our energy does not permit nucleation, the conservation law that we set up below for interfaces permits us to specify precisely the nature of nucleation.
Further, the kinetic equation described there is multiplied by $|\nabla\phi|$, which suppresses the kinetic evolution of $\phi$ when it is spatially-uniform away from an interface.
Hence, away from an interface, the only way that $\phi$ can evolve is when the nucleation term -- that can be a function of stress or any other field -- in the interface conservation law is activated.

We remark on some features of this energy:
\begin{enumerate}

	\item Using $\bfsigma_T$, the tangent stress, allows us to position the characteristic strains $\bfE_1,\bfE_2$ at any point in strain-space where the tangent modulus is positive-definite.  
	These do not need to correspond to stress-free strains, and this property is useful in modeling situations such as stress-induced martensite, where one phase is observed only under stress.
	
	\item We have used only two terms in the Taylor expansion around the characteristic strains.  
	Increased fidelity to $W$ may be possible with use of additional terms, but this requires care to retain convexity in $\bfE$.  
	Our reason to have convexity is loosely based on obtaining unique solutions, and preventing phase transformations that occur without the evolution of $\phi$.  
	It is possible that convexity can be too strong an assumption \cite{antman2005nonlinear}.  
	However, this is a larger issue beyond the scope of our work here.

	\item $\Wcirc$ is faithful to the original energy $W$ near the characteristic strains, but less so further away.  
	At the barriers, it is completely at odds with $W$, because $\Wcirc$ is convex for fixed $\phi$.  
	However, passage over the barrier is governed by nucleation and kinetics, hence we do not need to accurately model it through $\Wcirc$.  
	We further note that the  driving force on an interface in sharp-interface classical elasticity is $f_{class} \equiv \llbracket W \rrbracket - \langle \bfsigma \rangle : \llbracket \bfF \rrbracket = W(\bfF^+) - W(\bfF^-) - \half (\bfsigma(\bfF^+) + \bfsigma(\bfF^-)) : (\bfF^+ - \bfF^-)$, where $\bfF^\pm$ are the limiting deformation gradients on either side of the interface \cite{abeyaratne2006evolution}.  
	Therefore, $f_{class}$ does not depend on the details of the barrier for given $\bfF^\pm$.
		
	\item Stresses and other applied fields can lead to the usual elastic deformations through the elastic response of each phase in any part of the domain, both near and away from interfaces.
	
	\item The energy density of the body includes a contribution $\half \epsilon |\nabla\phi|^2$.  As in standard phase-field models, this prevents the formation of singularly-localized interfaces.  
	Therefore, the total energy written in the reference configuration is $\int_{\Omega_0} \left[ \Wcirc(\bfF,\phi) + \half \epsilon |\bfF^{-T} \nabla_{\bfx_0} \phi_0|^2 \right] \ d\Omega_0$, up to boundary terms.
	For simplicity, we approximate the gradient contribution in the reference by $\int_{\Omega_0} \half \epsilon |\nabla_{\bfx_0} \phi_0|^2 \ d\Omega_0$

\end{enumerate}

%%%%%%%%%%%%%%%%%%%%%
%%%%%%%%%%%%%%%%%%%%%
%%%%%%%%%%%%%%%%%%%%%
%%%%%%%%%%%%%%%%%%%%%

\subsection{Evolution Law}
\label{sec:evolution-law}

\begin{figure}[htb!]
	\begin{center}
	%\vspace{-3ex}
	\fbox{
	\begin{minipage}{155mm}
		\begin{center}
		\includegraphics[width=75mm]{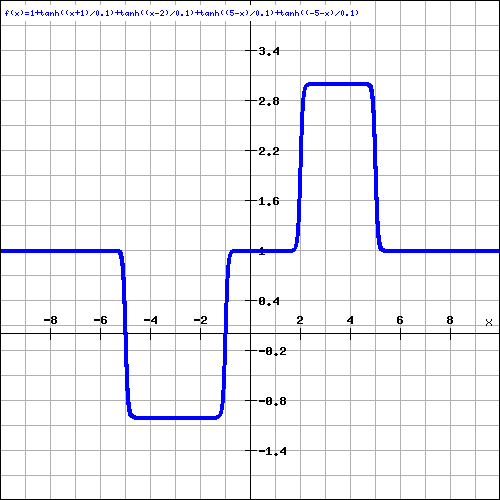}
		\includegraphics[width=75mm]{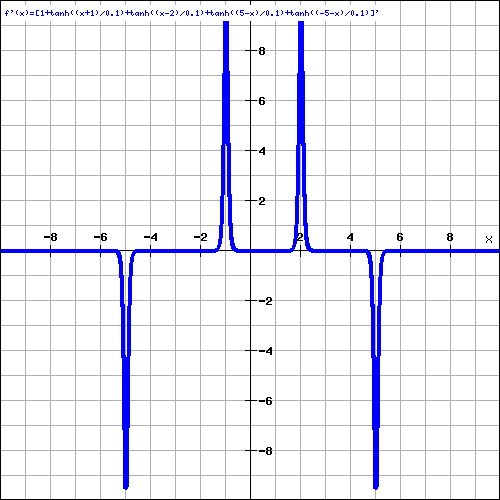}
	\caption{Left: a field $\phi$ with a number of interfaces.  Right: $\nabla\phi$ provides a measure of (signed) interface density per unit length.}
	\label{fig:interface-density-schematic}
		\end{center}
	\end{minipage}
	}
%	\vspace{-5ex}
	\end{center}
\end{figure}

%Our formulation makes an important departure in the kinetics of $\phi$.
Our starting point in formulating the evolution of $\phi$ is to note that $\nabla\phi$ provides, roughly, a measure of the number or ``strength'' of the interfaces in the $\phi$ field per unit length, Fig. \ref{fig:interface-density-schematic}.

In general, given a field $\phi(\bfx)$ with localized transitions between constant values, we can readily locate the interfaces in this field using $\nabla\phi$.
Further, if we pick any curve and integrate $\nabla\phi$ along this curve, the value that we obtain provides a measure of the net number of interfaces that we have traversed, assuming that all interfaces have the same ``strength''.
If the interfaces have different strengths, we obtain a measure of the net interface strength that we have traversed.
This physical picture provides the intuition behind what follows, but it also expresses the simple fact that if we have a single-valued field $\phi$, then integrating the gradient is simply the difference between $\phi$ at either end of the curve.

%To make this intuitive picture more precise, we appeal to the differential geometric notion of 
%That is, they are objects that are integrated along curves.
The geometric picture is roughly related to gradients of fields being so-called 1-forms, i.e., they are objects that are naturally integrated along curves \cite{marsden-hughes}.
Analogies of this are commonplace in elasticity: e.g., the divergence of a field is a 3-form and is naturally integrated over volumes, as is used in the conservation laws for mass, momentum, and energy.
The curl of a field is a 2-form and is naturally integrated over surfaces, as is used in proving the single-valuedness of a deformation field corresponding to $\curl \bfF = {\bf 0}$, as well as in dislocation mechanics where $\curl \bfF$ provides an areal density of dislocation line defects \cite{acharya2001model}.

\begin{figure}[htb!]
	\begin{center}
	%\vspace{-3ex}
	\fbox{
	\begin{minipage}{155mm}
		\begin{center}
		\includegraphics[width=75mm]{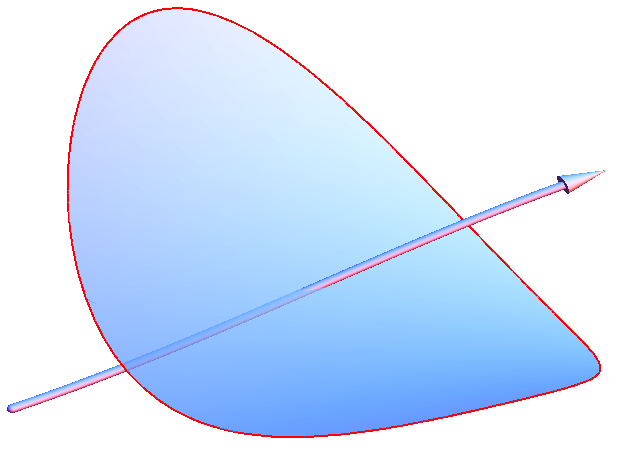}
		%\hspace{5mm}
		\includegraphics[width=75mm]{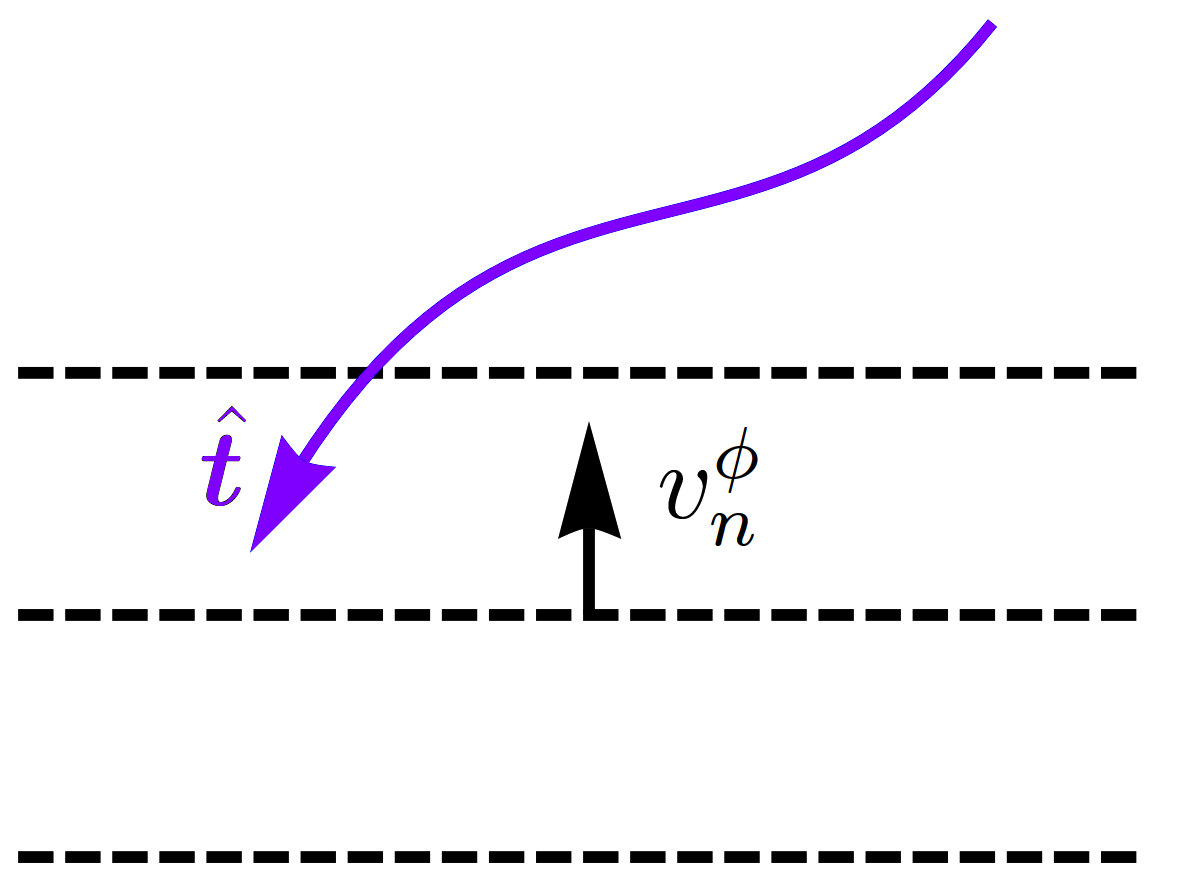}
		\caption{\small  Left: A schematic representation of an interface threading a curve.  Right: The flux of interfaces at one end of the curve. $\hat{\bft}$ is the tangent to the end of the curve, and $v_n^\phi$ is the interfacial normal velocity field.  The relative velocity of the interface with respect to the direction $\hat{\bft}$ is $\frac{v_n^\phi}{|\hat{\bfn} \cdot \hat{\bft}|}$, and is defined as the distance along the direction $\hat{\bft}$ traversed by the threading interface in unit time.}
		\label{fig:balance-law}
		\end{center}
	\end{minipage}
	}
%	\vspace{-5ex}
	\end{center}
\end{figure}

Given this notion of the interface density field $\nabla\phi$, we then formulate a balance law (see Fig. \ref{fig:balance-law}).
Let the interfaces have a normal velocity given by the field $v_n^\phi$; note that this velocity is distinct from the material velocity $\dot{\bfu}$.
Now consider a curve $C(t)$ in space.
This curve ``threads'' or passes through some number of interfaces.
Further, interfaces are entering or exiting at one end and leaving at the other end of the curve due to their motion given by the field $v_n^\phi$.
The conservation principle is that the net increase in the number of interfaces that are threaded by $C(t)$ is a balance between interfaces entering, interfaces exiting, and interfaces being created and destroyed by a sources and sinks.
\begin{align*}
	\deriv{}{t}\left\{
	\begin{array}{c}
		\text{Number of } \\
		\text{interfaces within} \\
		\text{the curve}
	\end{array} \right\}
	 = 
	\left\{
	\begin{array}{c}
		\text{Number of} \\
		\text{interfaces entering}\\
		\text{the curve}
	\end{array} 
	-
	\begin{array}{c}
		\text{Number of} \\
		\text{interfaces leaving} \\
		\text{the curve}
	\end{array} \right\} + 
	\left\{
	\begin{array}{c}
		\text{interfaces} \\
		\text{generated within}\\
		\text{curve}
	\end{array}
	\right\}
\end{align*}
Using that this must hold for every curve $C(t)$ enables us to localize the balance law.

The flux of interfaces through the ends of the curve $C(t)$ can be computed by referring to Fig. \ref{fig:balance-law}.
Let $\hat{\bft}$ be the unit tangent to the end of the curve, and $\hat{\bfn} \equiv \frac{\nabla\phi}{|\nabla\phi|}$ the unit normal to the interface.
Then the flux can be written
\begin{equation}
	|\nabla\phi| \frac{\nabla\phi\cdot\hat{\bft}}{|\nabla\phi\cdot\hat{\bft}|} \frac{v_n^\phi}{\hat{\bfn}\cdot\hat{\bft}} |\hat{\bft}\cdot\hat{\bfn}|
\end{equation}
The first term represents the strength of the interface; the second term is simply $+1$ if the interface enters and $-1$ if it leaves; the third term is the velocity of the interface projected onto the $\hat{\bft}$ direction to obtain the velocity relative {\em relative} to the curve direction, i.e. the distance along the $\hat{\bft}$ direction traversed by the interface in unit time; and the fourth term picks out only the portion of the flux that is threading the curve by moving along $\hat{\bft}$.

An alternate picture is to consider that $\nabla\phi \cdot \hat{\bft}$ is the (signed) interface density along the direction $\hat{\bft}$, and $\frac{v_n^\phi}{\hat{\bfn}\cdot\hat{\bft}}$ is the velocity of the interface relative to the direction $\hat{\bft}$, so the flux is simply:
\begin{equation}
	\nabla\phi \cdot \hat{\bft} \frac{v_n^\phi}{\hat{\bfn}\cdot\hat{\bft}}
\end{equation}
Both expressions for the flux are identical, and simplify to $|\nabla\phi| v_n$ when we substitute $\hat{\bfn} \equiv \frac{\nabla\phi}{|\nabla\phi|}$\footnote{A. Acharya gave a different argument for why the flux must have this final form that guided us,  and also many useful discussions on Section \ref{sec:evolution-law}.}.

Defining the interface density $\bfalpha := \nabla\phi$, we have:
\begin{equation}
	\underbrace{\left. |\bfalpha| v_n^\phi \right|_{C^-}^{C^+}}_{\text{net flux of interfaces}}
	= \underbrace{\deriv{}{t} \int\limits_{C(t)}{\bfalpha} \ d\bfx}_{\text{increase in number of interfaces threaded}}
	-  \underbrace{\int\limits_{C(t)} {\bfS } \ d\bfx}_{\text{source of new interfaces}}
\end{equation}
We can transform $\left. |\bfalpha| v_n^\phi \right|_{C^-}^{C^+} = \int \limits_{C(t)} \nabla(|\bfalpha| v_n^\phi) \ d\bfx$.

Using that $C(t)$ is a material curve, we can write the mapping $d\bfx = \bfF \ d\bfx_0 $ between the infinitesimal elements of $C(t)$ and its image $C_0$ in the reference.
Then, the time derivative can be transformed to $\deriv{}{t} \int\limits_{C_0}{\bfalpha\bfF \ d\bfX}  = \int\limits_{C(t)}{(\dot{\bfalpha} + \bfalpha\bfL) \ d\bfx}$ where $\bfL$ is the spatial velocity gradient.

This lets us localize to obtain:
\begin{equation}
	\label{eqn:balance-1}
	\dot{\bfalpha} = \nabla(|\bfalpha | v_n^\phi) + \bfS(\bfx,t) - \bfalpha\bfL
\end{equation}
Noting that the source is constrained by the above equation to be of the form, i.e. $\bfS = \nabla G + \bfalpha\bfL$, we can integrate the above equation to obtain:
\begin{equation}
 	|\nabla\phi| v_n^\phi + G = \dot{\phi}
\end{equation}

The nucleation / source term $G$ can be an arbitrary function of any of the fields in the problem, up to some weak limitations imposed by thermodynamics (discussed below).
The term $\bfS$ must be a gradient up to the term $ \bfalpha\bfL$, and represents the fact that in a single-valued field $\phi$, interfaces that nucleate must either terminate on the boundary or close on themselves but cannot end in the interior of the body.

We note certain important features of the evolution law that we have posed:
\begin{enumerate}

	\item The kinetics of existing interfaces is constitutively prescribed through the interface velocity field $v_n^\phi$, which can be a function of stress, strain, as well as any other relevant quantity, such as the work-conjugate to $\phi$ (the Eshelby / configurational force).
This makes it trivial to obtain complex kinetics; for instance, if the interface is pinned below a critical value of the stress, we simply prescribe that $v_n^\phi$ is zero at all spatial points where the stress is below the critical value.
Similarly, other kinds of nonlinear and complex kinetics can be readily incorporated.

	\item Nucleation of new interfaces is prescribed through the source term in the balance law, and provides precise control on the nucleation process.  For instance, we can prescribe that a source is activated only beyond some critical stress and stress rate; thus, for example, it is straightforward to model a nucleation process in which the critical nucleation stress is extremely sensitive to strain rate.  In addition, the activation of the source can be completely heterogeneous and vary vastly from point to point.

	\item The appearance of $|\grad\phi|$ in the evolution is important to separate kinetics from nucleation: if we have a large driving force in a uniform phase far away from an interface, $|\grad\phi|$ will remain $0$ and therefore will not allow the kinetic term to play a role irrespective of driving force, stress, etc.

\end{enumerate}

We note that an analogous idea to the conservation principle stated above is used in \cite{acharya2001model} to obtain an evolution law for dislocations, and recently in \cite{acharya-dayal-qam2014, pourmatin-dayal-qam2014} for disclination dynamics.
In  \cite{acharya2001model}, by connecting the dislocation density to $\curl \bfF$ and using the physical picture that these are line defects, a conservation law is posed by using that the rate of change of dislocations intersecting an arbitrary area element is related to the net flux of intersecting dislocations and the creation of intersecting dislocations.
The conservation law that we have posed in this work builds on this picture, and the key point of departure of our work is the idea that the balance principle from \cite{acharya2001model} can be extended from line defects detected by $\curl \cdot$ to interfacial defects that are detected by  $\grad \cdot$.
A further use of this approach are the standard continuum balances of mass, momentum, energy all work with volumetric densities, and the appropriate quantity to be integrated over a volume is $\divergence \cdot$.

In Appendix \ref{sec:noether}, we examine the relation between this conservation principle and Noether's theorem.

\subsubsection{Balance Law in the Reference Configuration}

The entire argument above was posed in the current configuration.
Since the field $\phi$ relates to the state of material particles, it would be physically reasonable to alternatively pose the balance principle in the reference configuration.
We examine this approach briefly.

In this section, quantities with subscripts of $0$ denote referential objects.
$\bfx_0$ is the referential pre-image of the material particle $\bfx(\bfx_0,t)$.
We make the natural transformation that $\phi$ is the same in the reference and the current for a given material particle at a given time: $\phi_0(\bfx_0(\bfx,t),t) = \phi(\bfx,t)$.
From standard manipulations of continuum mechanics, it follows that $\bfalpha \bfL + \dot{\bfalpha} = \bfF^{-T} \dot{\bfalpha}_0$.
We further make the identification that $\bfS = \bfF^{-T}\bfS_0 \Leftrightarrow G = G_0$.

Substituting in the balance principle (\ref{eqn:balance-1}), we can write:
\begin{equation}
	\bfF^{-T} \dot{\bfalpha}_0 = \bfF^{-T}\bfS_0 + \bfF^{-T}\nabla_{\bfx_0} (|\bfalpha | v_n^\phi)
\end{equation}
We have also used above that $\nabla_{\bfx} = \bfF^{-T} \nabla_{\bfx_0}$.

The natural transformation induced on the interface velocity field is obtained by requiring $|\bfalpha | v_n^\phi = |\alpha_0| v_{n0}^\phi$.
The result is the non-standard transformation $v_{n0}^\phi \hat{\bfn}_0 = \bfF^{-1} v_n^\phi \hat{\bfn}$.
This is deceptively simple, because the transformation between $\hat{\bfn}_0$ to $\hat{\bfn} \equiv \nabla\phi / |\nabla\phi|$ is not as a standard normal to a material surface.
The final result can be compactly written $v_{n0}^\phi = v_n^\phi \left( \hat{\bfn}_0 \bfF^{-1}  \bfF^{-T} \hat{\bfn}_0 \right)^\half$.

Using this further transformation of $v_n^\phi$, we obtain the interface balance in the reference configuration:
\begin{equation}
	\dot{\bfalpha}_0 = \bfS_0 + \nabla_{\bfx_0} (|\bfalpha_0| v_{n0}^\phi)
\end{equation}
This can be readily integrated once to obtain
\begin{equation}
	\dot{\phi}_0 = G_0 + |\bfalpha_0| v_{n0}^\phi
\end{equation}

There are two practical, though minor, advantages to the referential form of the balance principle.
First, frame-indifference is readily seen to be satisfied.
Second, the interpretation of $\bfS_0 = \nabla_{\bfx_0} G_0$ is simpler without the additional terms from the material derivative.

%%%%%%%%%%%%%%%%%%%%%
%%%%%%%%%%%%%%%%%%%%%
%%%%%%%%%%%%%%%%%%%%%
%%%%%%%%%%%%%%%%%%%%%

\subsection{Thermodynamics and Dissipation}
\label{sec:dissipation}

Following established ideas, we use the statement of the second law that the dissipation must be non-negative for every motion of the body to find the thermodynamic conjugate driving forces for kinetics and nucleation.
The dissipation is defined as the deficit between the rate of external work done and the increase in stored energy:
\begin{equation}
	\mathcal{D} = \text{External working} 
		- \deriv{}{t} \left( \int_{\Omega_0} \left[ \Wcirc(\bfF,\phi) + \half \epsilon \frac{\partial \phi_0}{\partial x_{0i}} \frac{\partial \phi_0}{\partial x_{0i}} \right] \ d\Omega_0 + \half \int_{\Omega_0} \rho_0 V_{0i} V_{0i} \ d\Omega_0 \right)
\end{equation}
This can be manipulated to find the conjugates to $v_n^\phi$ and $G$.
\begin{equation}
	\mathcal{D} 
		 = \text{External working} 
			- \int_{\Omega_0} \left[ \frac{\partial \Wcirc}{\partial F_{ij}} \deriv{F_{ij}}{t}  +  \frac{\partial \Wcirc}{\partial \phi} \deriv{\phi}{t} + \epsilon \frac{\partial \phi_0}{\partial x_{0i}} \deriv{}{t} \frac{\partial \phi_0}{\partial x_{0i}} \right] \ d\Omega_0 
			- \int_{\Omega_0} \rho_0 V_{0i} \dot{V}_{0i} \ d\Omega_0
\end{equation}
Using $\deriv{F_{ij}}{t} = \frac{\partial V_{0i}}{\partial x_{0j}}$ and integration-by-parts:
\begin{equation}
\begin{split}
	\mathcal{D} 
		 = & \quad \text{External working} 
			- \int_{\Omega_0} \frac{\partial }{\partial x_{0j}} \left( \frac{\partial \Wcirc}{\partial F_{ij}} V_{0i} \right)  \ d\Omega_0
			+  \int_{\Omega_0} V_{0i} \left( \frac{\partial }{\partial x_{0j}} \frac{\partial \Wcirc}{\partial F_{ij}} - \rho_0 \dot{V}_{0i} \right)  \ d\Omega_0 \\
			& \quad -   \int_{\Omega_0} \left[ \frac{\partial \Wcirc}{\partial \phi} \deriv{\phi}{t} + \epsilon \frac{\partial \phi_0}{\partial x_{0i}} \deriv{}{t} \frac{\partial \phi_0}{\partial x_{0i}} \right] \ d\Omega_0 
\end{split}
\end{equation}
The first integral above is exactly balanced by the external work done by boundary tractions\footnote{We assume that there is no work done on $\phi$ at the boundary for now.  We revisit this in the section on boundary kinetics.}.
The second integral is identically zero from balance of linear momentum.
Therefore, the dissipation simplifies to:
\begin{equation}
	\mathcal{D} 
		 =  \int_{\Omega_0} \left[ -\frac{\partial \Wcirc}{\partial \phi}  + \epsilon \frac{\partial^2 \phi_0}{\partial x_{0i} \partial x_{0i}} \right] \deriv{\phi_0}{t} \ d\Omega_0 
		 	- \int_{\partial \Omega_0} \epsilon \frac{\partial \phi_0}{\partial x_{0i}} N_{0i} \deriv{\phi}{t} \ d\partial\Omega_0 
\end{equation}
For now, we assume the boundary condition $\nabla_{\bfx_0} \phi_0 \cdot \bfN_0 = 0$ thereby removing the boundary contribution to $\mathcal{D}$.
We substitute the balance law $\dot{\phi}_0 = G_0 + |\bfalpha_0| v_{n0}^\phi$, to get:
\begin{equation}
	\mathcal{D} 
		 =  \int_{\Omega_0} \left[ -\frac{\partial \Wcirc}{\partial \phi}  + \epsilon \frac{\partial^2 \phi_0}{\partial x_{0i} \partial x_{0i}} \right] \left( G_0 + |\nabla_{\bfx_0} \phi| v_{n0}^\phi \right) \ d\Omega_0 
\end{equation}
Defining the driving force $f := -\left[ -\frac{\partial \Wcirc}{\partial \phi}  + \epsilon \frac{\partial^2 \phi_0}{\partial x_{0i} \partial x_{0i}} \right]$, we get:
\begin{equation}
	\mathcal{D} 
		 =  \int_{\Omega_0}  f \left( G_0 + |\nabla_{\bfx_0} \phi| v_{n0}^\phi \right) \ d\Omega_0 
\end{equation}
To ensure that dissipation is always non-negative, we need both $f G_0$ and $f  |\nabla_{\bfx_0} \phi| v_{n0}^\phi$ to be non-negative.
These are fairly easy conditions to satisfy in a material model.
For kinetics, we choose the constitutive response of the form $v_{n0}^\phi = \frac{f}{|f|} \hat{v}_{n0}^\phi(|f|, \ldots)$, where the constitutive response function $\hat{v}_{n0}^\phi$ can be any non-negative function of the arguments, and the list of arguments can consist of any of the field variables, as well as possibly nonlocal quantities.
A similarly weak requirement holds for nucleation.

%%%%%%%%%%%%%%%%%%%%%
%%%%%%%%%%%%%%%%%%%%%
%%%%%%%%%%%%%%%%%%%%%
%%%%%%%%%%%%%%%%%%%%%

\subsection{Formal Sharp-Interface Limit}
\label{sec:sharp_limit}

We consider briefly the formal limit of the driving force in the sharp-interface limit $\epsilon = 0$.
We emphasize that this is not rigorous, as the limit $\epsilon \rightarrow 0$ involves the delicate singular perturbation of a nonlinear hyperbolic equation.

We start with the dissipation expression ignoring the nucleation contribution:
\begin{equation}
	\mathcal{D} = \int_{\Omega_0} -\frac{\partial \Wcirc}{\partial \phi} |\nabla_{\bfx_0} \phi| v_{n0}^\phi  \ d\Omega_0 
				= \int_{\Omega_0} -\frac{\partial \Wcirc}{\partial \phi} \nabla_{\bfx_0}\phi \underbrace{\left(\frac{ v_{n0}^\phi \nabla_{\bfx_0}\phi}{| \nabla_{\bfx_0}\phi |} \right)}_{=: \bfv_{n0}^\phi} \ d\Omega_0 
\end{equation}
Adding and subtracting $\int_{\Omega_0} -\frac{\partial \Wcirc}{\partial \bfF} : \nabla_{\bfx_0}\bfF \cdot \bfv_{n0}^\phi  \ d\Omega_0$, and also using that $\nabla_{\bfx_0} \Wcirc = \frac{\partial \Wcirc}{\partial \bfF} : \nabla_{\bfx_0}\bfF + \frac{\partial \Wcirc}{\partial \phi} \nabla_{\bfx_0}\phi$, we have
\begin{equation}
	\mathcal{D} = - \int_{\Omega_0} \nabla_{\bfx_0} \Wcirc \cdot \bfv_{n0}^\phi  \ d\Omega_0 
				+\int_{\Omega_0} \frac{\partial \Wcirc}{\partial \bfF} : \nabla_{\bfx_0}\bfF \cdot \bfv_{n0}^\phi  \ d\Omega_0 
\end{equation}

We now further assume the following:
(i) the evolution is quasistatic, i.e. inertia is negligible;
(ii) the phase boundary is flat and the fields are one-dimensional; and
(iii) $\bfv_{n0}^\phi$ is constant in space.
The assumptions (i) and (ii) allow us to assume that $\bfsigma = \frac{\partial \Wcirc}{\partial \bfF}$ is constant in space and can be pulled out of the integral.
Assumption (iii) is a direct consequence of assuming steady motion of the phase boundary as a traveling wave (Section \ref{sec:trav-waves}), and allows us to pull $\bfv_{n0}^\phi$ out of the integral.

With these assumptions, we can write:
\begin{equation}
	\mathcal{D} = - \left( \underbrace{\int_{\Omega_0} \nabla_{\bfx_0} \Wcirc \ d\Omega_0}_{\llbracket \Wcirc \rrbracket} - \bfsigma : \underbrace{\int_{\Omega_0} \nabla_{\bfx_0}\bfF \ d\Omega_0}_{\llbracket \bfF \rrbracket } \right) \cdot \bfv_{n0}^\phi
\end{equation}
which is identical to the driving force obtained by Abeyaratne and Knowles \cite{abeyaratne2006evolution} in the quasistatic setting.
Consequently, it is reasonable to further expect that the Maxwell stress is correctly captured by our energy.

%%%%%%%%%%%%%%%%%%%%%
%%%%%%%%%%%%%%%%%%%%%
%%%%%%%%%%%%%%%%%%%%%
%%%%%%%%%%%%%%%%%%%%%

\section{Traveling Waves in One Dimension}
\label{sec:trav-waves}

We investigate the behavior of traveling wave solutions in our model.
These correspond to steadily moving interfaces.

For simplicity, we use a one-dimensional setting with linearized kinematics.
For $\Wcirc$, we use the form:
\begin{equation}
	\Wcirc(u_x,\phi) = \Big(1-H_l(\phi-0.5)\Big)\half C (u_x-\eps_1)^2 + H_l(\phi-0.5) \half C (u_x - \eps_2)^2
\end{equation}
We use $\eps_1 = 0$ and $\eps_2 = 1$.
The stress is $\sigma = \parderiv{\Wcirc}{\left(u_x\right)} = C \left( u_x - H_l (\phi-0.5)\right)$, and the driving force is $f = \delta_l(\phi-0.5) \cdot \left(u_x - 0.5\right) + \epsilon \phi_{xx}$.

We search for traveling wave solutions of the form $u(x,t) = U(x-Vt)$ and $\phi(x,t) = \Phi(x-Vt)$ for a few different given kinetic relations.
We substitute these into the balance of linear momentum and the evolution equation.
Below, $U'$ and $\Phi'$ denote derivatives of $U$ and $\Phi$.
We assume that the kinetic response $\hat{v}_n^\phi (\ldots)$is a function of only the driving force $f$, and further that it is a monotone function and hence invertible.

From momentum balance, we obtain:
\begin{equation}
\begin{split}
\label{eq:twave-gradu}
	\rho u_{tt} = \sigma_x  \Rightarrow & V^2 \rho U'' = C \Big\{ U'' - \delta_a(\Phi-0.5)\Phi' \Big\}  \Rightarrow (1-M^2)U'' = \delta_a(\Phi-0.5)\Phi' \\
				& \Rightarrow U' = \frac{H_a(\Phi-0.5) + \tilde{c}_2}{1-M^2} =  \frac{H_a(\Phi-0.5)}{1-M^2} + c_2 
\end{split}
\end{equation}
 where $c_2 \equiv \frac{\tilde{c}_2}{1-M^2}$ is a constant of integration, and $M$ is the Mach number.
We see that as $M \to 1$, the derivative is unbounded unless $H_a(\phi-0.5) + \tilde{c}_2 = 0$.
The latter condition requires that $\Phi$ is constant in space, implying that only elastic waves and not phase interfaces are permitted at $M=1$.
As expected, this limitation is a consequence of momentum balance alone.

Next, from the evolution equation, we obtain:
\begin{equation}
 \label{eq:twave-phi}
	\dot{\phi} = |\phi_x| \hat{v}_n^\phi(|f|) \Rightarrow -V \Phi' = |\Phi'| v_n^\phi
\end{equation}
(\ref{eq:twave-phi}) implies that the interface velocity field $v_n^\phi$ is constant in space and time.

Using further that the kinetic response is a monotone function of $f$ implies that the driving force field has to be a constant in space and time.
Therefore, $f = \delta_l(\Phi-0.5) \cdot \left(U' - 0.5\right) + \epsilon \Phi'' = \const$, and substituting for $u'$ from \eqref{eq:twave-gradu} gives an ODE in $\Phi$:
\begin{equation} 
\label{eq:twave-E5}
	f = \const = \epsilon \Phi'' + \delta_a(\Phi-0.5) \cdot \Big( \frac{H_a(\Phi-0.5)}{1-M^2} - \half -c_2 \Big)
\end{equation}
Given a value of $M$ or alternatively $V$, we can solve this equation to obtain $\Phi$.
Also, given $M$, the value of $f$ is obtained from the assumed kinetic response.

\eqref{eq:twave-E5} is a nonlinear ODE because of $H_a(\Phi-0.5)$ and $\delta_a(\Phi-0.5)$.
So we seek to find approximate solutions numerically using finite differences and least-squares minimization following \cite{dayal-bhatta-jmps2006}.
Divide the domain of length $L = 1$ into $N$ elements each of length $\Delta x= L/N$; the $N+1$ grid points are denoted $x_i$.
Discretize the ODE with as:
\begin{equation}
	g(x_i):=\epsilon \frac{\Phi(x_{i+1}) - 2\Phi(x_i) + \Phi(x_{i-1})}{(\Delta x)^2} + \delta_a(\Phi(x_i)-0.5)\Big\{ \frac{H_a(\Phi(x_i)-0.5)}{1-M^2} - \half -c_2 \Big\} - f
\end{equation}
Define the residue $\calR := \sum\limits_{i=2}^{N}|g(x_i)|^2$. 
To find $\Phi$, we minimize $\calR$ with respect to the nodal values $\Phi_i := \Phi(x_i)$; at the completion of minimization, we evaluate $\calR$ to ensure that it is near $0$ and we have not found a local minimum.
To prevent the solution algorithm from finding trivial single-phase solutions, we fix $\Phi |_{x=0.5} = 0.5$.

We note that there is an additional constant $c_2$ that is unknown.
From classical sharp-interface analyses, we expect that the combination of momentum balance and kinetic relation should give us a unique solution in \eqref{eq:twave-E5} once $M$ is fixed.
Examining \eqref{eq:twave-gradu}, we can infer that it is related to the strains / stresses at $\pm\infty$, but it is not clear how exactly to find this explicitly.
Therefore, we simply treat $c_2$ as an additional variable over which to minimize $\calR$.

\begin{figure}[htb!]
\begin{center}
	\fbox{
	\begin{minipage}{185mm}
		\begin{center}
			\includegraphics[width = 60mm]{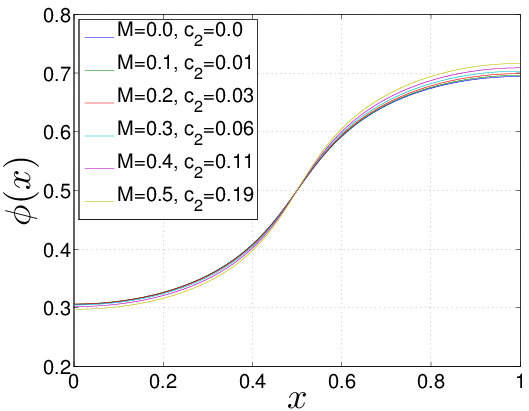}
			\includegraphics[width = 60mm]{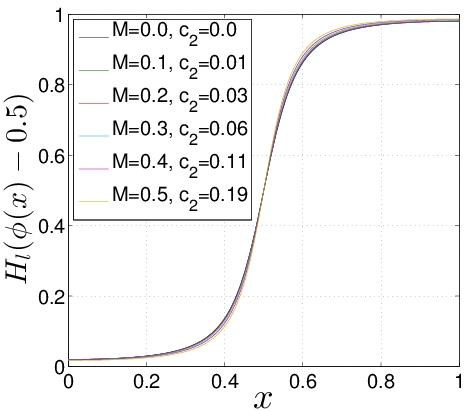}
			\includegraphics[width = 60mm]{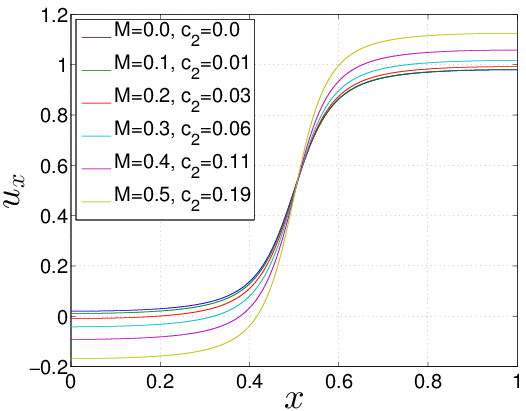}
			\caption{Plots of $\Phi,H_l(\Phi-0.5),U'$ respectively  for traveling wave solutions with different values of $M$ using linear kinetics.}
			\label{fig:tw-profiles}
		\end{center}
	\end{minipage}
	}
\end{center}
\end{figure}

Fig. \ref{fig:tw-profiles} plots the solutions for $U$ and $\Phi$ for a linear kinetic relation.
Qualitatively similar profiles are obtained for a quadratic kinetic relation.

\begin{figure}[htb!]
\begin{center}
	\fbox{
	\begin{minipage}{125mm}
		\begin{center}
			\includegraphics[width = 120mm]{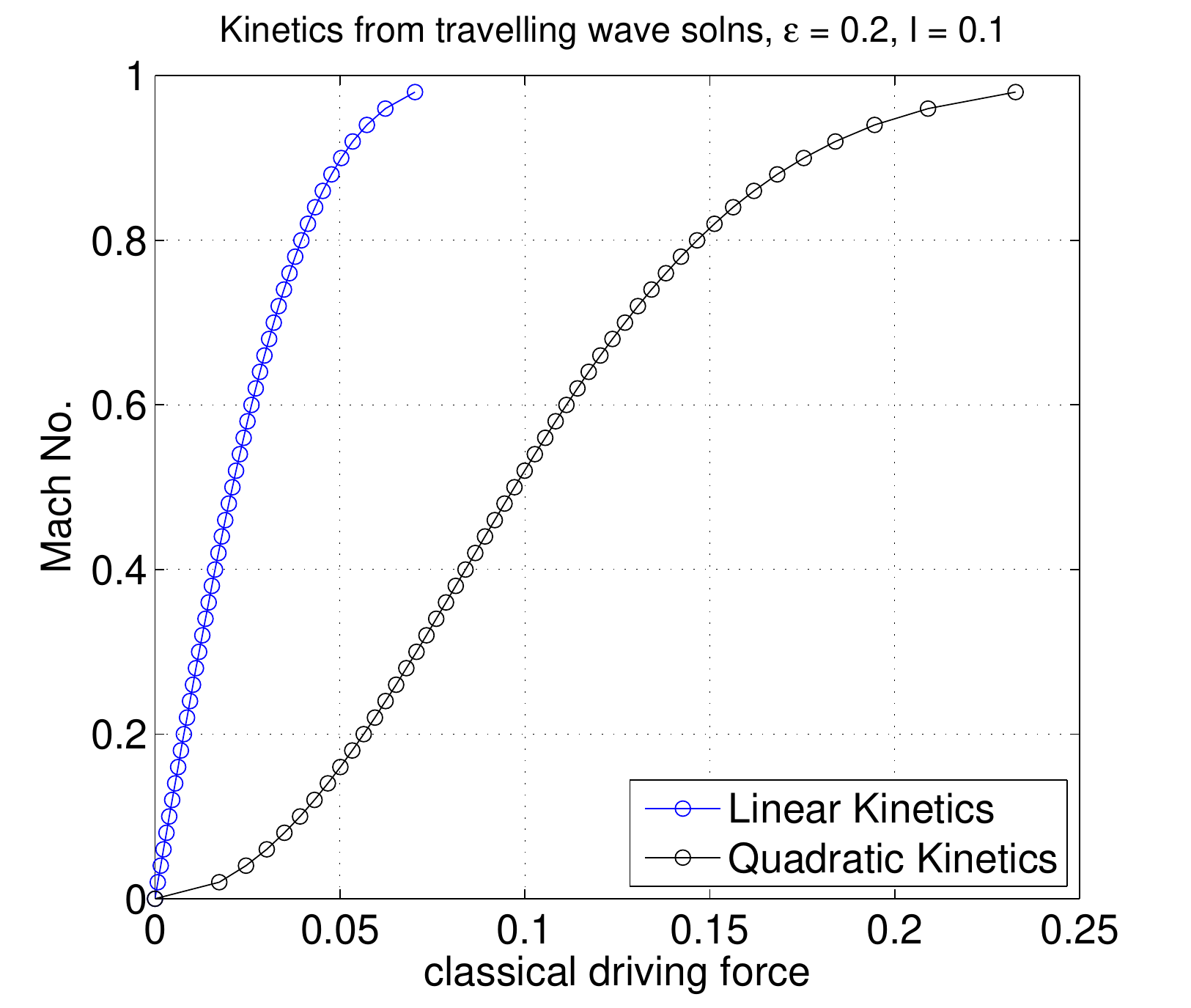}
			\caption{$M$ vs. classical driving force for linear kinetics and quadratic kinetics, derived from the traveling wave solutions. The classical driving force is $f=\jump{U} - \langle \sigma \rangle \jump{u_x}$, and is evaluated using the values of fields at the edges of the domain.}
			\label{fig:tw-lin-quad}
		\end{center}
	\end{minipage}
	}
\end{center}
\end{figure}

We extract $(U')^\pm$, the limiting constant strains far from the interface, and use these to evaluate the classical driving force $\jump{\Wcirc} - \langle \sigma \rangle \jump{U'}$.
Fig. \ref{fig:tw-lin-quad} plots the {\em classical} driving force (not $f$) against $M$ for solutions obtained for linear and quadratic kinetic relations.
We find that the kinetic response that is specified through the response function $\hat{v}_n^\phi$ is reproduced in terms of the classical driving force.
This supports the belief that our model provides the advantages of both the sharp-interface and the regularized-interface models without the disadvantages of either.
We note that the classical kinetic relation deviates from the kinetic response function as $M \to 1$, but this is expected from linear momentum balance.

We emphasize an interesting difference between our model and existing phase-field models.
In our model, the driving force field and the interface velocity field are both constant in space.
Therefore, the relation between them is a simple relation between two scalar quantities, and the notion of a kinetic relation is well-defined.
In existing phase-field models, the driving force field is large near an interface and goes to zero away from the interface, i.e. it is a function of location.
Therefore, there is no obvious unique scalar measure of the driving force that one can extract from this field; one could use the maximum value, or the mean value in some region, and so on.
In this perspective, our model has the advantage that it has a closer link to the classical continuum model because there is a unique and obvious relation between driving force and interface velocity.

%%%%%%%%%%%%%%%%%%%%%
%%%%%%%%%%%%%%%%%%%%%
%%%%%%%%%%%%%%%%%%%%%
%%%%%%%%%%%%%%%%%%%%%

\section{Dynamics of Interfaces in One Dimension}
\label{sec:1d-dynamics}

We examine the kinetics of phase interfaces through direct dynamic simulations.
We solve linear momentum balance along with the evolution equation for $\phi$ in various configurations and with various choices for the kinetic response.

\subsection{Examining Various Induced Kinetic Relations}
\label{sec:1d-kin-relns}

We work with the elastic material as follows:
\begin{align}
	\Wcirc(u_x,\phi) &= \Big(1-H_a(\phi-0.5)\Big)\half C (u_x-\eps_1)^2 + H_a(\phi-0.5) \half C (u_x - \eps_2)^2 \\
	\sigma &= \parderiv{\Wcirc}{(u_x)} = \Big(1-H_a(\phi-0.5)\Big) C (u_x-\eps_1) + H_a(\phi-0.5) C (u_x - \eps_2) \\
	f &= \delta_a(\phi-0.5)\Big( \half C (u_x-\eps_1)^2 - \half C (u_x-\eps_2)^2 \Big) + \epsilon \phi_{xx} \\
	\rho\ddot{u} &= \sigma_x \\
	\dot{\phi} &= |\phi_x|v_n^\phi
\end{align}
The stored energy density near each well is taken to be quadratic with wells at $\varepsilon_1 = 0$ and $\varepsilon_2 = 1$.

We test three different kinetic laws:
\begin{equation}
\label{eqn:3-kin-laws}
	\hat{v}_n^\phi = \left\{ \begin{array}{l l}
			\mathrm{sign}(f) \kappa  |f| \quad &\text{linear kinetics} \\
			 \mathrm{sign}(f)\kappa|f|^2 \quad &\text{quadratic kinetics} \\
			0 \text{ if $|f|<f_0$ } \text{ else } \mathrm{sign}(f) \kappa \cdot (|f|-f_0) \quad &\text{stick-slip kinetics} 
		\end{array}\right.
\end{equation}
and examine if the direct dynamic simulations show a similar relation between interface velocity and classical driving force.

The configuration is a 1D bar with a phase interface at the center of the bar. 
The bar is fixed at the left end and a constant load $P$ is applied at the right end.
This causes an elastic wave to head towards the left from the right end.
When the elastic wave hits the phase interface, it causes the interface to begin moving.
Repeated calculations over a range of applied loads causes interfaces to propagate at a range of velocities.

\begin{figure}[htb!]
\begin{center}
	\fbox{
	\begin{minipage}{185mm}
		\begin{center}
			\includegraphics[width = 90mm]{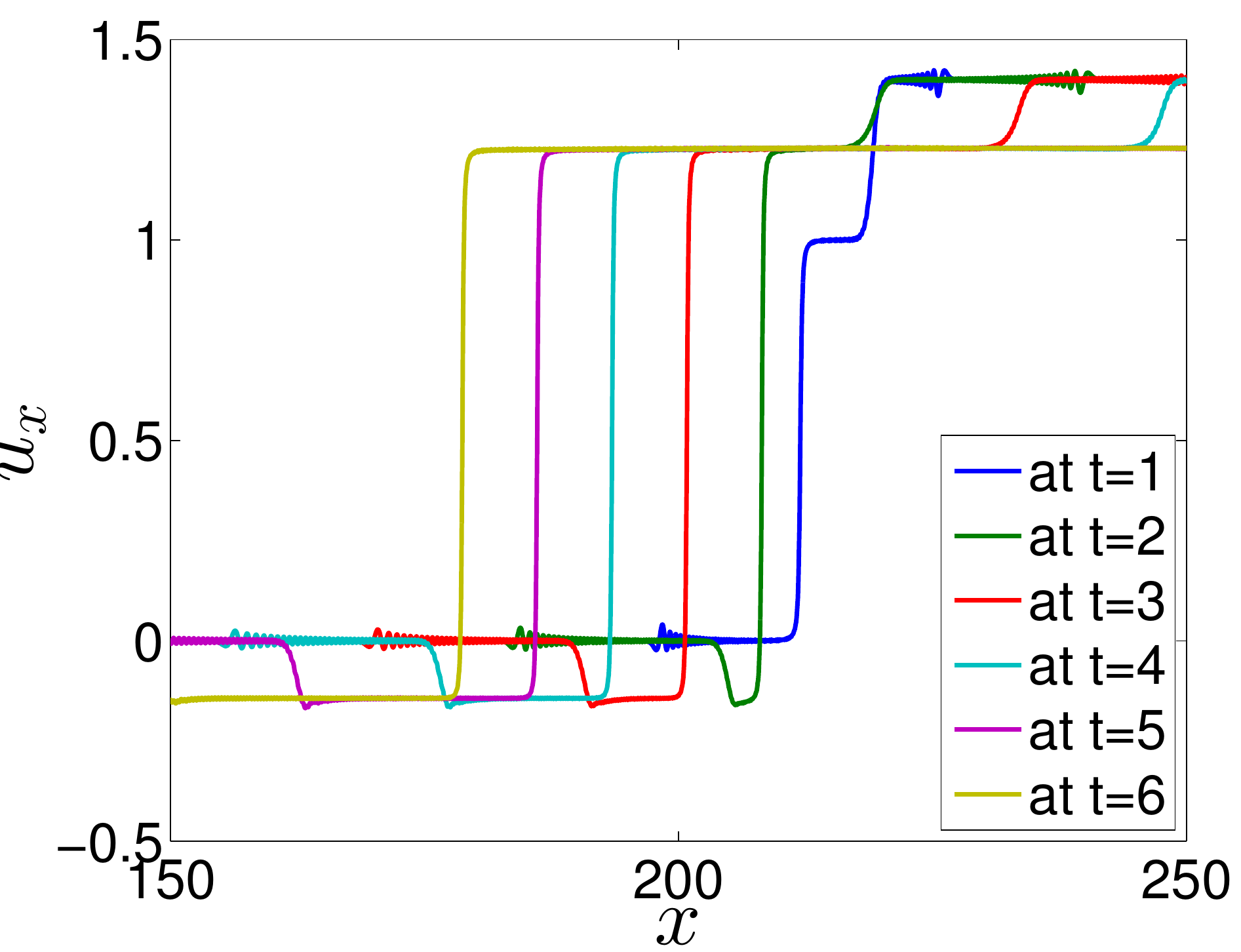}
			\includegraphics[width = 90mm]{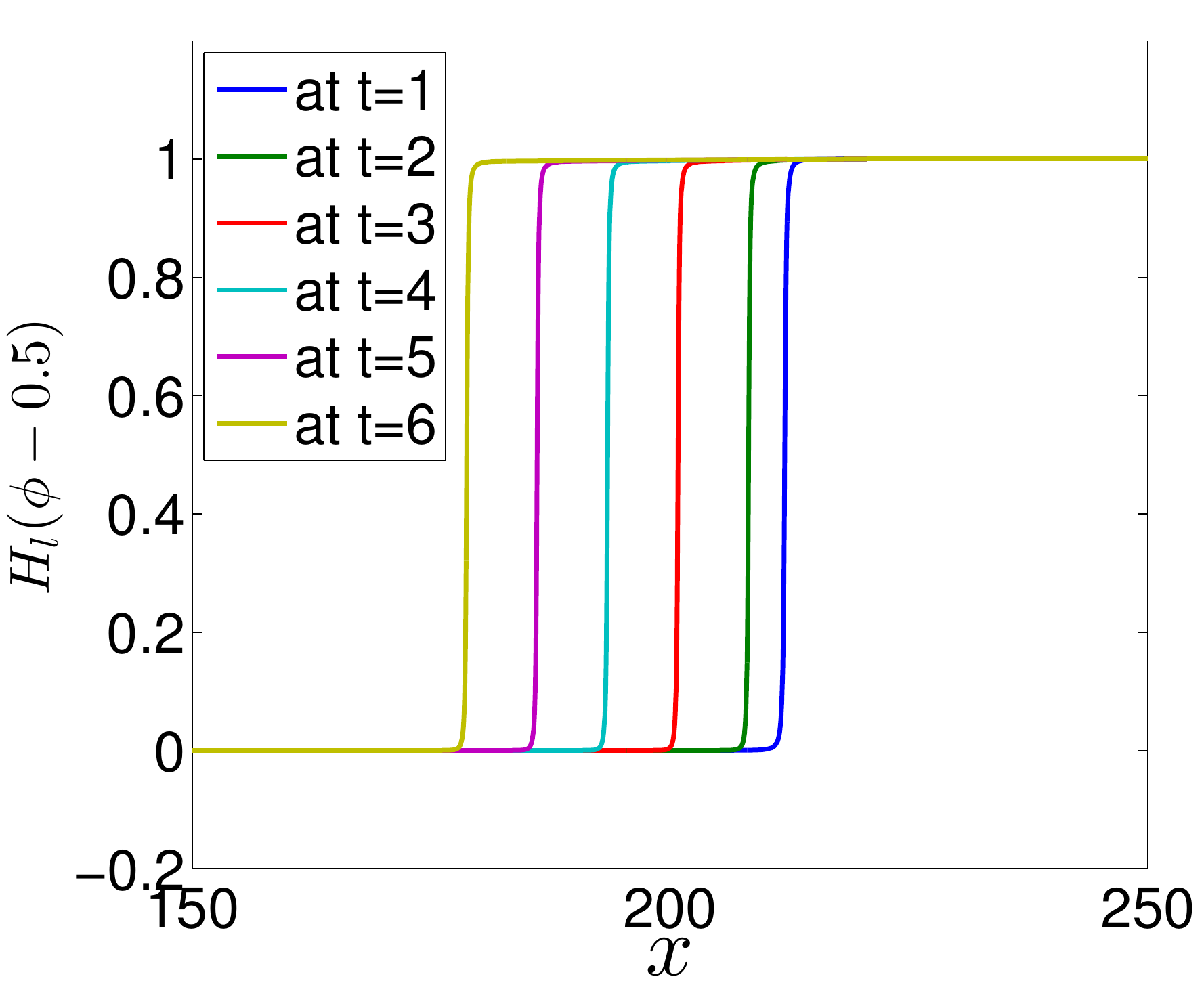}
			\caption{$u_x$ (left) and $H_l(\phi(x) - 0.5)$ at different times after the elastic wave hits the phase interface, showing the steady state evolution of the interface.}
			\label{fig:1d-profiles}
		\end{center}
	\end{minipage}
	}
\end{center}
\end{figure}

Fig. \ref{fig:1d-profiles} shows the evolution of the interface after the elastic wave hits it in the case of linear kinetics.
It can be seen that the solution quickly reaches a steady-state evolution.
The quadratic kinetics and the stick-slip kinetics above the sticking threshold display qualitatively similar evolution.

\begin{figure}[htb!]
\begin{center}
	\fbox{
	\begin{minipage}{185mm}
		\begin{center}
			\includegraphics[width = 90mm]{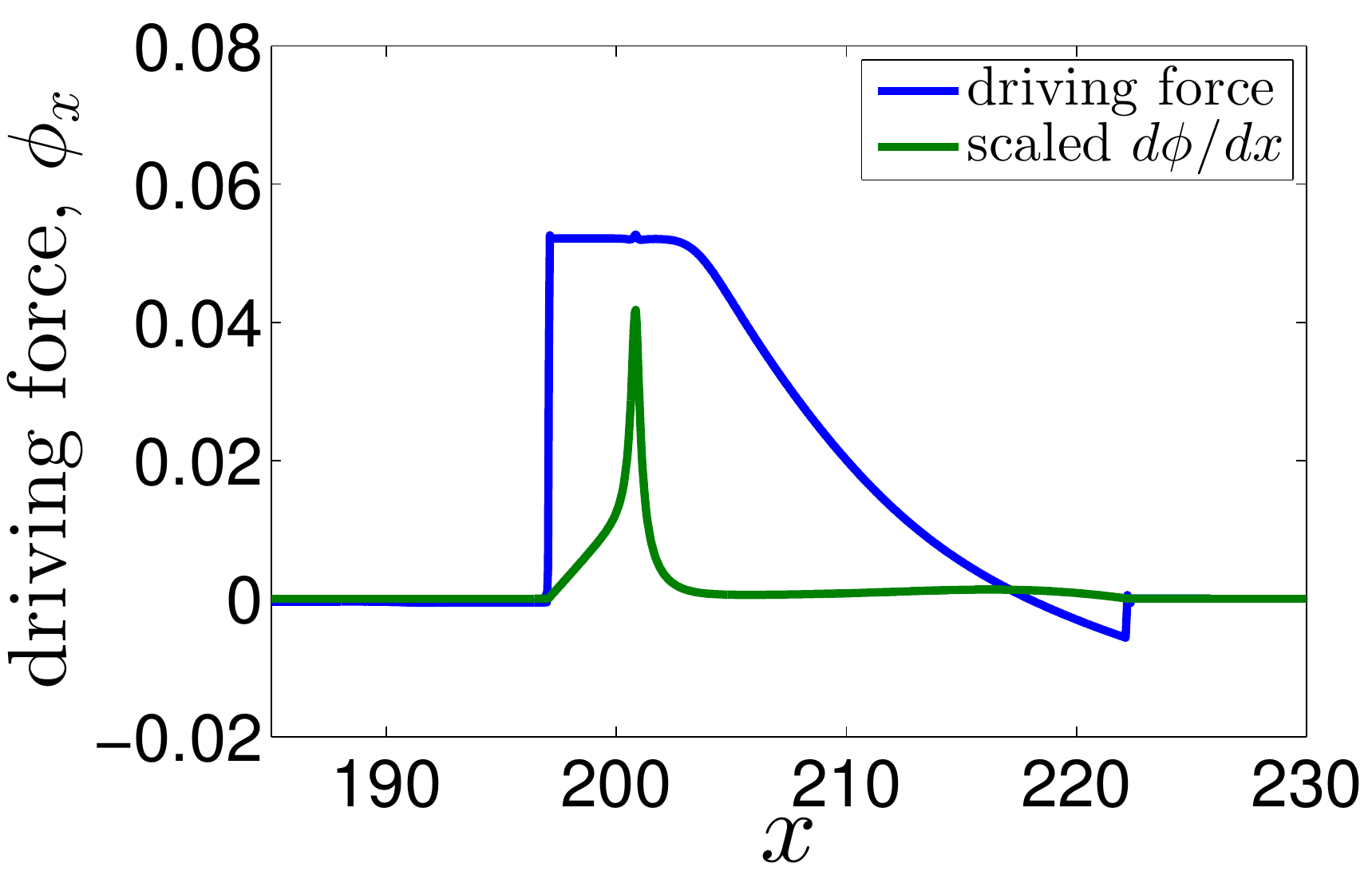}
			\includegraphics[width = 90mm]{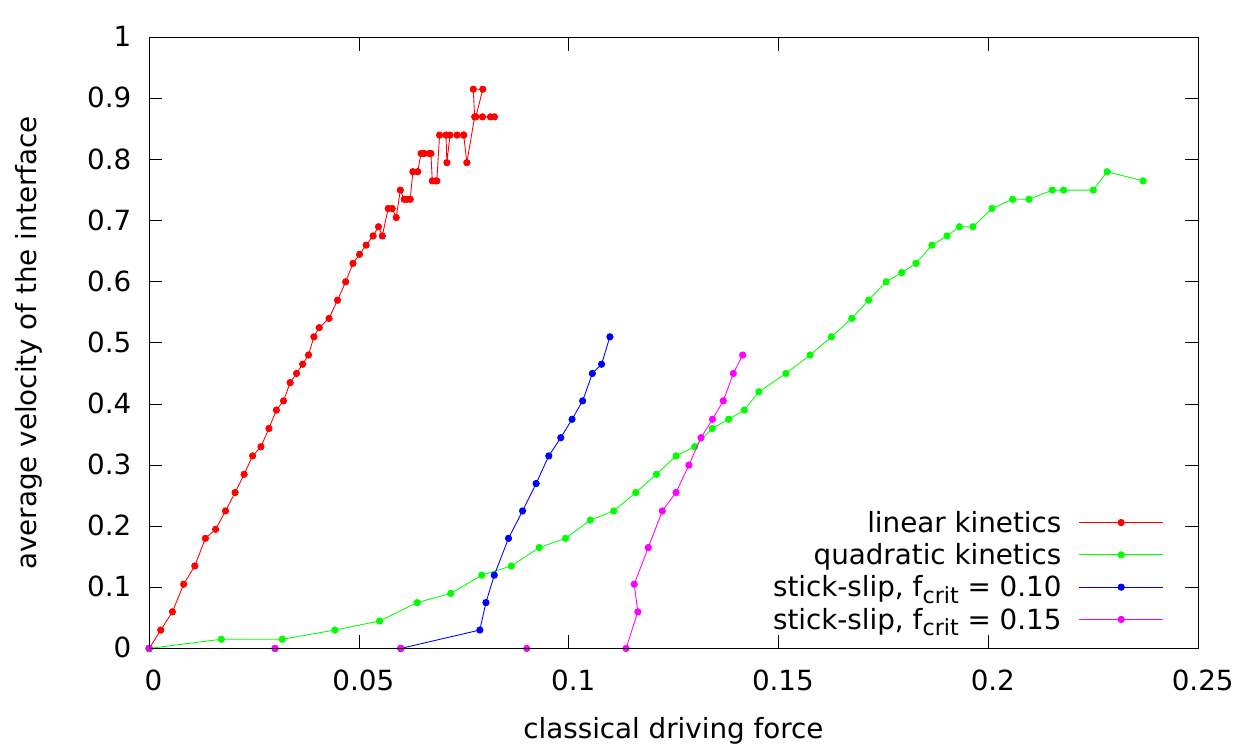}
			\caption{Left: Driving force in the vicinity of the interface, showing that it constant.  The interface is moving towards the left.  Right: Interface velocity vs. classical driving force for different kinetic laws.}
			\label{fig:1d-df-kinetics}
		\end{center}
	\end{minipage}
	}
\end{center}
\end{figure}

We note in Fig. \ref{fig:1d-df-kinetics} (left) the attractive feature of the model that the driving force field is constant in the vicinity of the interface.
Consequently,$v_n^\phi$ is constant in that region, enabling the transport of the interface density without distortion of the interface shape.
This also enables clear physical interpretations of the notion of driving force and interface velocity.
To find the induced kinetics, we compute the classical driving force and plot it against the interface velocity. 
Fig. \ref{fig:1d-df-kinetics} (right) shows the induced kinetics for the kinetic response functions in \eqref{eqn:3-kin-laws}.

The induced kinetic relation follows quite well the kinetic response functions in \eqref{eqn:3-kin-laws} but the agreement gets worse as $M\to 1$.
This is to be expected since balance of linear momentum does not permit supersonic interfaces irrespective of the driving force.

The stick-slip kinetic response permits evolution only if driving force exceeds a threshold value, and we see the same induced behavior in terms of classical driving force.
The precise threshold value is different, but the ratio of the threshold value is preserved for the two stick-slip kinetic laws that were tested in Fig. \ref{fig:1d-df-kinetics} (right).

%%%%%%%%%%%%%%%%%%%%%
%%%%%%%%%%%%%%%%%%%%%

\subsection{Supersonic Interface Velocity in a Cubic Material}

Phase interfaces can be supersonic with respect to one or both of the phases if the elastic response in each phase is nonlinear.
Supersonic is typically defined with respect to the sonic velocity at the well.
In the case of a cubic strain energy, the speed of sound at the well is $0$.
Hence, it is trivially possible to get interfaces that are supersonic.
We ensure here that our model reproduces this feature.

\begin{figure}[htb!]
\begin{center}
	\fbox{
	\begin{minipage}{175mm}
		\begin{center}
			\includegraphics[width = 85mm]{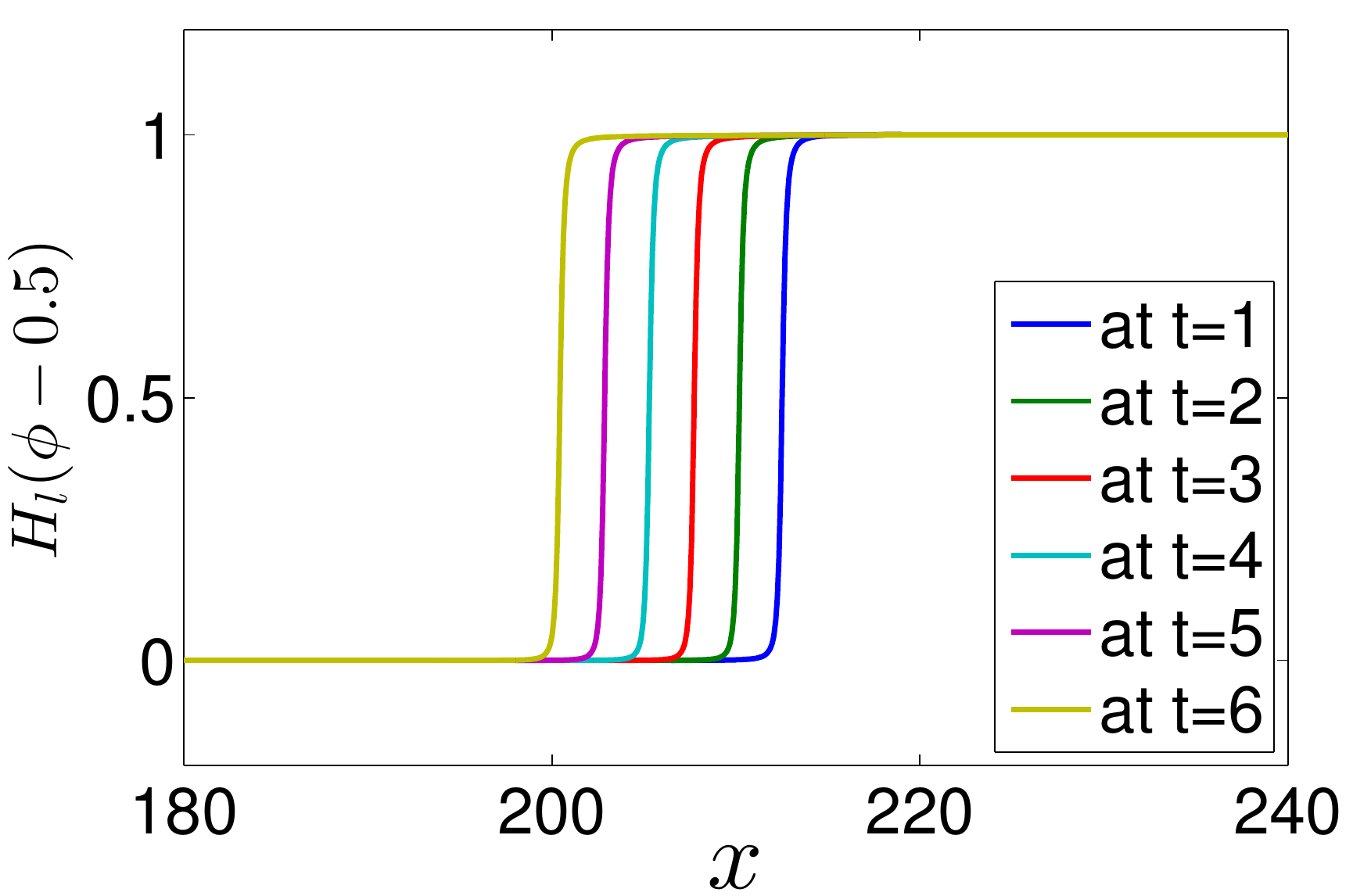}
			\includegraphics[width = 85mm]{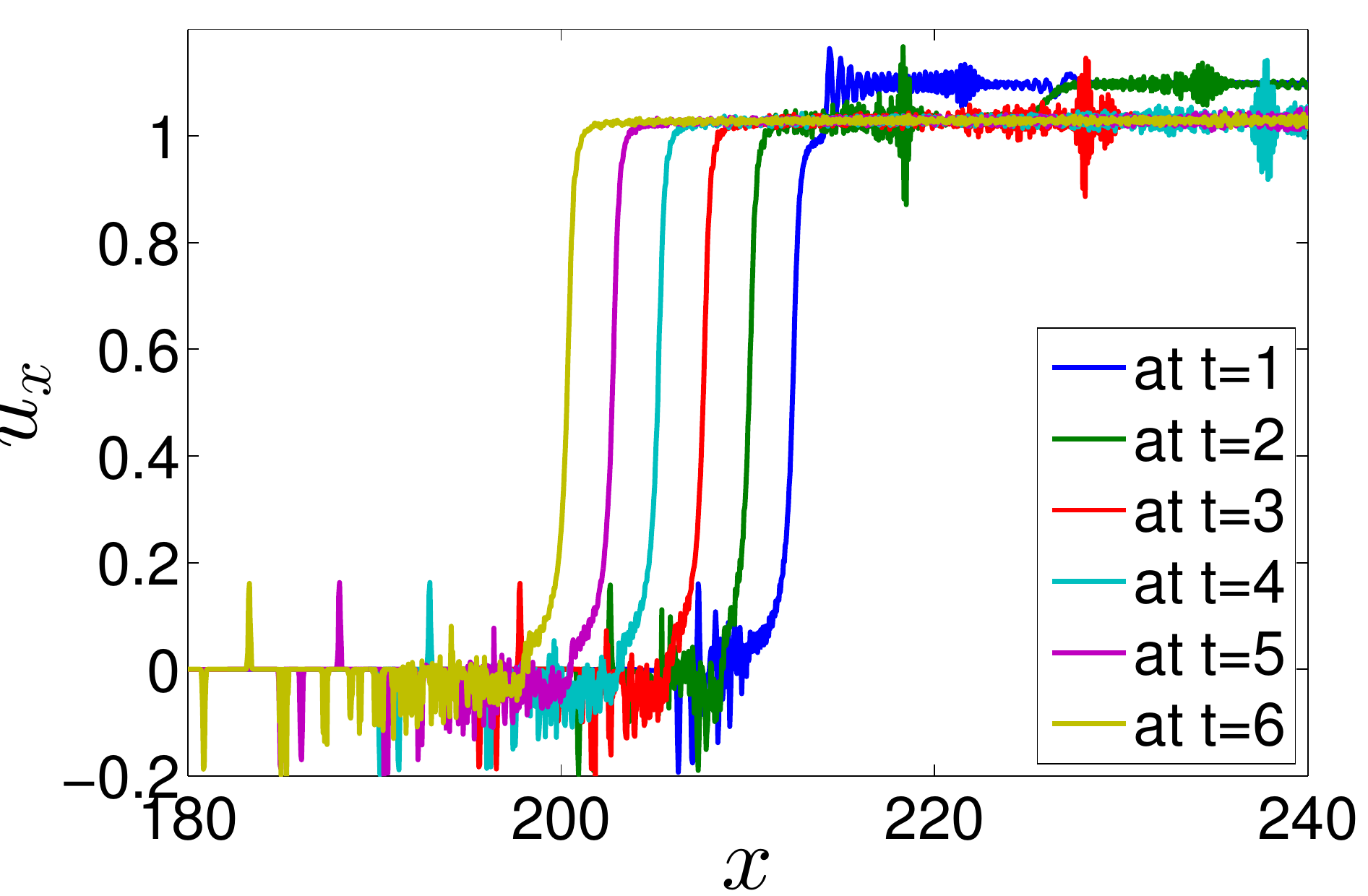}
			\caption{Supersonic evolution of $H_l(\phi(x) - 0.5)$ (left) and $u_x$ (right) in a cubic material.}
			\label{fig:cubic-supersonic}
		\end{center}
	\end{minipage}
	}
\end{center}
\end{figure}

We use the linear kinetic response from Section \ref{sec:1d-kin-relns}, but modify the elastic energy to be cubic in each phase:
\begin{equation}
	\Wcirc(u_x,\phi) = \Big(1-H_a(\phi-0.5)\Big)\half C (u_x-\eps_1)^3 + H_a(\phi-0.5) \half C (u_x - \eps_2)^3
\end{equation}
Fig. \ref{fig:cubic-supersonic} shows the evolution of $\phi$ and $u_x$ which are supersonic but typical in all other ways.

%%%%%%%%%%%%%%%%%%%%%
%%%%%%%%%%%%%%%%%%%%%
%%%%%%%%%%%%%%%%%%%%%
%%%%%%%%%%%%%%%%%%%%%

\section{Effect of the Small Parameter $l$}
\label{sec:small-parameter}

In addition to the constitutive input in terms of $\Wcirc, \hat{v}_n^\phi, G_0$, our model contains two small parameters: $\epsilon$, the coefficient of $|\nabla\phi|^2$, and $l$, the parameter in the regularized Heaviside-like function $H_l$ (see Fig. \ref{fig:energy-landscape}).
There is a good physical understanding of $\epsilon$ as being related to the thickness of phase interfaces.
In this section, we probe the role of $l$ by examining the induced kinetic relations for different values of $l$.
We examine this both using traveling waves with linear kinetics, and using dynamic calculations with a stick-slip kinetic response.
We use $\Wcirc$ as in Section \ref{sec:1d-kin-relns}.

\begin{figure}[htb!]
\begin{center}
	\fbox{
	\begin{minipage}{185mm}
		\begin{center}
			\includegraphics[width = 90mm]{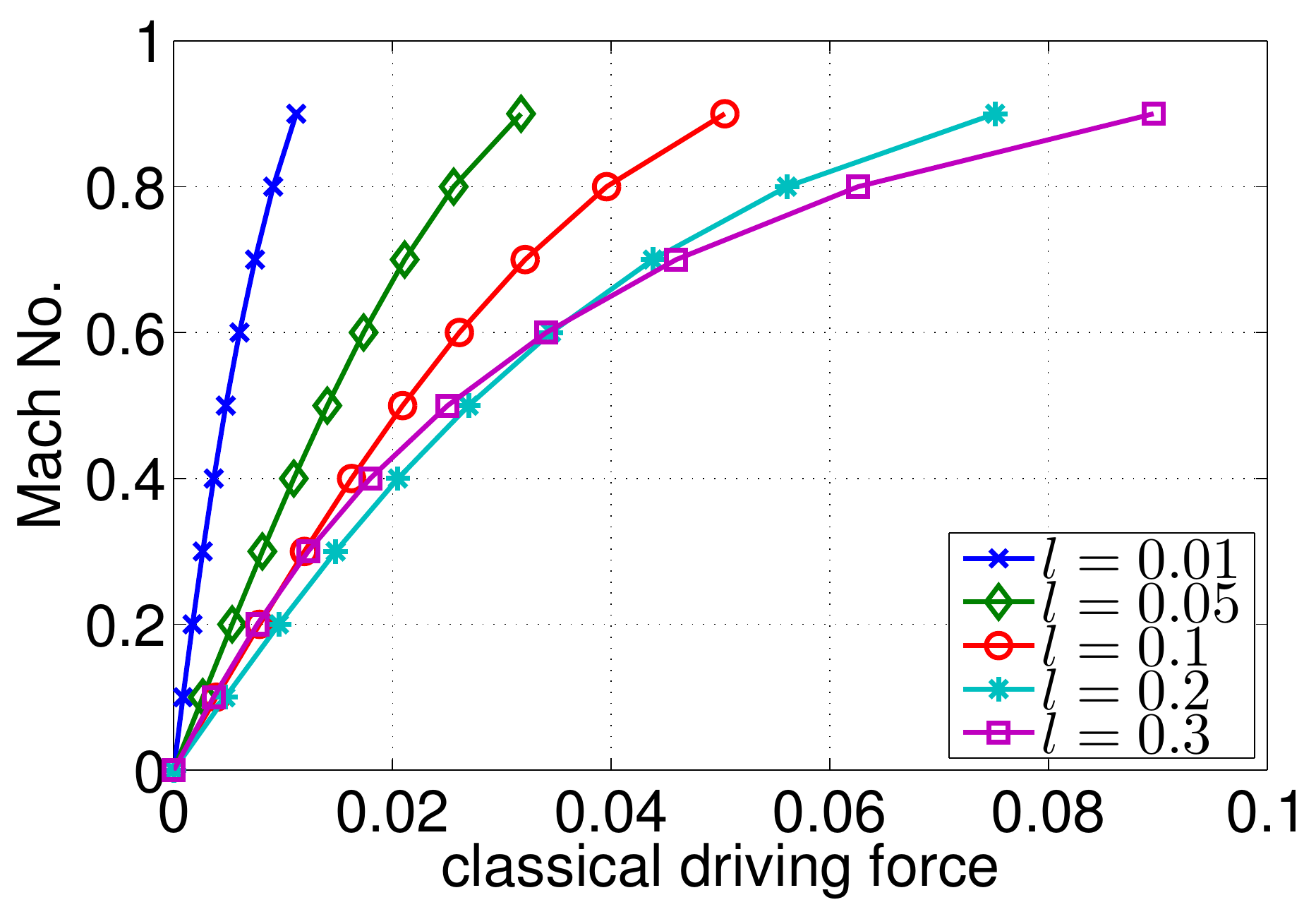}
			\includegraphics[width =90mm]{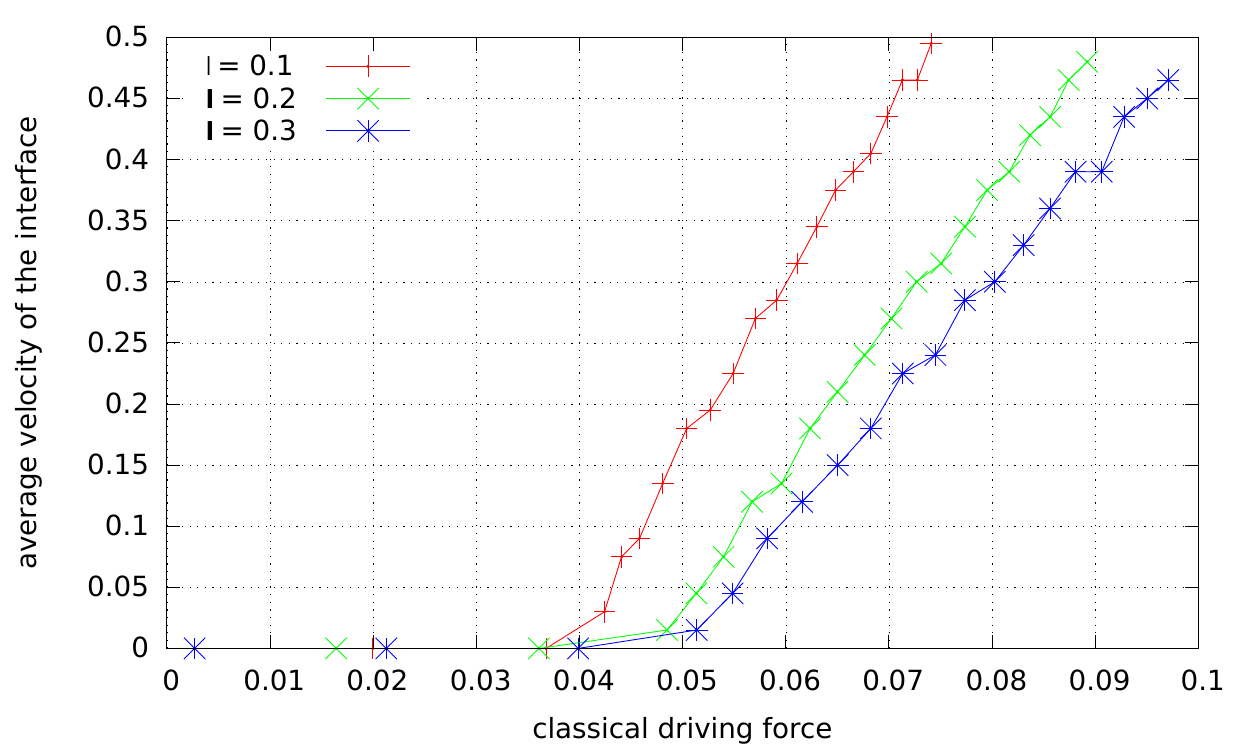}
			\\
			\includegraphics[width = 90mm]{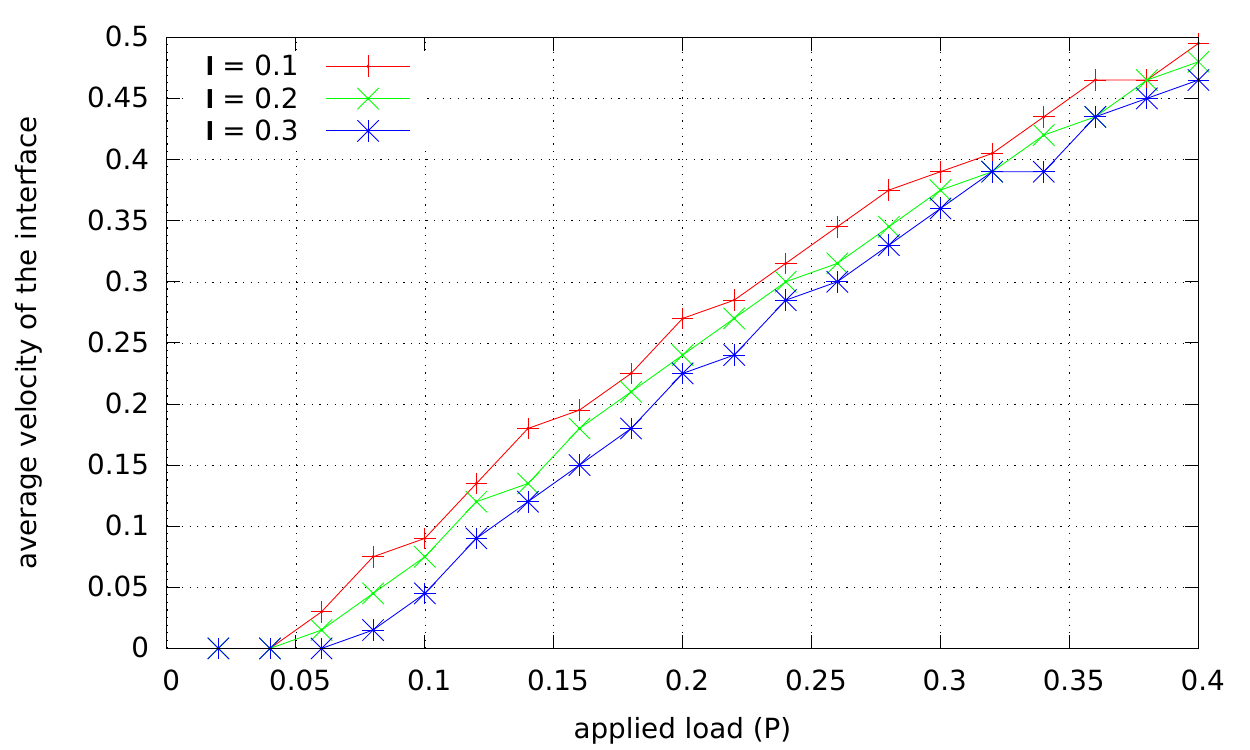}
			\caption{Left: Interface velocity vs. classical driving force with linear kinetics computed using traveling waves for different values of $l$. Right: Interface velocity vs. classical driving force with stick-slip kinetics computed using dynamic calculations for different values of $l$. Below: Interface velocity vs. applied end load for the same dynamic calculations.}
			\label{fig:role-of-l}
		\end{center}
	\end{minipage}
	}
\end{center}
\end{figure}

Fig. \ref{fig:role-of-l} shows that the kinetics is quite sensitive to $l$.
Ideally, we would like to see if there is convergence in any sense as $l \to 0$, but the energy is extremely steep as $l$ becomes smaller and does not permit numerical simulations with confidence.
From the calculations that we could confidently carry out, there appears to be no such convergence.
However, while the kinetics is sensitive to $l$, the essential effect seems to be as a pre-multiplying coefficient that does not affect the shape of the kinetic response function.
Therefore, a simple strategy to deal with this is to fix a given value of $l$ that allows easy numerical simulations, and then calibrate the pre-multiplier in the kinetic response function to the desired value based on this fixed value of $l$.
In other words, treat $l$ as a fixed material parameter.

Additionally, an interesting observation from the dynamic calculations is that the relation between interface velocity and applied end load is fairly insensitive to $l$.

%%%%%%%%%%%%%%%%%%%%%
%%%%%%%%%%%%%%%%%%%%%
%%%%%%%%%%%%%%%%%%%%%
%%%%%%%%%%%%%%%%%%%%%

\section{Formulation of a Two-Dimensional Energy for Twinning}
\label{sec:2D-energy}

We perform a number of 2D calculations for twinning in the subsequent sections.
Here, we briefly the common aspects of all those calculations, such as the formulation of $\Wcirc(\bfE,\phi)$.

The form of $\Wcirc$ is:
\begin{equation}
	\Wcirc(\bfE,\phi) = 
		\left(1-H_l(\phi-0.5)\right) \half (\bfE-\bfE_1):\bfC:(\bfE-\bfE_1) + H_l(\phi-0.5)\half (\bfE-\bfE_2):\bfC:(\bfE-\bfE_2)
\end{equation}
Both $\bfE_1$ and $\bfE_2$ are the stress-free states because $\bfsigma_T=0$.
We have further assumed that both wells are at the same height, i.e., $\Wcirc(\bfE_1,0) = \Wcirc(\bfE_2,1) = 0$, and that the moduli $\bfC$ are the same.
This is appropriate for twinning, but perhaps not so for other transformations.
The relative height of the wells is important because it appears directly in the driving force, and is trivial to change if appropriate.

For twinning, consider the transformation stretch tensors:
\begin{align}
	\bfU_1 = 
		\begin{bmatrix} 
			\alpha & 0 \\
			0 & \beta
		\end{bmatrix},
	\bfU_2 = 
		\begin{bmatrix}
			\beta & 0 \\
			0 & \alpha 
		\end{bmatrix}, 
		\quad \alpha = 1-0.1042 , \beta = 1+0.09659
\end{align}
$\bfE_1$ and $\bfE_2$ are computed from $\bfF_1 = \bfU_1$ and $\bfF_2 = \bfU_2$. 
The stress-free compatible interfaces between these wells have normals $\frac{1}{\sqrt{2}}\begin{pmatrix} 1 \\ 1 \end{pmatrix}$ and $ \frac{1}{\sqrt{2}}\begin{pmatrix} 1 \\ -1 \end{pmatrix}$.

In certain cases, we rotate the specimen by $\pi / 12$ radians with respect to the coordinate axes.
In that case, the compatible interfaces are oriented with normal $\frac{1}{2}\begin{pmatrix} \sqrt{3} \\ 1 \end{pmatrix}$ and $\frac{1}{2}\begin{pmatrix} -1 \\ \sqrt{3} \end{pmatrix}$.
The reason is that an interface that is inclined at a large angle will feel the effects of the loading on the right boundary of the domain quite differently at different points along its length.
On the other hand, an interface aligned perfectly normal to the top and bottom boundaries of the domain will not have any shear stress in the direction tangent to the interface at the junction between the interface and the domain boundary.
Some level of shear stress is required for evolution on the boundary/interface junction (Section \ref{sec:boundary-kinetics}).
The angle that we have chosen balances between these competing reasons.

%%%%%%%%%%%%%%%%%%%%%
%%%%%%%%%%%%%%%%%%%%%
%%%%%%%%%%%%%%%%%%%%%
%%%%%%%%%%%%%%%%%%%%%

\section{Non-Monotone Kinetic Laws}
\label{sec:non-monotone}

Non-monotone kinetic laws have been predicted to show extremely complex and interesting behavior, e.g. \cite{rosakis-knowles-non-monotone}.
However, in that literature, the driving force is taken as a function of interface velocity.
In the case where it is non-monotone, it is not invertible to obtain the interface velocity as a function of driving force.
In our case, we assume that the interface velocity is a non-monotone function of driving force.
We briefly present the results of calculations in 1D and 2D, but the summary is that there is no complex and unexpected behavior as observed in \cite{rosakis-knowles-non-monotone}.
We have verified after the calculations that the the level of driving force was appropriate to access the non-monotonic portion of the kinetic response.

%%%%%%%%%%%%%%%%%%%%%
%%%%%%%%%%%%%%%%%%%%%

\subsection{1D Non-Monotone Kinetics}

$\Wcirc$ follows Section \ref{sec:1d-kin-relns}, but the kinetic response is chosen to be non-monotonic:
\begin{equation}
\label{eqn:non-mono-kinetics}
	\hat{v}_n^\phi = \left\{\begin{array}{l l}
				|f| \cdot (0.1-|f|) & \text{if} \quad |f|<0.075 \\
				0.075 \cdot (0.1-0.075) & \text{if} \quad |f| \geq 0.075
			\end{array} \right.
\end{equation}

\begin{figure}[htb!]
\begin{center}
	\fbox{
	\begin{minipage}{175mm}
		\begin{center}
			\includegraphics[width = 85mm]{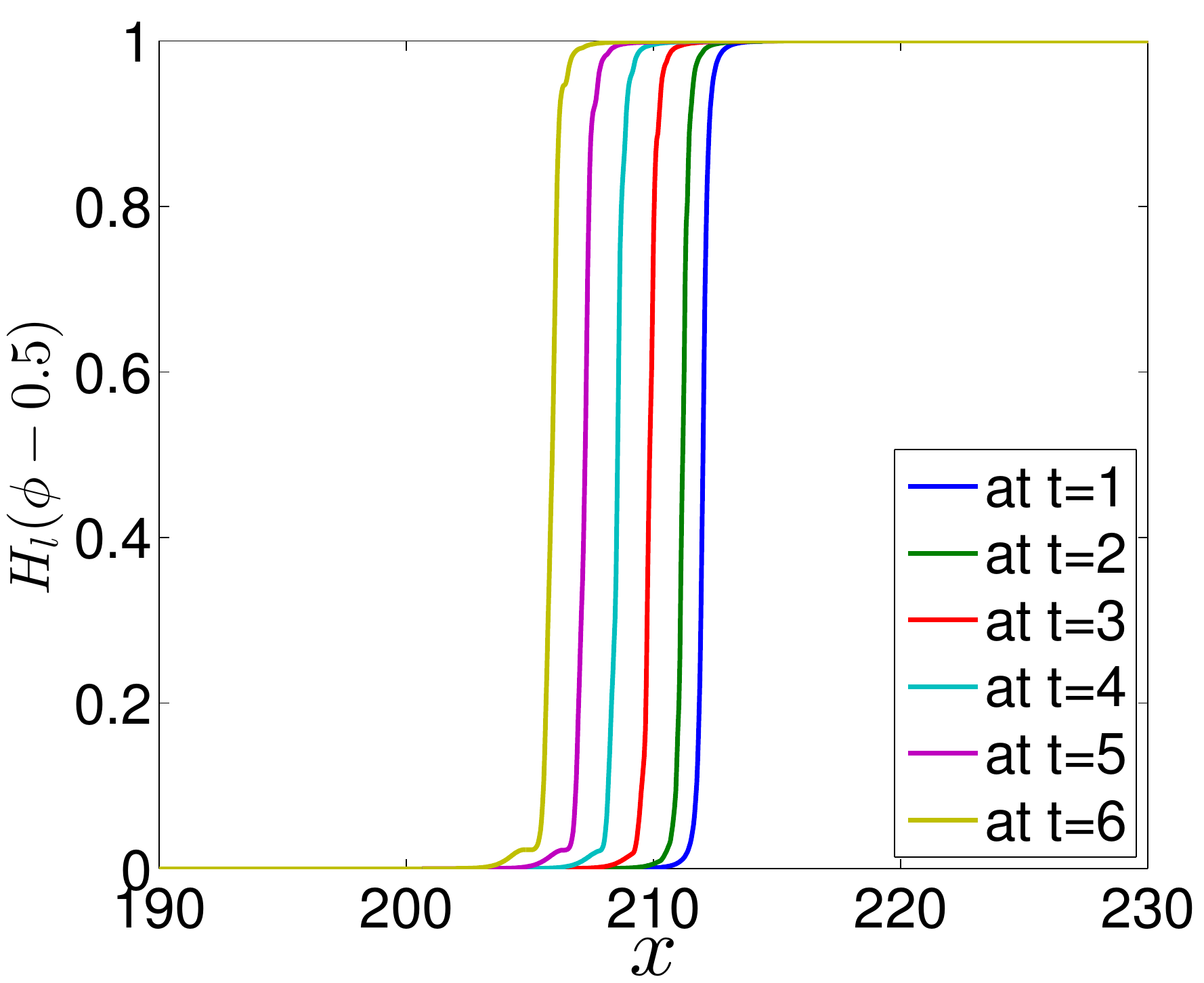}
			\includegraphics[width = 85mm]{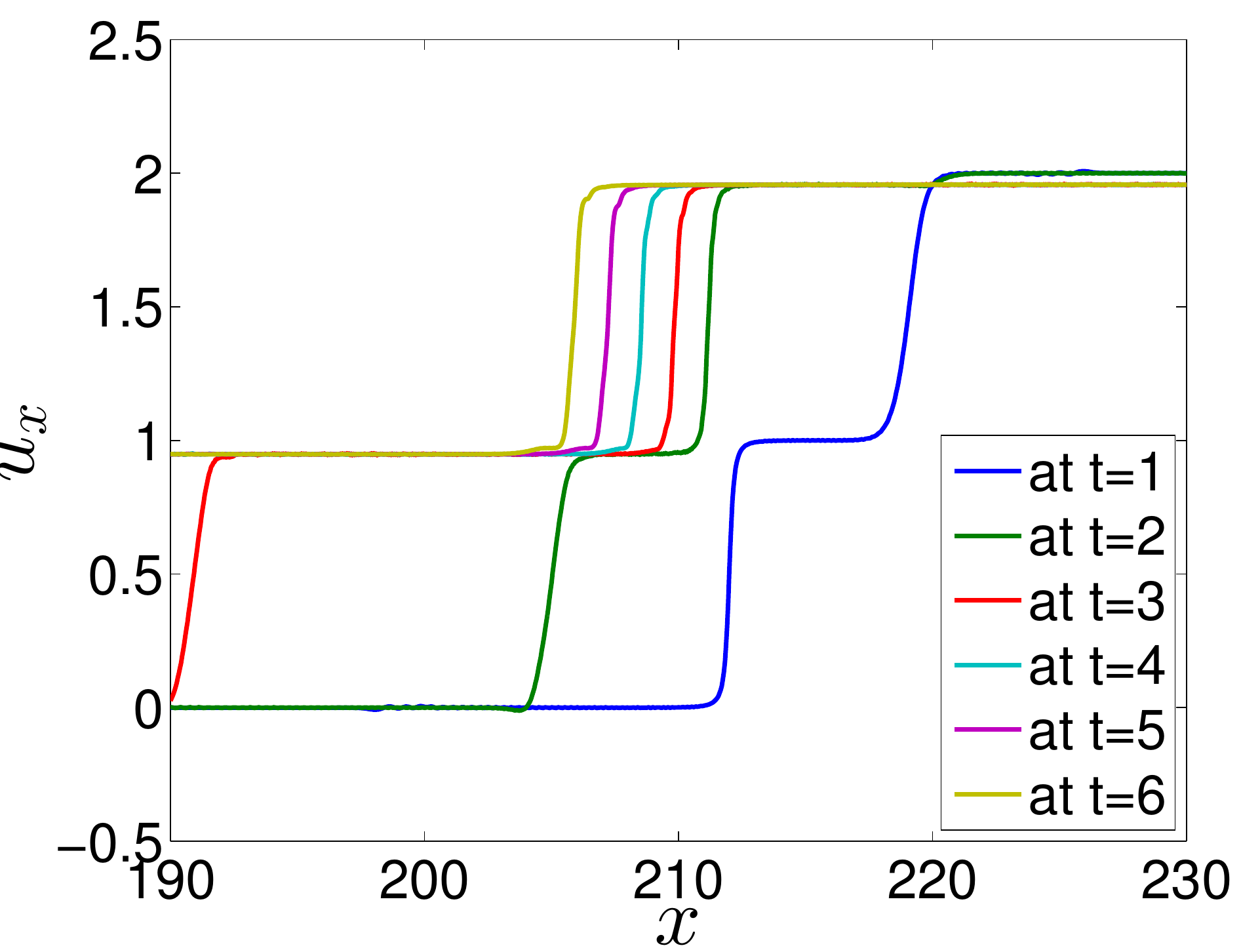}
			\caption{Evolution of $H_l(\phi(x) - 0.5)$ (left) and $u_x$ (right) with a non-monotone kinetic response.}
			\label{fig:1d-non-monotone}
		\end{center}
	\end{minipage}
	}
\end{center}
\end{figure}

Fig. \ref{fig:1d-non-monotone} shows the evolution of $\phi$ and $u_x$ which are qualitatively similar to the calculations with simpler kinetic response functions.

%%%%%%%%%%%%%%%%%%%%%
%%%%%%%%%%%%%%%%%%%%%

\subsection{2D Non-Monotone Kinetics}

We examine 2 settings with non-monotone kinetics; first, a problem with a stress-free compatible interface, and second, where there is necessarily stress around the interface.
The reasoning to test both cases is that it is possible that elastic compatibility will dominate the evolution, and therefore testing both cases will let us compare the role of kinetics.
We compare a linear kinetic response and a non-monotone kinetic response.
For the latter, we use the same kinetic response as in 1D from \eqref{eqn:non-mono-kinetics}.

\begin{figure}[htb!]
\begin{center}
	\fbox{
	\begin{minipage}{155mm}
		\begin{center}
			\includegraphics[width = 75mm]{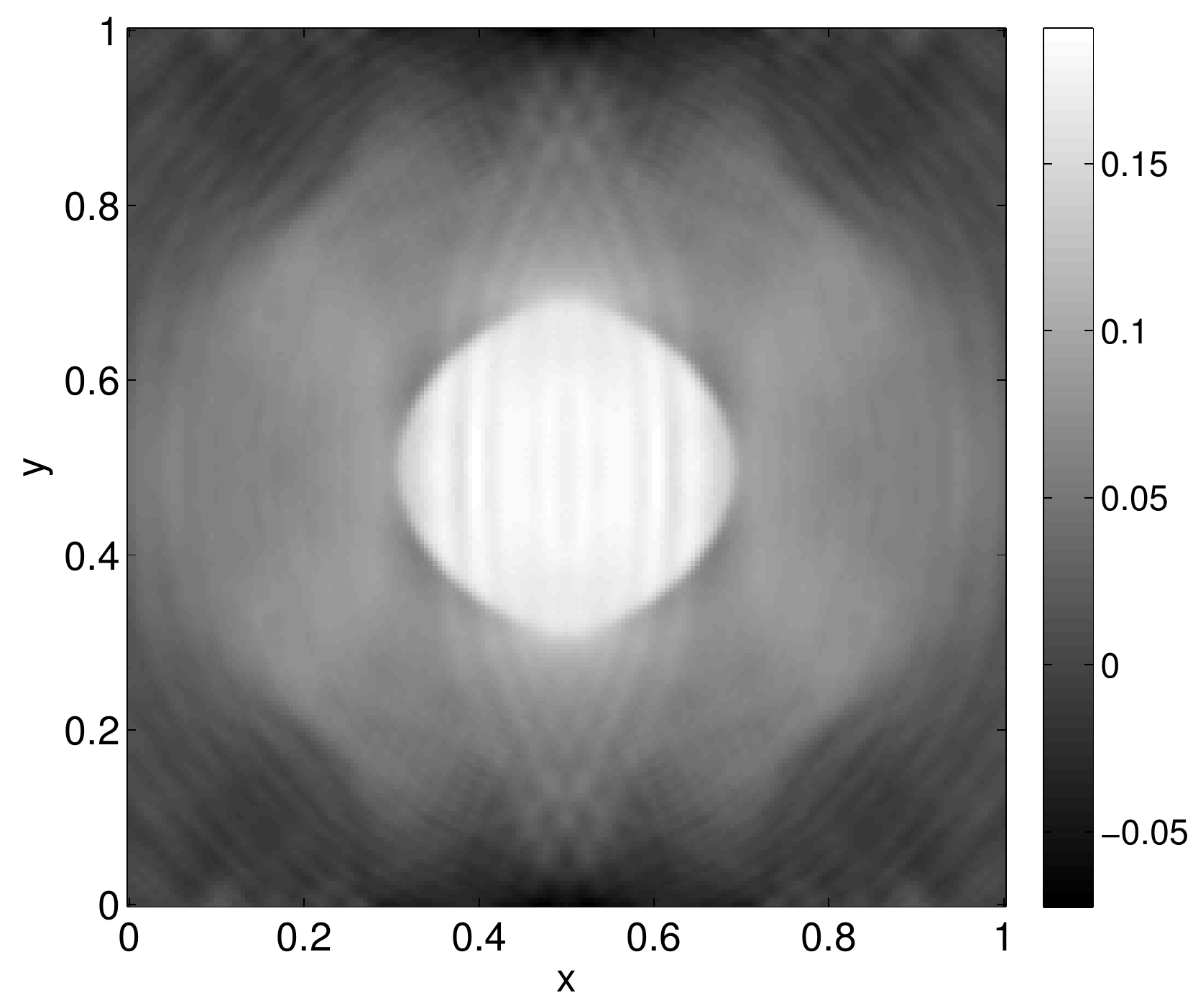}
			\includegraphics[width = 75mm]{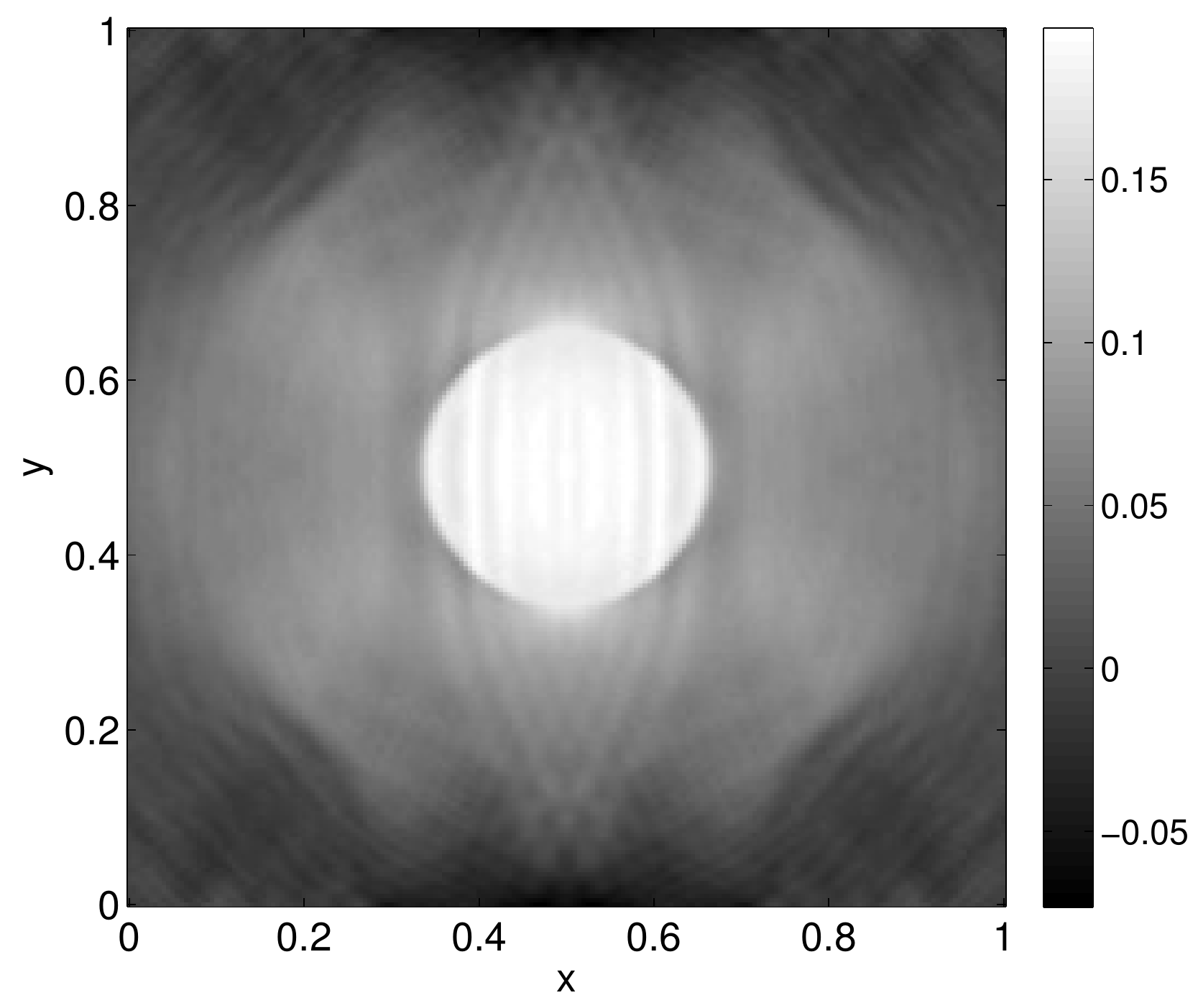}
			\\
			\includegraphics[width = 75mm]{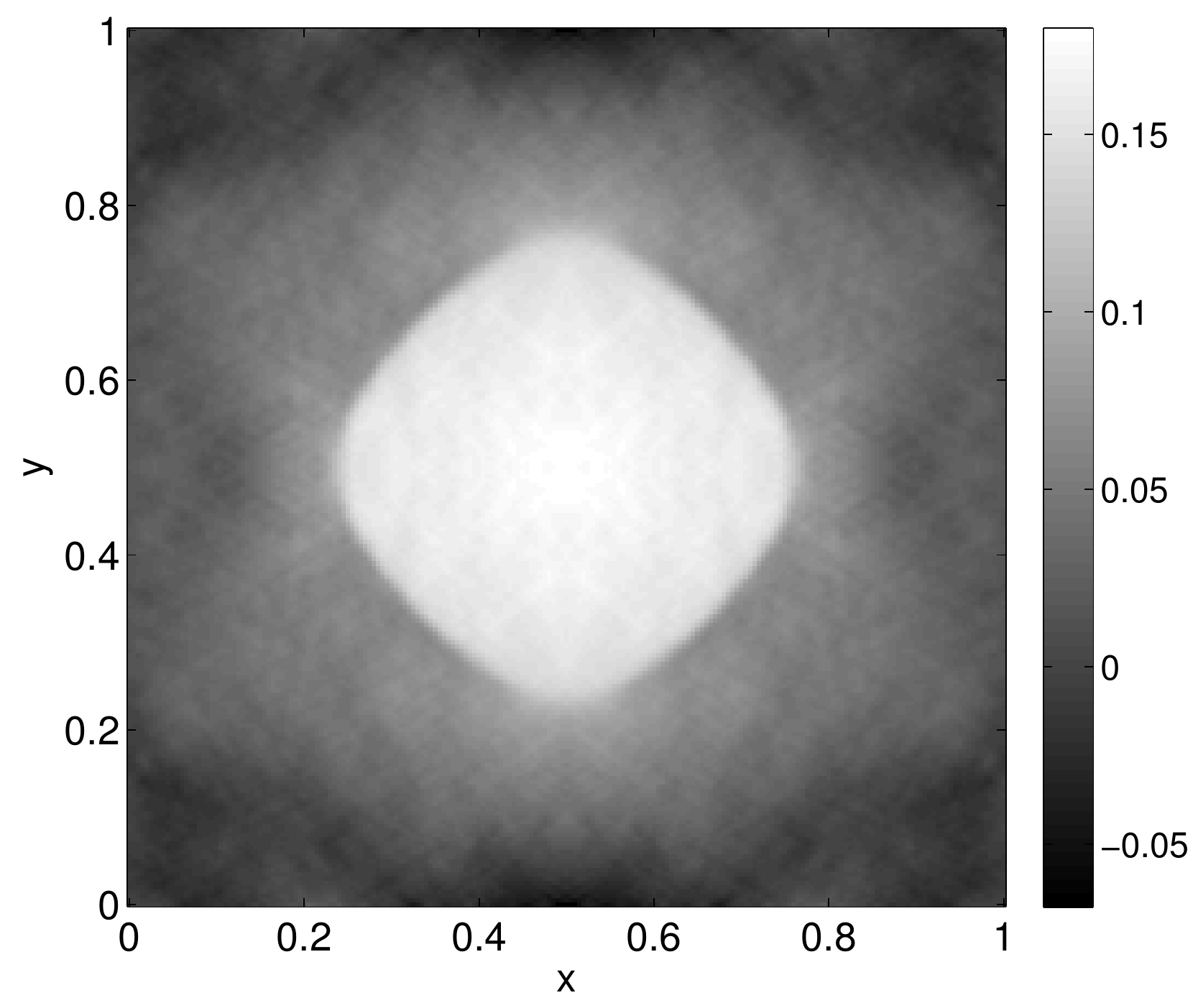}
			\includegraphics[width = 75mm]{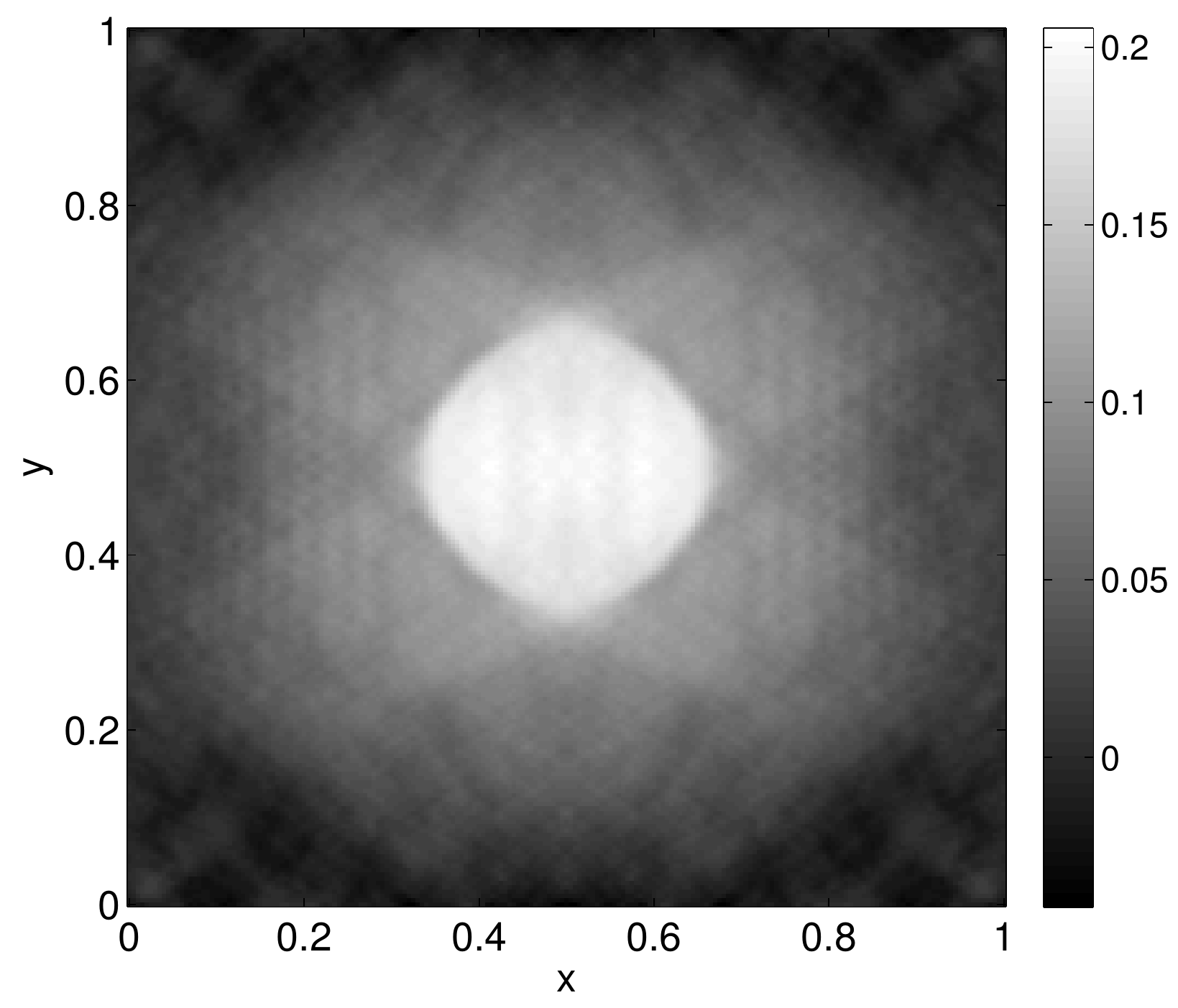}
			\\
			\includegraphics[width = 75mm]{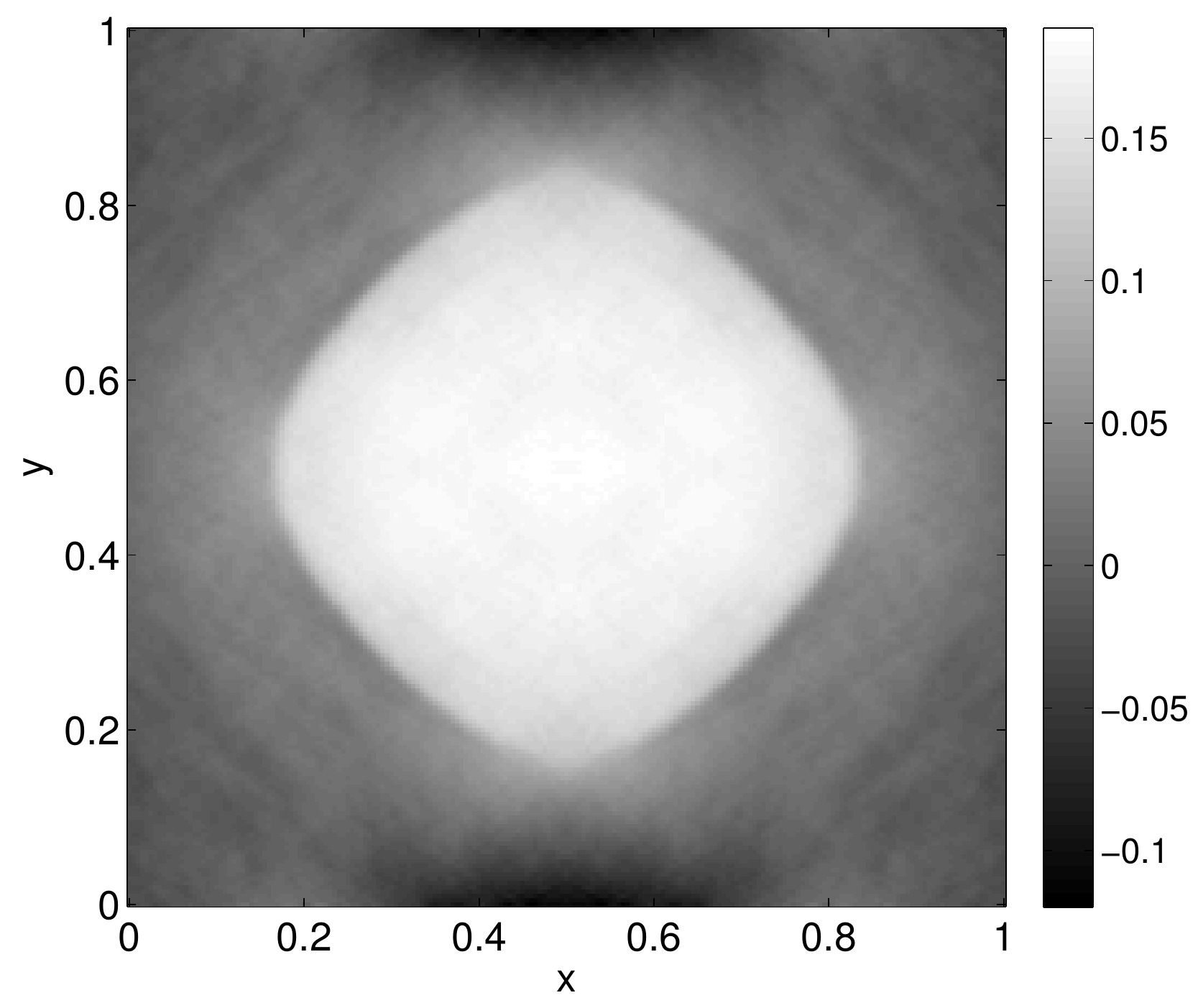}
			\includegraphics[width = 75mm]{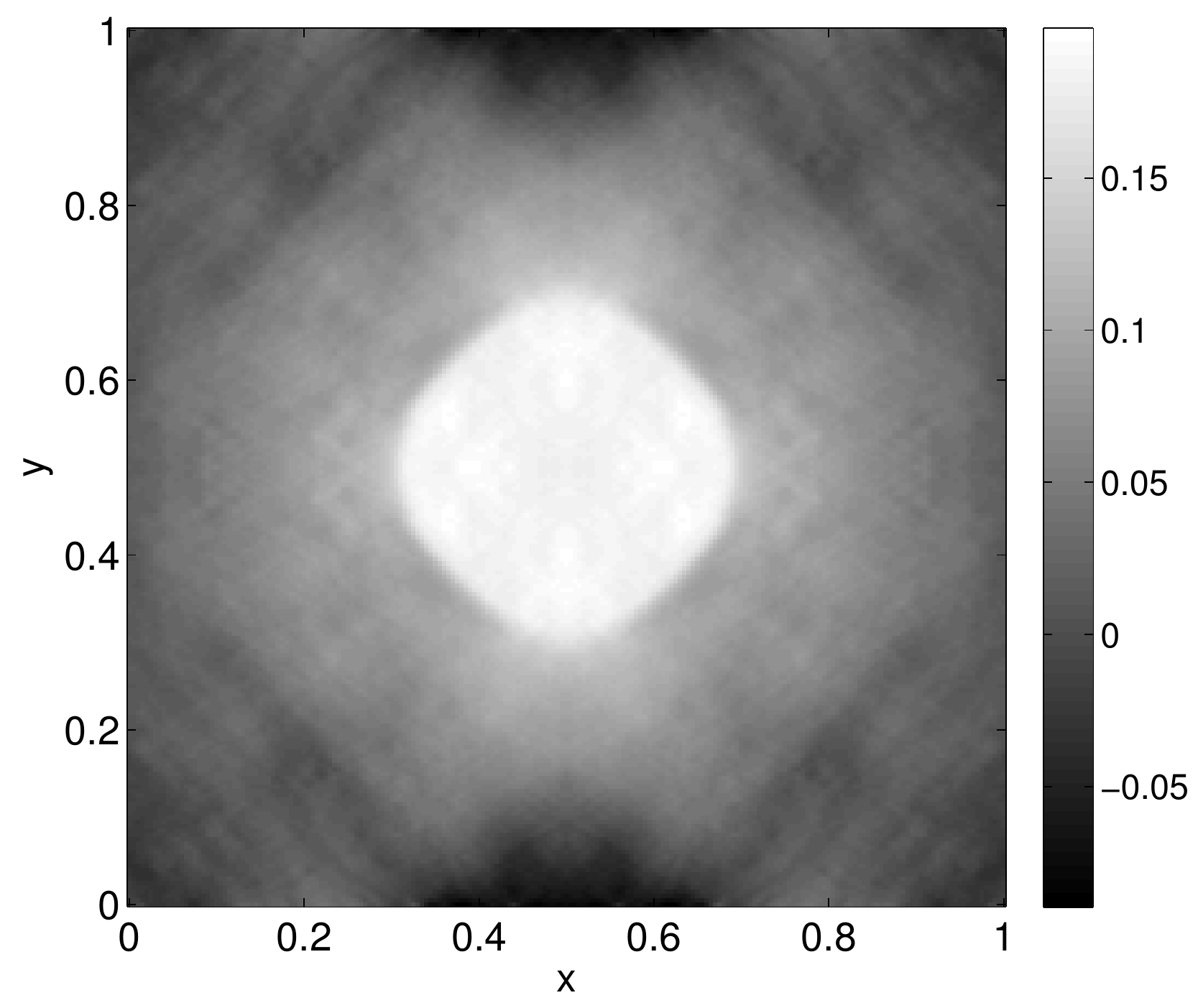}
			\caption{The left column is the evolution with linear kinetics, and the right column is the evolution with non-monotone kinetics, with snapshots of the $F_{11} - 1$ field taken at the same time for both processes.  The phases are not stress-free compatible.  The top row, at small times, is fairly similar.  As evolution progresses, there are quantitative but not qualitative differences.}
			\label{fig:2d-non-monotone}
		\end{center}
	\end{minipage}
	}
\end{center}
\end{figure}

We first examine the case of stressed interfaces using a square plate where a circular region near the center has a second phase.
The energy is described in Section \ref{sec:2D-energy}, and for the incompatible wells we use:
\begin{align}
	\bfU_1 = 
		\begin{bmatrix} 
			1 & 0 \\
			0 & 1
		\end{bmatrix},
	\bfU_2 = 
		\begin{bmatrix}
			\beta & 0 \\
			0 & \alpha 
		\end{bmatrix}, 
	\quad \alpha = 1-0.1042 , \beta = 1+0.09659
\end{align}
Fig. \ref{fig:2d-non-monotone} shows the evolution through $F_{11} - 1$ at various times for the linear and non-monotone kinetic responses respectively.
While there are quantitative differences, they are no obvious qualitative differences, and further there is no complex behavior in the non-monotone case.

\begin{figure}[htb!]
\begin{center}
	\fbox{
	\begin{minipage}{175mm}
		\begin{center}
			\includegraphics[width = 85mm]{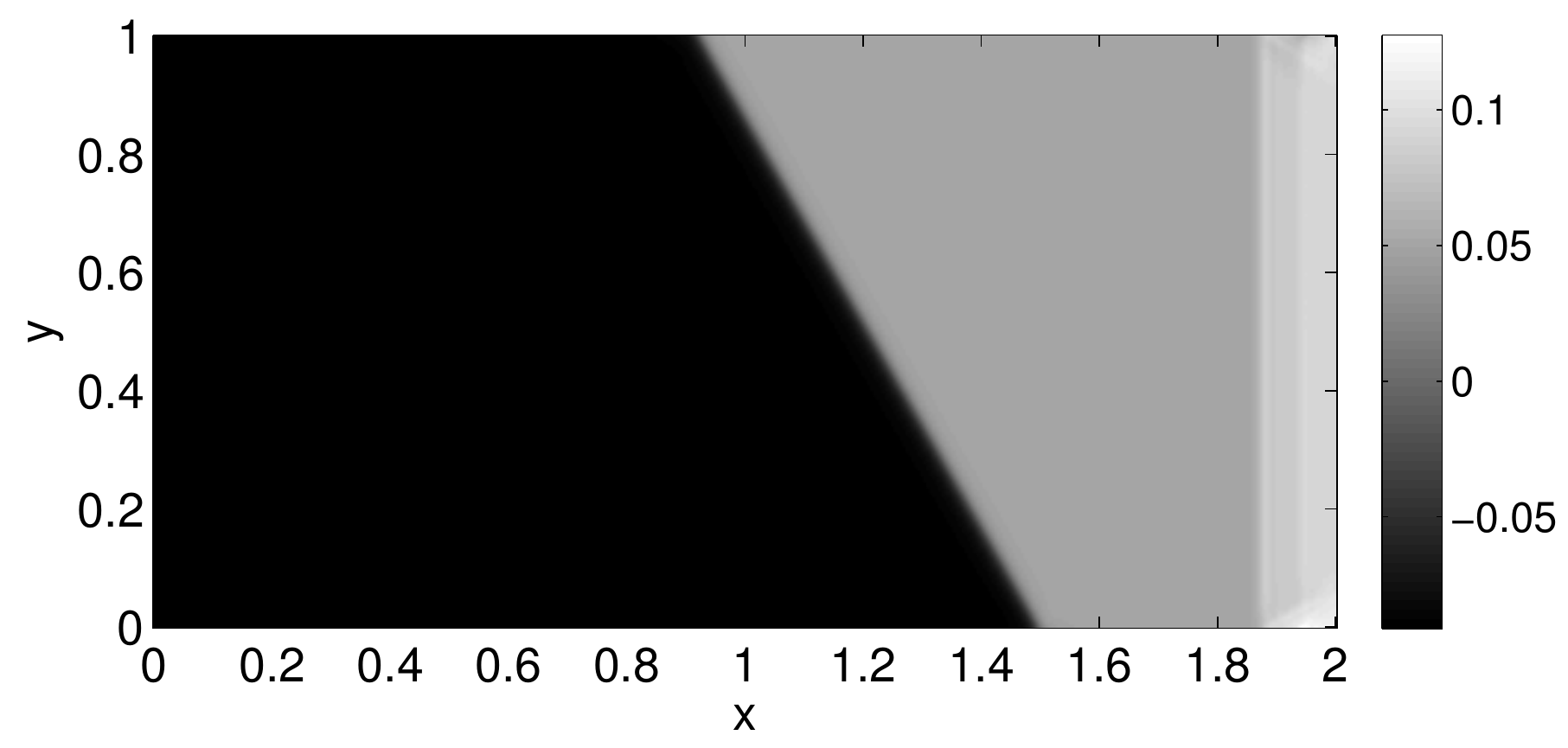}
			\\
			\includegraphics[width = 85mm]{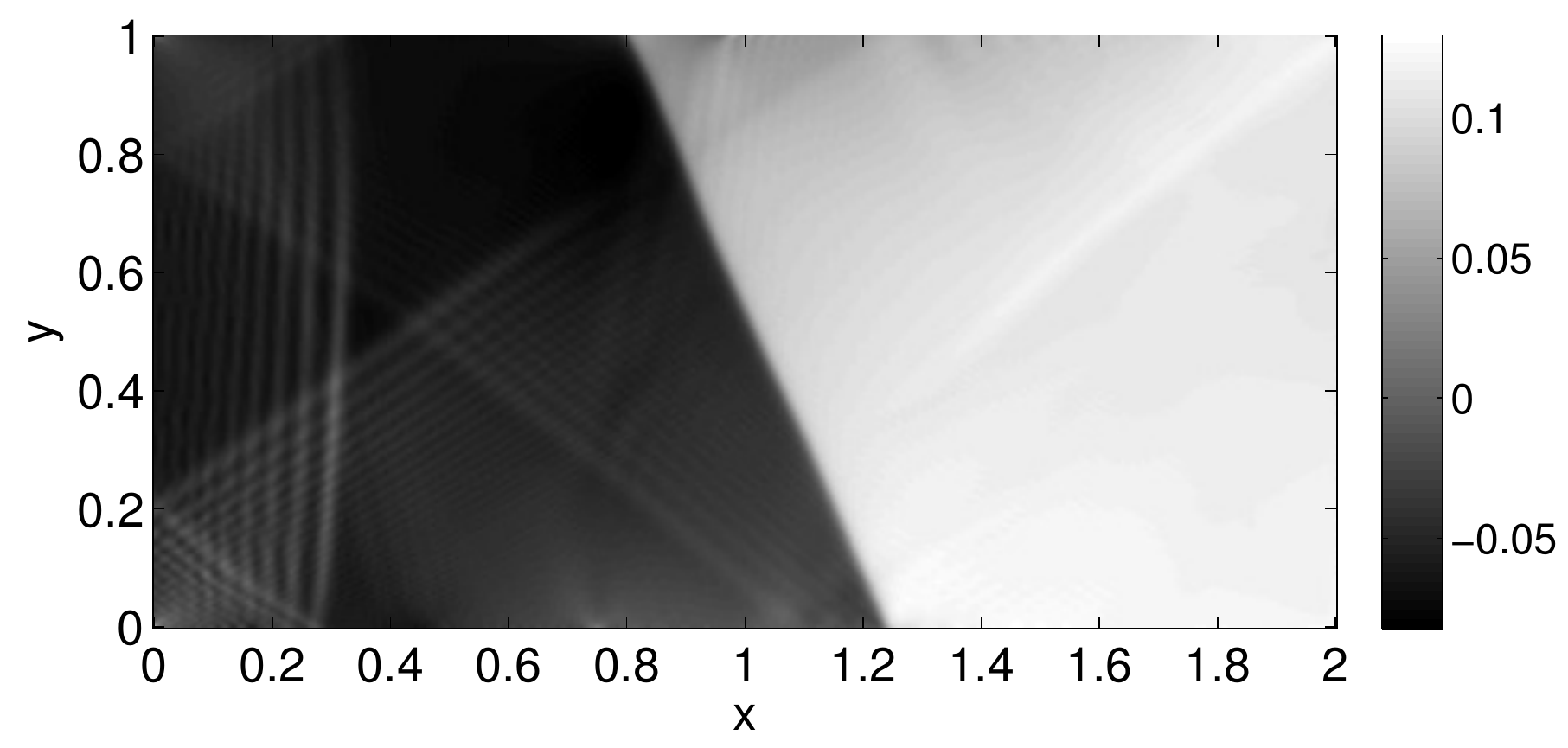}
			\includegraphics[width = 85mm]{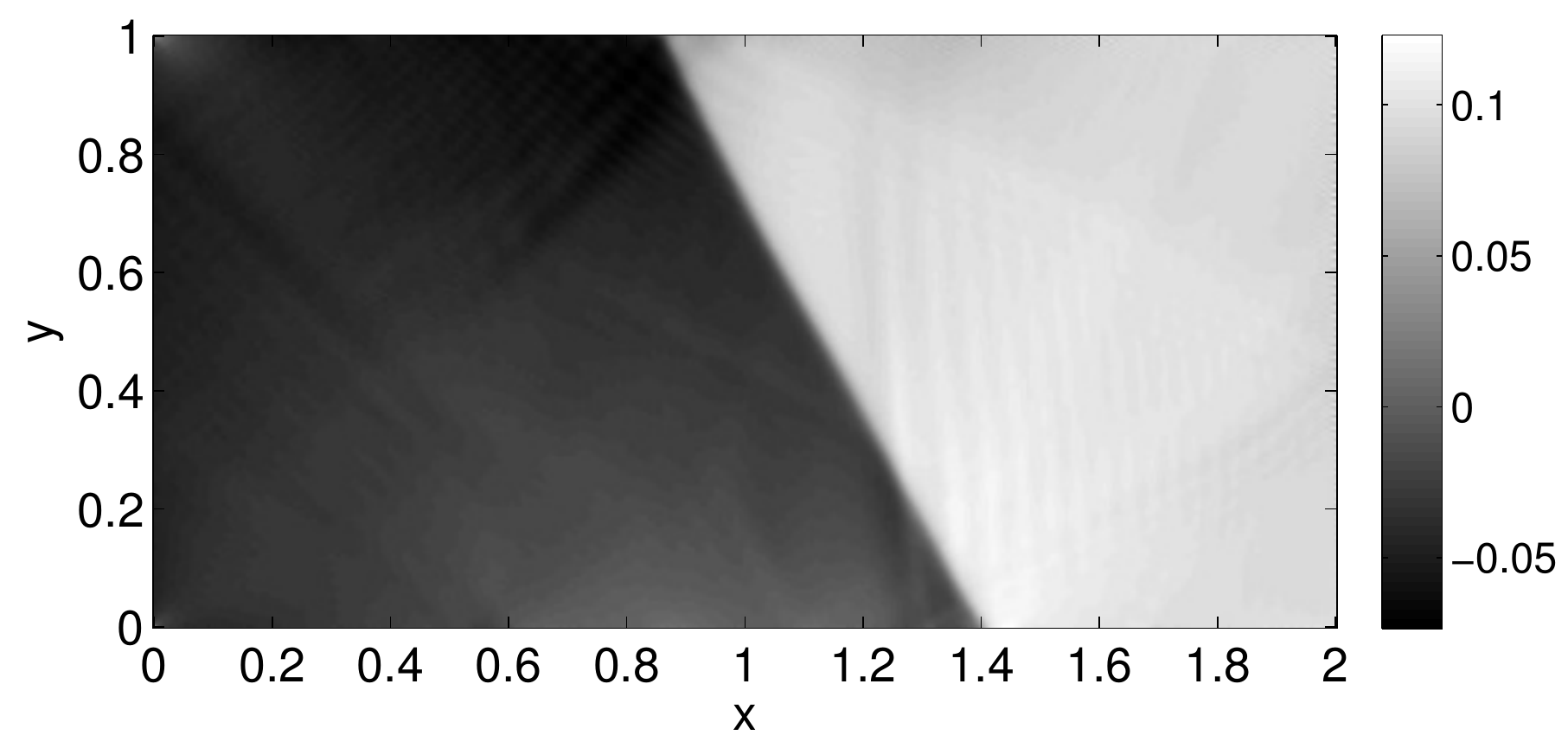}
			\caption{The top figure shows the initial configuration, the lower left figure shows the configuration after some time using linear kinetics, and the lower right figure shows the configuration after some time using non-monotone kinetics.  All plots are of the $F_{11} - 1$ field.  As in the incompatible case, there are quantitative but no obvious qualitative differences.}
			\label{fig:2d-non-monotone-compat}
		\end{center}
	\end{minipage}
	}
\end{center}
\end{figure}

The previous calculation leaves open the possibility that the kinetics is possibly complex but that momentum balance simply dominates due to the stresses that are necessarily present.
Therefore, we briefly examine a problem with a stress-free compatible interface.
The energy is described in Section \ref{sec:2D-energy}, and we use $\bfE_1$ and $\bfE_2$ as described there, with the rotated sample.
We consider a 2D rectangular plate fixed at the left edge, and traction-free at the top and bottom edges. 
A stress free compatible phase interface exists in the plate initially.
A constant tensile load is then applied at the right edge. 
Fig. \ref{fig:2d-non-monotone-compat} shows the initial configuration, and the configuration after some evolution has occurred for both linear and non-monotone kinetics.

%%%%%%%%%%%%%%%%%%%%%
%%%%%%%%%%%%%%%%%%%%%
%%%%%%%%%%%%%%%%%%%%%
%%%%%%%%%%%%%%%%%%%%%

\section{Anisotropic Kinetics in Two Dimensions}
\label{sec:aniso-kinetics}

We consider the role of anisotropic kinetics in a 2D transformation.
To isolate the role of anisotropy in the kinetics, we keep all other effects isotropic.
Therefore, we use the energy described in Section \ref{sec:2D-energy}, but with stress-free wells:
\begin{equation}
	 \bfU_1 = \alpha \bfI \quad \text{and} \quad \bfU_2 = \beta\bfI, \quad \text{ with } \alpha = 0.05, \beta = 0.1
\end{equation}
These wells not stress-free compatible.

We consider a square domain with a hydrostatic loading applied on the boundary.
The material is entirely in a single phase, but the new phase nucleates as the loading increases.
To force the nucleation to occur away from the boundaries, we make the source term heterogeneous, and of the form:
\begin{equation}
	G(\phi,\bfsigma,\bfx) = \left\{
  		\begin{array}{l l}
    			A_{0-1} H_l \big(1-\phi\big) & \quad \text{if $|\bfsigma_{11}+\bfsigma_{22}|>\sigma_0$ and $|\bfx| < 0.1$}\\
    			0 & \quad \text{otherwise}
		\end{array} \right.
\end{equation}
$A_{0-1}$ is a constant characteristic of the $0-1$ reaction, and $H_l \big(1-\phi\big)$ ensures that the $1$-phase is created only when we are not already in the $1$-phase, i.e., the nucleation term turns off when the $1$-phase has nucleated.
The nucleation term is only active in a circle centered around the middle of the domain, and is only active when the hydrostatic stress $|\bfsigma_{11}+\bfsigma_{22}|$ is above a critical stress $\sigma_0$.
We have not allowed for the reverse transformation, but it is trivial to add this if required.
The nucleation term and the loading both maintain the circular symmetry in the problem.

The kinetic response is $\hat{v}_n^\phi (f) = \mathrm{sign}(f) \kappa\ |f| \ |\nabla\phi\cdot\bfd_m |$, where $\bfd_m = \frac{\sqrt{3}}{2} \bfe_1 + \half\bfe_2$ is a distinguished material direction that sets the anisotropy.
Recall that $\nabla\phi$ sets the normal to the interface.
Therefore, the kinetic response depends on the relative orientation of the interface to $\bfd_m$, and interface velocity goes to $0$ along directions normal to $\bfd_m$.

\begin{figure}[htb!]
\begin{center}
	\fbox{
	\begin{minipage}{175mm}
		\begin{center}
			\includegraphics[width = 85mm]{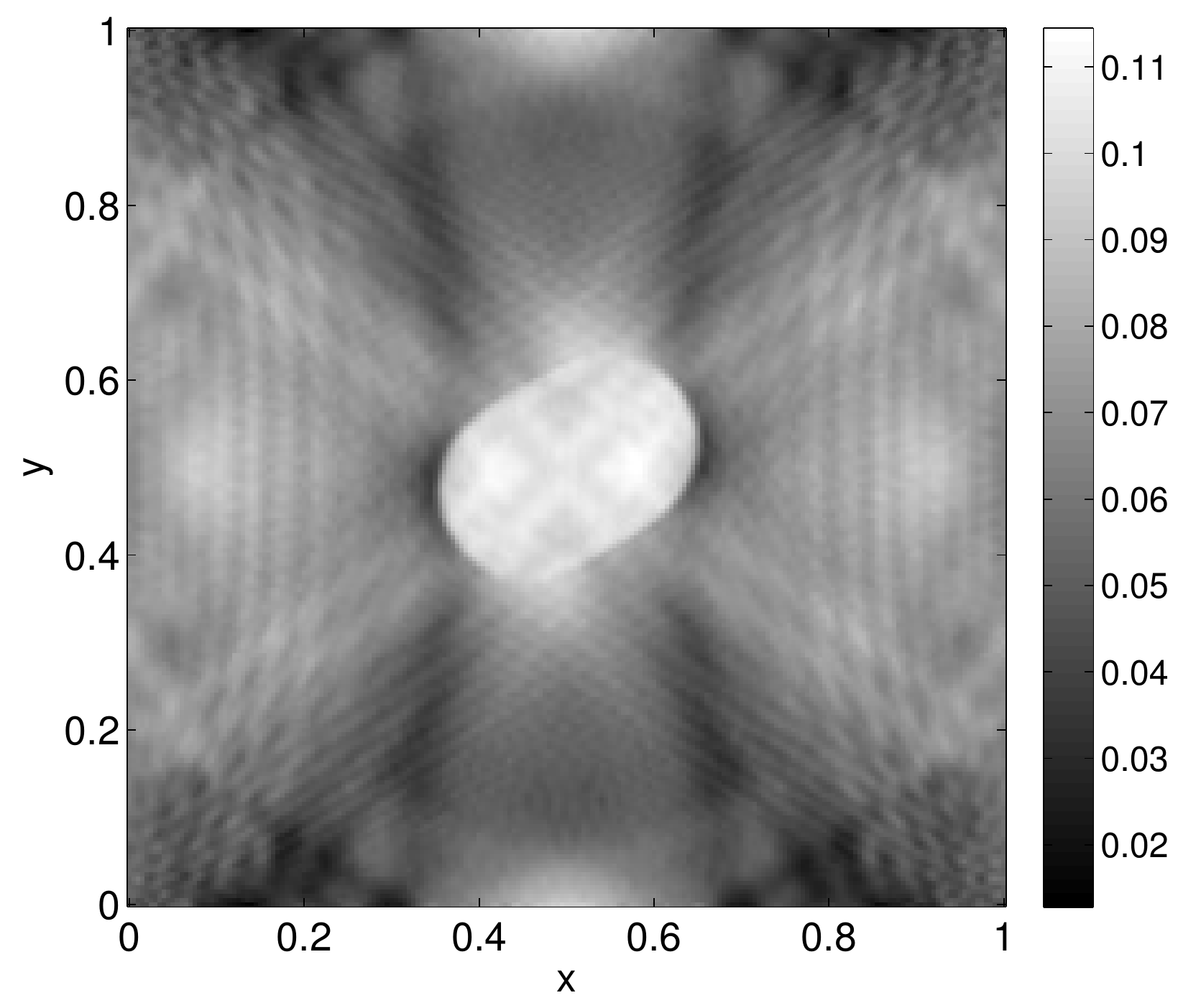}
			\includegraphics[width = 85mm]{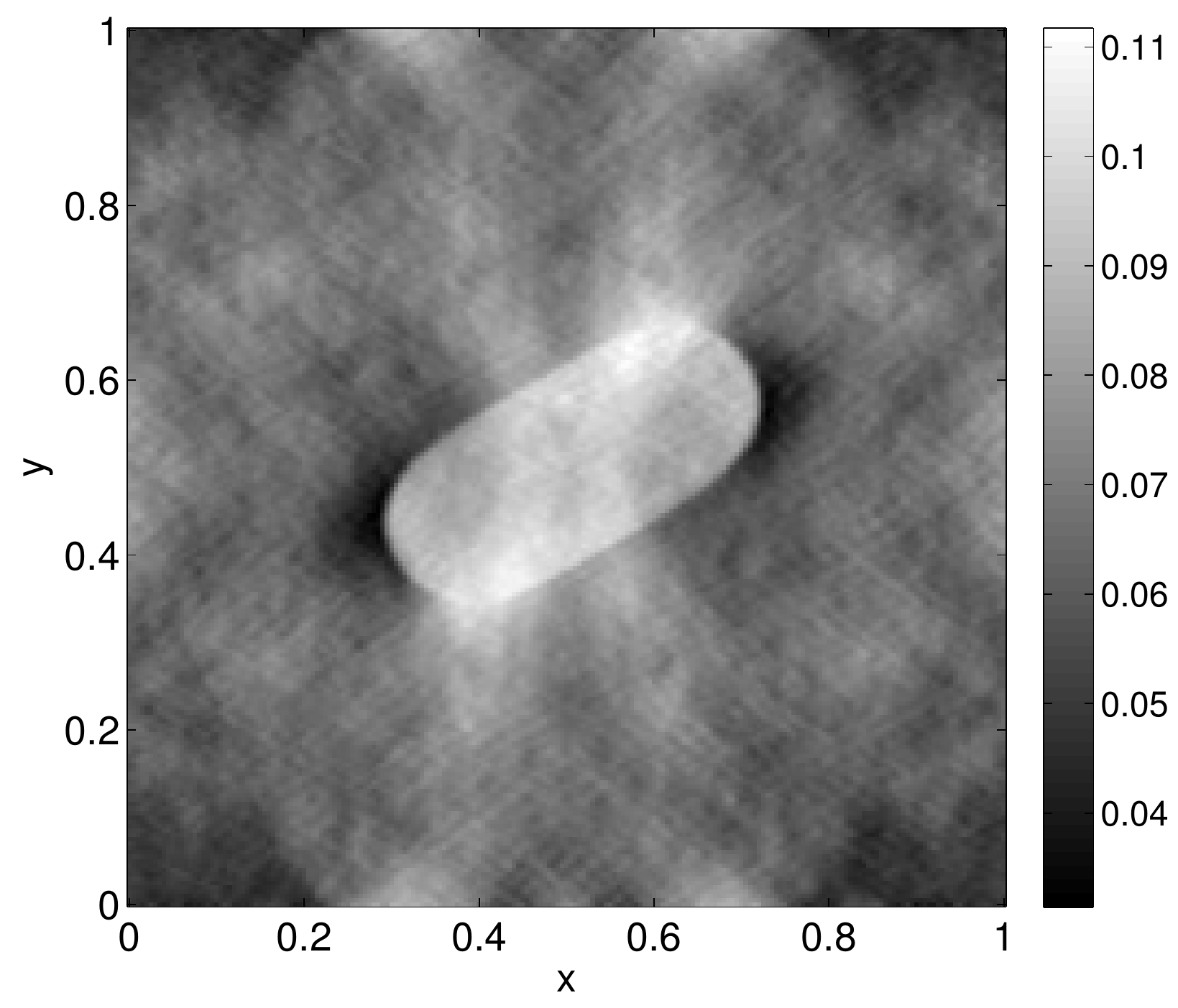}
			\caption{The $F_{11} - 1$ field at different times for an anisotropic kinetic response.  Nucleation takes place within a circle of radius $0.1$ at the center of the domain.  The interface velocity is anisotropic, and the effect can clearly be seen because everything else in the problem maintains circular symmetry.  It can be seen that regions where the interface normal is perpendicular to $\bfd_m$ do not show any motion / growth after the initial nucleation stage.}
			\label{fig:aniso-evolution}
		\end{center}
	\end{minipage}
	}
\end{center}
\end{figure}

The evolution of the deformation is shown in Fig. \ref{fig:aniso-evolution}.

%%%%%%%%%%%%%%%%%%%%%
%%%%%%%%%%%%%%%%%%%%%
%%%%%%%%%%%%%%%%%%%%%
%%%%%%%%%%%%%%%%%%%%%

\section{Stick-slip Twinning Kinetics}
\label{sec:2d-stick-slip}

We examine the evolution of twinning interfaces in 2D using stick-slip kinetics using the energy described in Section \ref{sec:2D-energy} with the rotated specimen.
We consider a 2D rectangular plate fixed at the left edge, and traction-free at the top and bottom edges. 
A stress-free compatible phase interface exists in the plate initially.
A constant tensile load is then applied at the right edge. 

\begin{figure}[htb!]
\begin{center}
	\fbox{
	\begin{minipage}{175mm}
		\begin{center}
			\includegraphics[width = 85mm]{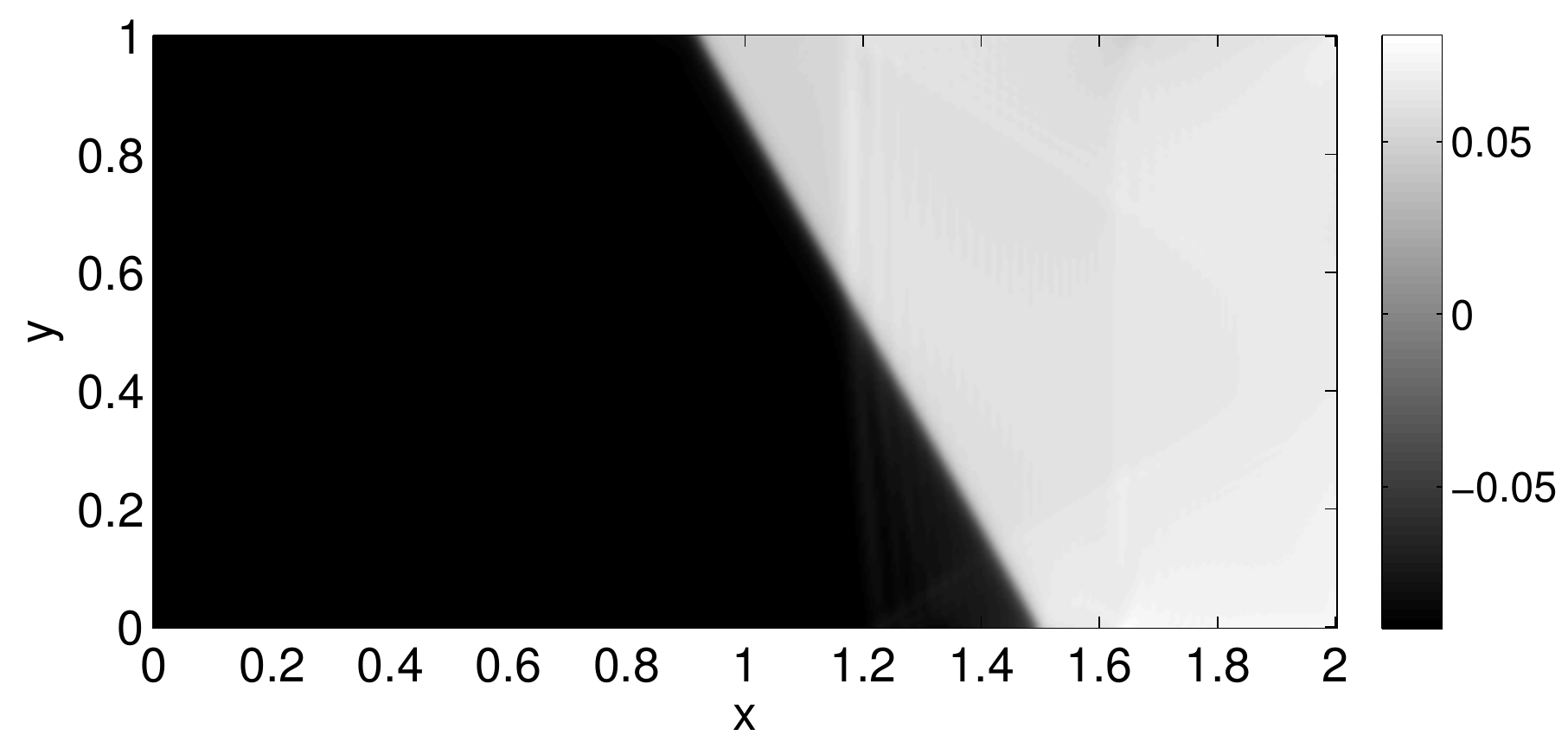}
			\\
			\includegraphics[width = 85mm]{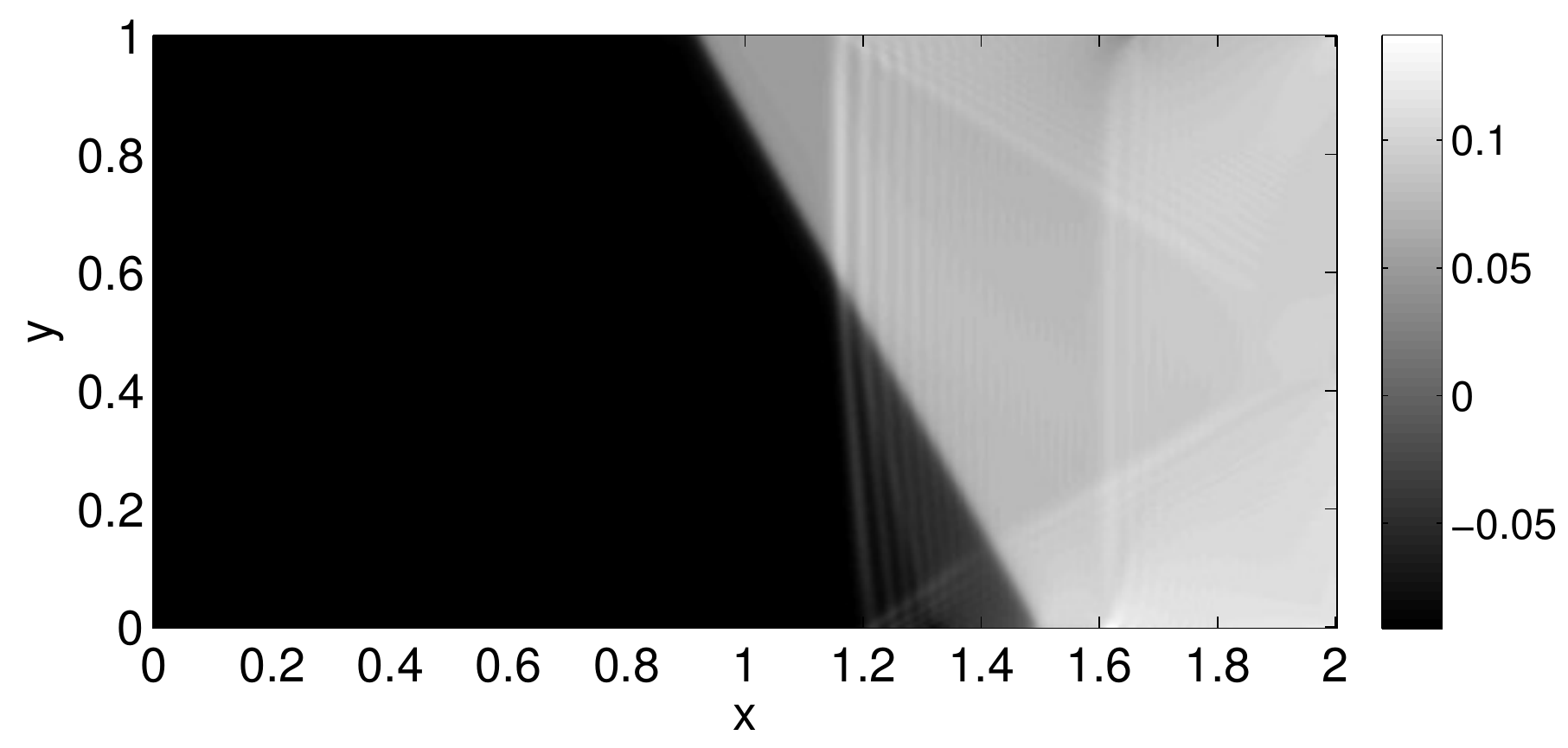}
			\includegraphics[width = 85mm]{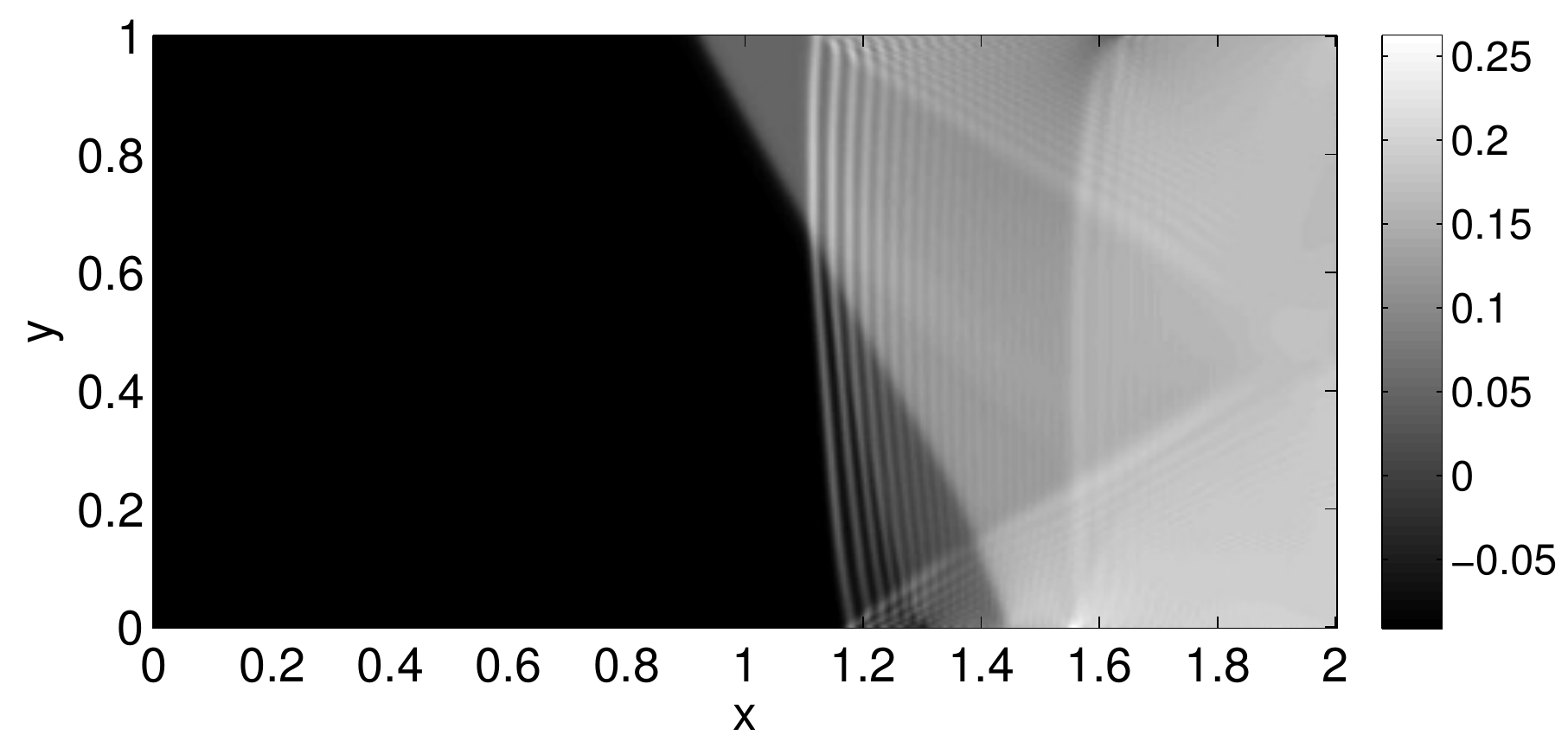}
			\caption{The $F_{11} - 1$ field at a given time for varying applied load levels.  In the first case, well below the critical load, there is no evolution; in the second case, just above the critical load, there is very slow evolution; in the third case, well above the critical load, there is rapid evolution.}
			\label{fig:2d-stick-slip}
		\end{center}
	\end{minipage}
	}
\end{center}
\end{figure}

We use a stick-slip kinetic response: $\hat{v}_n^\phi = 0 \text{ if } |f| < f_0$ and $\hat{v}_n^\phi = \kappa \mathrm{sign}(f) \ (|f|-f_0) \text{ if } |f| \geq f_0$.
We examine the evolution of the interface for a number of applied load levels.
Fig. \ref{fig:2d-stick-slip} shows the $F_{11} - 1$ field some time after evolution has commenced, for different load levels.
We find that we are able to impose stick-slip easily and effectively through the kinetic response listed above in 2D as well.

%%%%%%%%%%%%%%%%%%%%%
%%%%%%%%%%%%%%%%%%%%%
%%%%%%%%%%%%%%%%%%%%%
%%%%%%%%%%%%%%%%%%%%%

\section{Rate-Dependent and Asymmetric Nucleation}
\label{sec:rate-asym-nucleation}

We examine the prescription of complex nucleation rules in our formulation.
In this section, we demonstrate two key features: (1) that we transparently incorporate complex nucleation behavior such as rate-dependence; and (2) that we can tailor the nucleation stress easily without any modifications to the energy but purely through the activation of the source term $G$.
We perform calculations in 1D and 2D to show the ability of the model to separate kinetics and energetics from nucleation.
An additional feature that we demonstrate is that we can independently prescribe the forward and reverse nucleation stresses; i.e., given a completely symmetric energy landscape, we are able to induce nucleation in one direction at a certain critical stress, but the reverse transformation is induced at a completely independent critical value of the stress.

This gives us a powerful approach to prescribe complex nucleation criteria.
For instance, in the 1D calculation described below, we are able to change the nucleation stress by a factor of $3$ for a change of $4\%$ in the loading rate.

%%%%%%%%%%%%%%%%%%%%%
%%%%%%%%%%%%%%%%%%%%%
\subsection{Rate-Dependent and Asymmetric Nucleation in One Dimension}

We use the same energy $\Wcirc$ as in Section \ref{sec:1d-kin-relns} with linear kinetics.
This energy is quadratic around the stress-free strains $0$ and $1$, corresponding to phase $1$ and phase $2$.
The elastic moduli is the same in both phases.

The nucleation criterion is specified as follows:
\begin{equation}
	G(x) = 
		\left\{\begin{array}{l l}
			5.0  (1-H_l(\phi-0.6)) & \text{if } \sigma(x) > \sigma_{1\to 2} \\
			5.0  H_l(\phi-0.4) & \text{if } \sigma(x) < \sigma_{2\to 1} \\
			0 & \text{else}
		\end{array} \right.
\end{equation}
The term $(1-H_l(\phi-0.6))$ ensures consistency with thermodynamics in that it turns off the nucleation source for the transformation $1\to 2$ when we are in phase $2$.
Similarly, $H_l(\phi-0.4)$ turns off the source for the reverse transformation $2\to 1$ when we are in phase $1$.

Asymmetry in the forward and reverse transformations is readily prescribed by using different values for the forward and reverse threshold stresses $\sigma_{1\to 2}$ and $\sigma_{1\to 2}$.

Rate-dependence is prescribed by making the threshold stresses functions of the loading rate.
We use:
\begin{equation}
	\sigma_{1\to 2} = 
		\left\{ \begin{array}{l l}
			0.06 & \text{if } \dot{\sigma} < 5.1 \times 10^{-4} \\
			0.2 & \text{if } \dot{\sigma} \geq 5.1 \times 10^{-4}
		\end{array}	\right.
	\qquad \text{ and }
	\sigma_{2\to 1} = 
		\left\{ \begin{array}{l l}
			-0.03 & \text{if } \dot{\sigma} < 5.1 \times 10^{-4} \\
			-0.1 & \text{if } \dot{\sigma} \geq 5.1 \times 10^{-4}
		\end{array} \right.
\end{equation}

\begin{figure}[htb!]
\begin{center}
	\fbox{
	\begin{minipage}{175mm}
		\begin{center}
			\includegraphics[width = 85mm]{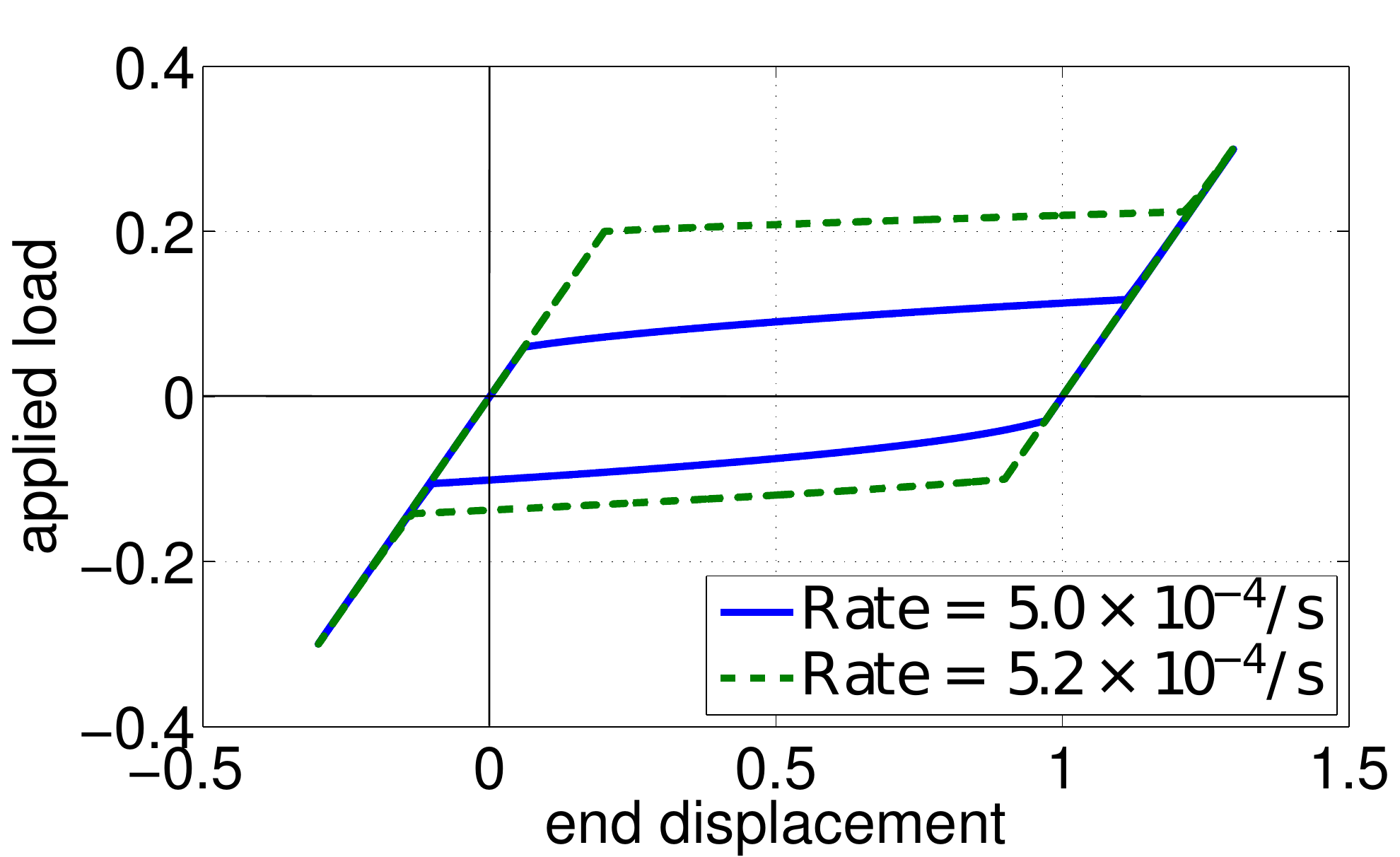}
			\includegraphics[width = 85mm]{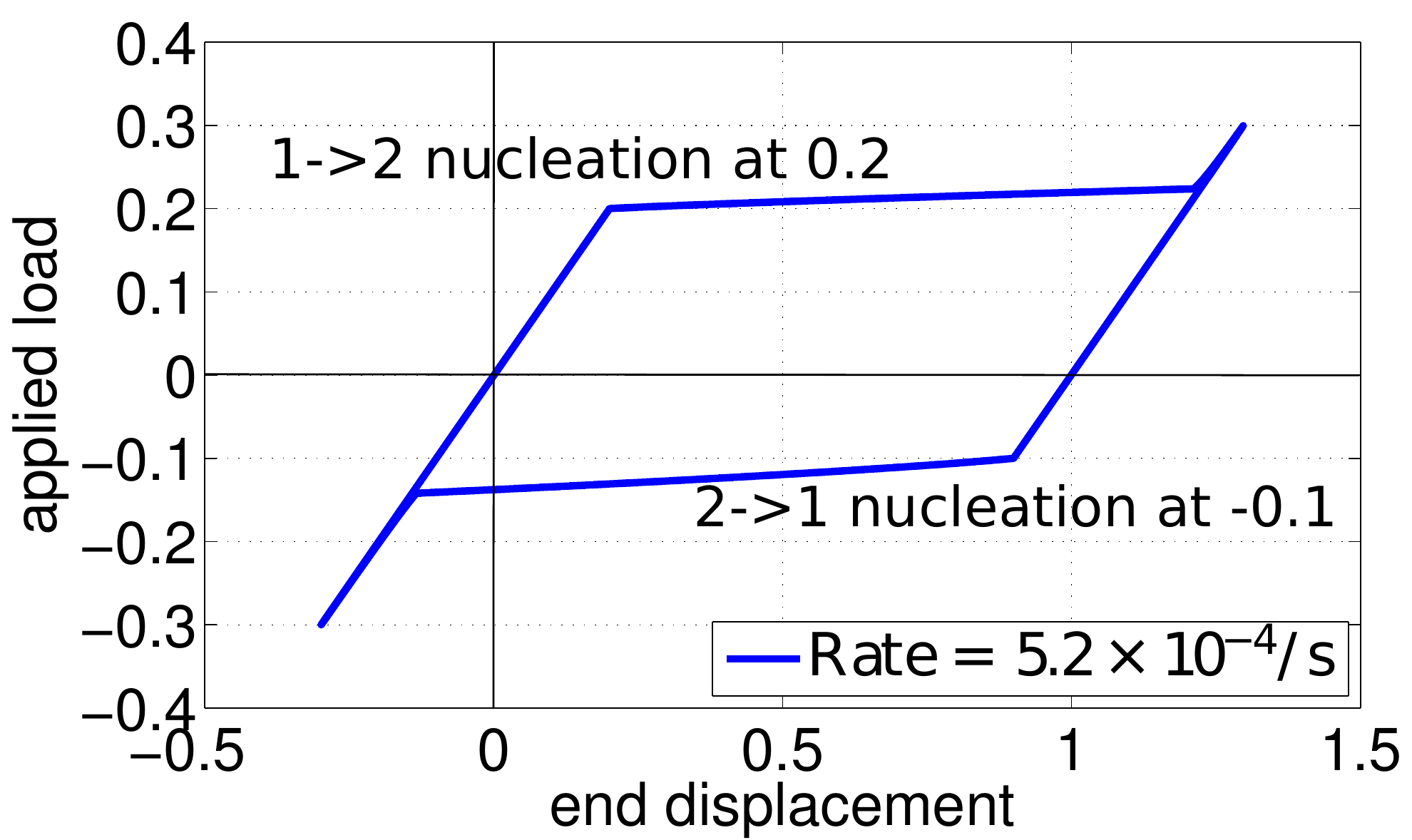}
			\caption{Left: Hysteresis curve for two loading rates demonstrating effect of rate-dependent nucleation. Right: Hysteresis curve demonstrating asymmetric nucleation, i.e., forward and reverse nucleation criteria are independently prescribed without changing the energy $\Wcirc$.  The bold horizontal line marks the stress around which the energy $\Wcirc$ is symmetric.}
			\label{fig:1d-hysteresis}
		\end{center}
	\end{minipage}
	}
\end{center}
\end{figure}

We demonstrate this nucleation criterion by computing an entire hysteresis loop, starting with the forward transformation from phase $1$ to phase $2$, and then the reverse transformation back to phase $1$ (Fig. \ref{fig:1d-hysteresis}).
This loop is computed by simply evolving the kinetic equation for $\phi$ while solving the quasistatic (i.e., without inertia) momentum balance as the evolution proceeds.

As we notice from Fig. \ref{fig:1d-hysteresis}, we are able to precisely set the nucleation stress in a direct and transparent manner.
In addition, we are able to change the nucleation stress by a factor of $3$ for a change in loading rate of $2\%$. 

%%%%%%%%%%%%%%%%%%%%%
%%%%%%%%%%%%%%%%%%%%%
\subsection{Rate-Dependent and Asymmetric Nucleation in Twinning}

We demonstrate here the possibility of allowing rate-dependence and asymmetry in nucleation for a twinning transformation.
While twinning is typically expected to be symmetric, this calculation serves as a demonstration that asymmetry in nucleation can be readily introduced in 2D even when eveything else in the problem is symmetric.

We use the energy for twinning described in Section \ref{sec:2D-energy} with linear kinetics.
The nucleation criterion is specified as follows:
\begin{equation}
	G = 
		\left\{\begin{array}{l l}
			5.0  (1-H_l(\phi-0.8)) & \text{if } \bfn\cdot\bfsigma\bfm > \sigma_{1\to 2} \\
			5.0  H_l(\phi-0.2) & \text{if } \bfn\cdot\bfsigma\bfm < \sigma_{2\to 1} \\
			0 & \text{else}
		\end{array} \right.
\end{equation}
It resembles closely the 1D version.
The key difference from 1D is that the scalar stress is here replaced by the shear traction magnitude on the planes oriented at $\pi/12$.
Hence, this is essentially a critical resolved shear stress (CRSS) type of nucleation criterion, with the difference being that we can set $\sigma_{2\to 1}$ and $\sigma_{1\to 2}$ to completely distinct values.

Asymmetry and rate-dependence are prescribed by following exactly the 1D approach of making $\sigma_{2\to 1}$ and $\sigma_{1\to 2}$ different from each other and rate-dependent.
For simplicity, we set the rate-dependence through the $11$-component of the nonlinear strain measure.
For realistic calculations, a model based on an objective rate would be appropriate.

\begin{figure}[htb!]
\begin{center}
	\fbox{
	\begin{minipage}{125mm}
		\begin{center}
			\includegraphics[width = 120mm]{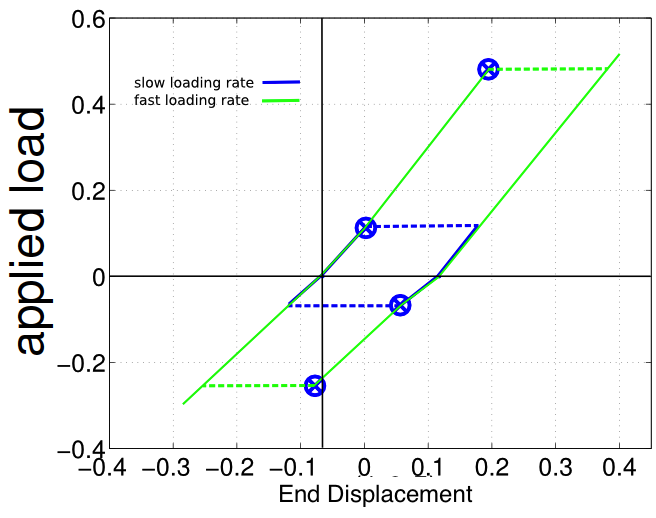}
			\caption{A hysteresis loop for twinning demonstrating the ability to apply rate-dependent and asymmetric nucleation criteria in 2D.  For computational efficiency, we have only computed the nucleation points (in the vicinity of the circles), and interpolated between them.}
			\label{fig:2d-hysteresis}
		\end{center}
	\end{minipage}
	}
\end{center}
\end{figure}

Fig. \ref{fig:2d-hysteresis} shows the hysteresis loop that demonstrates both asymmetry and rate-dependence of our evolution law.

%%%%%%%%%%%%%%%%%%%%%
%%%%%%%%%%%%%%%%%%%%%
%%%%%%%%%%%%%%%%%%%%%
%%%%%%%%%%%%%%%%%%%%%
\section{Imposing Hydrostatic- and Shear- Dependent Nucleation for Twinning}
\label{sec:2d-hydro-nucleation}

Defect nucleation and kinetics often does not depend exclusively on the driving force.
E.g., recent studies of dislocation motion in BCC metals show that other stress components can be important in addition to the shear-related driving force \cite{groger2008multiscale-1, groger2008multiscale-2}.
Motivated by these studies, we examine twin nucleation within our model, where the nucleation criterion has hydrostatic stress dependence in addition to the critical resolved shear stress (CRSS) for twin nucleation.
While this nucleation criterion is not motivated by any specific observations in experiments or atomistics, it serves as a toy model to demonstrate the transparent specification of complex nucleation criteria.

The elastic energy is as described in Section \ref{sec:2D-energy}, with the unrotated specimen.
We use simple linear kinetics.
Our nucleation criterion is as follows:
\begin{equation}
\label{eqn:2d-hydro-nucleation}
	G(\phi,\bfsigma) = 
		\left\{\begin{array}{l l}
			A_{0\to1} H_l\big(0.6-\phi\big) & \quad \text{if $|\bfb\cdot\bfsigma\bfa|>\tau_{CRSS}$ \text{ and } $\max \{ |\sigma_{principal}|\} > \sigma_0$ } \\
			0 & \quad \text{otherwise}
		\end{array} \right.
\end{equation} 
$A_{0\to1}$ is a constant which controls the rate of nucleation.
$H_l\big(0.6-\phi\big)$ causes nucleation to be active when $\phi<1$ and ``turns of'' when $\phi$ is close to 1. 
To prevent nucleation near the boundaries, $G$ is active only in a circle of radius $0.35$ centered at the middle of the domain and $0$ elsewhere.

The CRSS character is built into the nucleation stress through the condition $|\bfb\cdot\bfsigma\bfa|>\tau_{CRSS}$.
The vectors $\bfa  = \frac{1}{\sqrt{2}}\begin{pmatrix} 1 \\ 1 \end{pmatrix}$ and $\bfb = \frac{1}{\sqrt{2}}\begin{pmatrix} -1 \\ 1 \end{pmatrix}$ are the twin normal and shear direction, thereby resolving the shear stress in the direction appropriate for twin nucleation.
We note that it is direct and transparent to have the nucleation stress have CRSS character as in the equation.
It is equally direct to add a dependence on the hydrostatic stress state if required; in our example, we have taken a simple dependence that requires a minimum level of hydrostatic stress.
More complex pressure-sensitive nucleation responses can be easily incorporated in our model.

We examine the behavior of this nucleation response through the interaction of an elastic wave with the residual stress field around an inclusion.
A square 2D single-phase domain is considered with a circular non-transforming elastic inclusion at the center. 
The elastic inclusion is stress-free at $\bfF=\bfI$ but the surrounding matrix has $\bfF=\bfU$; therefore, the interface between the inclusion and the matrix sets up a stress field.
A shear at -- or above -- the critical level required by the CRSS criterion would not cause nucleation unless there is also sufficient hydrostatic stress.
The residual stress field around the misfitting inclusion provides this stress.

The specimen is fixed on the left boundary, traction-free at the top and bottom, and a constant load is applied to the right.
This load results in an elastic wave which interacts with the stress field around the inclusion and leads to nucleation when the critical nucleation conditions are satisfied.

The inclusion is elastic in the sense that it has a finite elastic modulus.
However, it has only a single stress-free deformation state at $\bfF = \bfI$.
Therefore, $\Wcirc$ is independent of $\phi$, and the energy within the inclusion has only the term $\half\epsilon|\nabla\phi|^2$ that depends on $\phi$.
Therefore, the evolution of $\phi$ is diffusion-driven within the inclusion, and interacts with the matrix through the value of $\phi$ on the boundary.
Physically the evolution of $\phi$ within the inclusion has no meaning, but it nonetheless affects the evolution in the matrix through the interaction on the inclusion-matrix boundary.
Therefore, the ideal strategy would be to not define $\phi$ at all within the inclusion and use a separate boundary kinetic equation that reflects the physics of the interface interacting with the inclusion-matrix boundary (see Section \ref{sec:boundary-kinetics}).
A simpler approach may be to define the kinetics within the inclusion in such a way as to mimic the correct physics at the inclusion-matrix boundary.
For now, we ignore these issues and simply evolve $\phi$ following the bulk driving force.

\begin{figure}[htb!]
\begin{center}
	\fbox{
	\begin{minipage}{175mm}
		\begin{center}
			\includegraphics[width = 85mm]{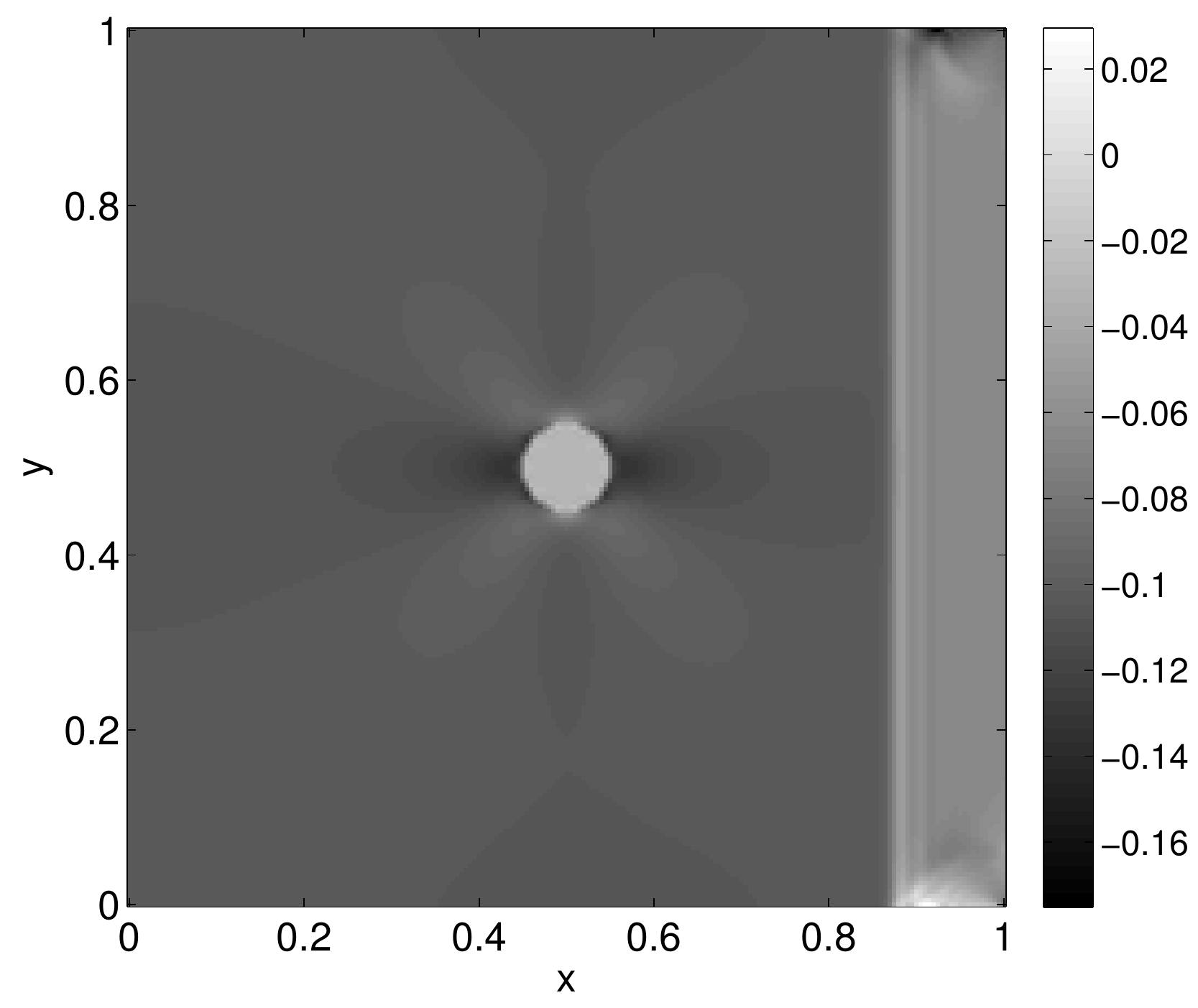}
			\includegraphics[width = 85mm]{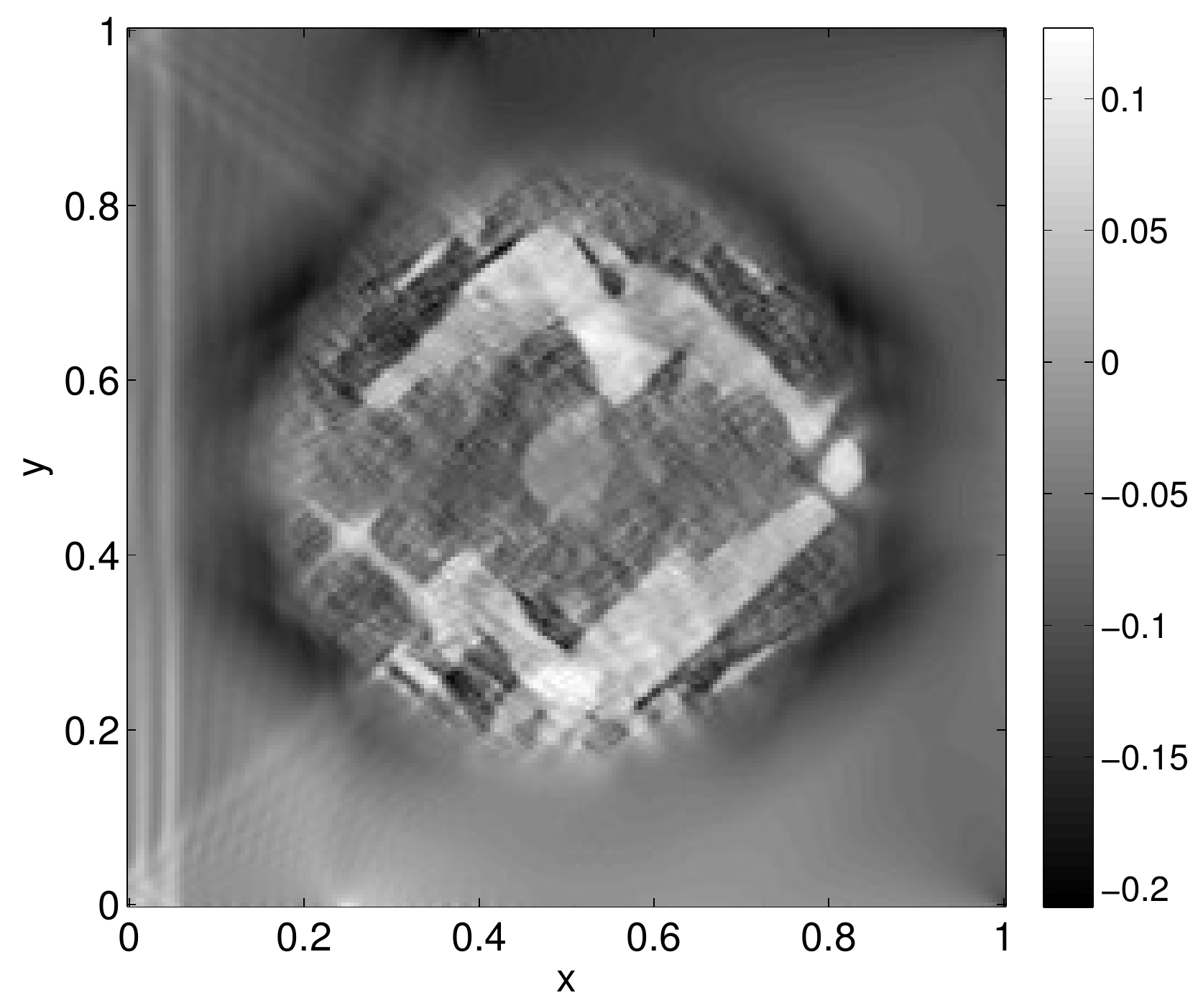}
			\caption{The $F_{11} - 1$ field before the elastic waves interacts with the stress field around the inclusion (left), and after (right).  Complex nucleation patterns can be seen, with the elastic stresses driving the interfaces -- through the driving force in the kinetic response -- to incline at $\pm \pi/4$.  Elastic waves that are created during the process of nucleation are also seen.}	
			\label{fig:2d-hydro-nucleation}
		\end{center}
	\end{minipage}
	}
\end{center}
\end{figure}

%%%%%%%%%%%%%%%%%%%%%
%%%%%%%%%%%%%%%%%%%%%
%%%%%%%%%%%%%%%%%%%%%
%%%%%%%%%%%%%%%%%%%%%
\section{Competition Between Nucleation and Kinetics in Twinning}
\label{sec:kin-vs-nucleation}

We examine the competition between kinetics and nucleation in the evolution of microstructure.
We use precisely the same material model, specimen geometry and inclusion, and loading conditions as in Section \ref{sec:2d-hydro-nucleation}.
Recalling that $G$ is a nucleation rate, we examine the effect of a large nucleation rate and a small nucleation rate, while keeping the kinetic response the same in both cases.
Varying the nucleation rate simply involves changing the value of $A_{0\to1}$ in \eqref{eqn:2d-hydro-nucleation}.

\begin{figure}[htb!]
\begin{center}
	\fbox{
	\begin{minipage}{155mm}
		\begin{center}
			\includegraphics[width = 75mm]{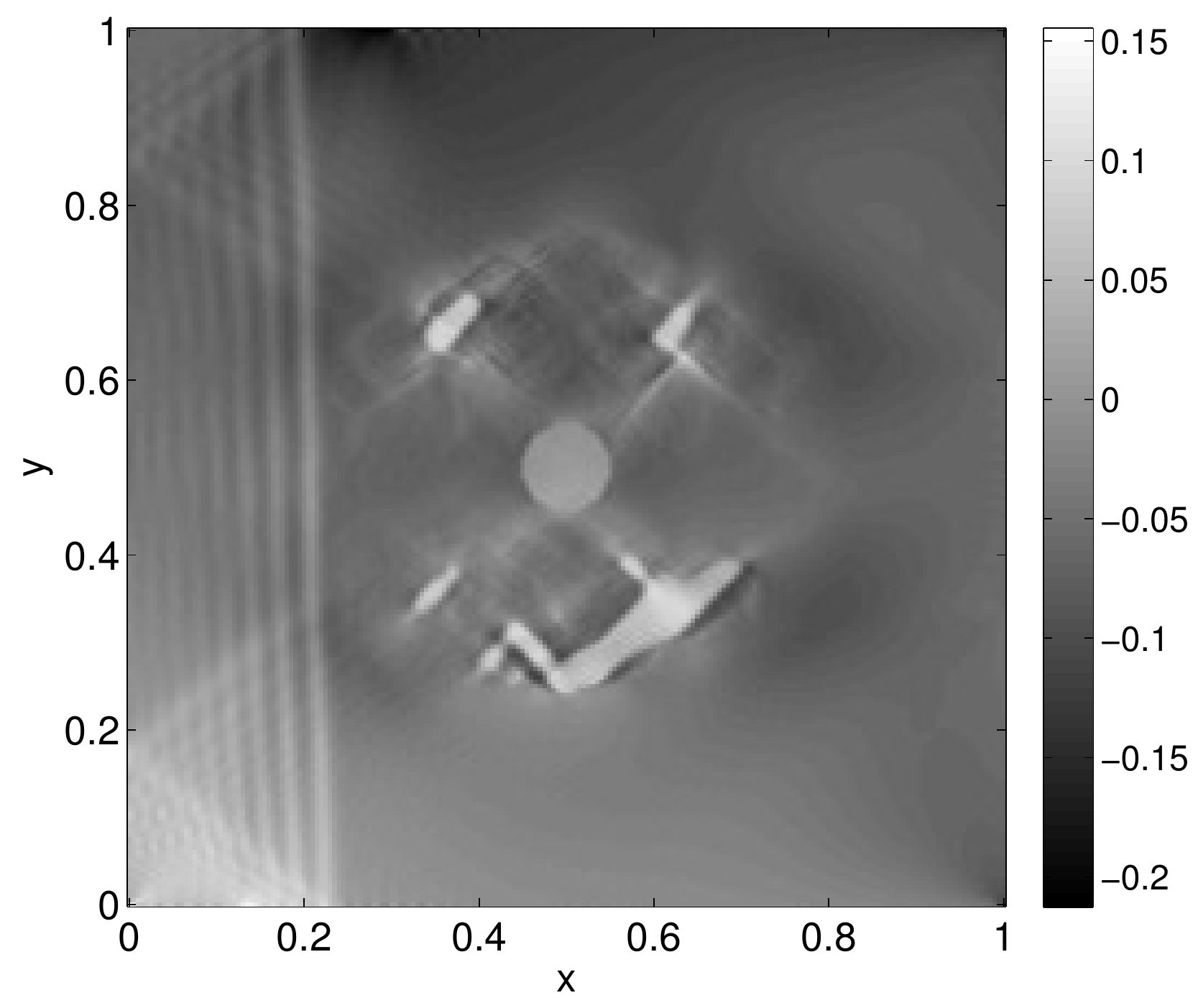}
			\includegraphics[width = 75mm]{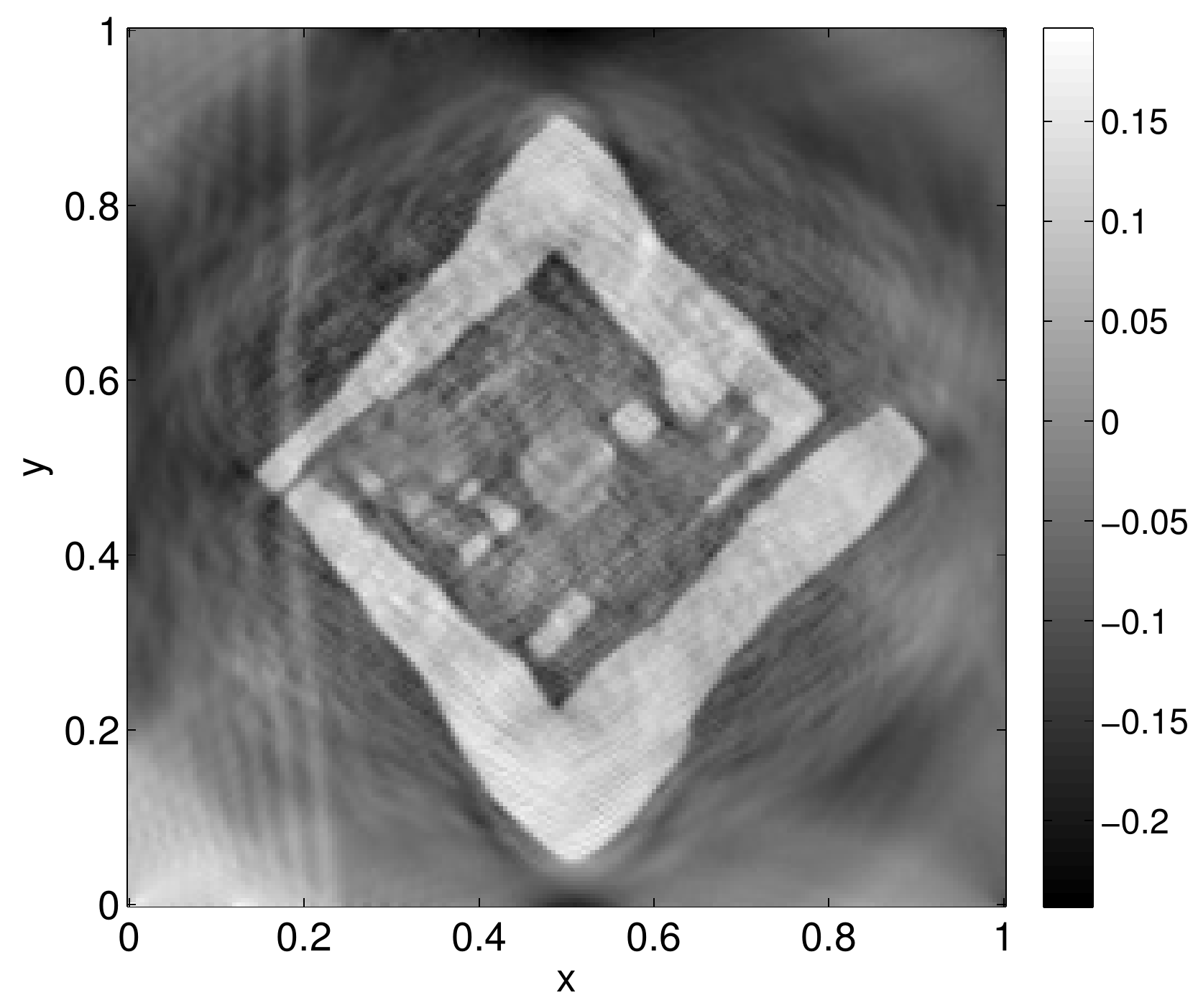}
			\\
			\includegraphics[width = 75mm]{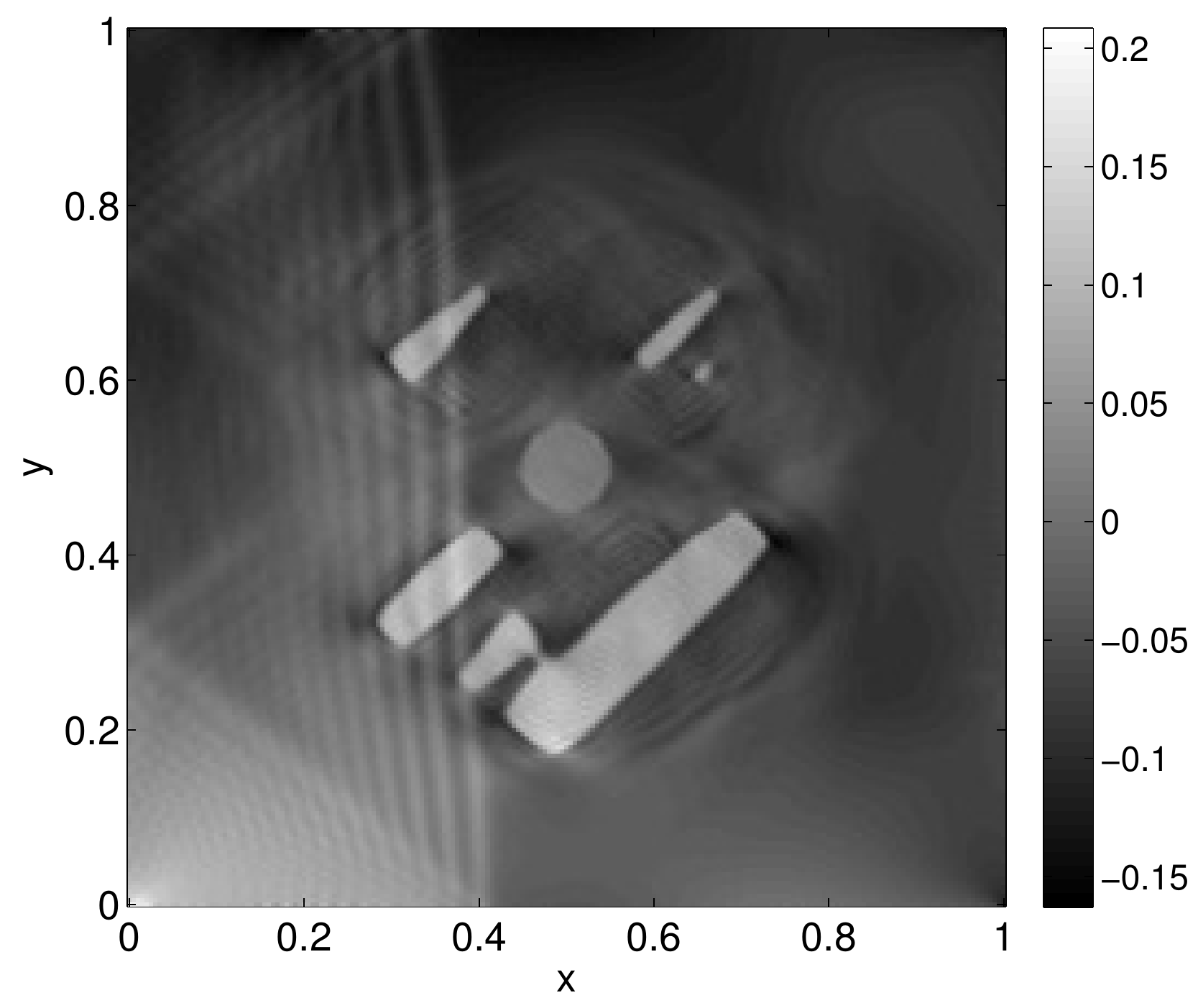}
			\includegraphics[width = 75mm]{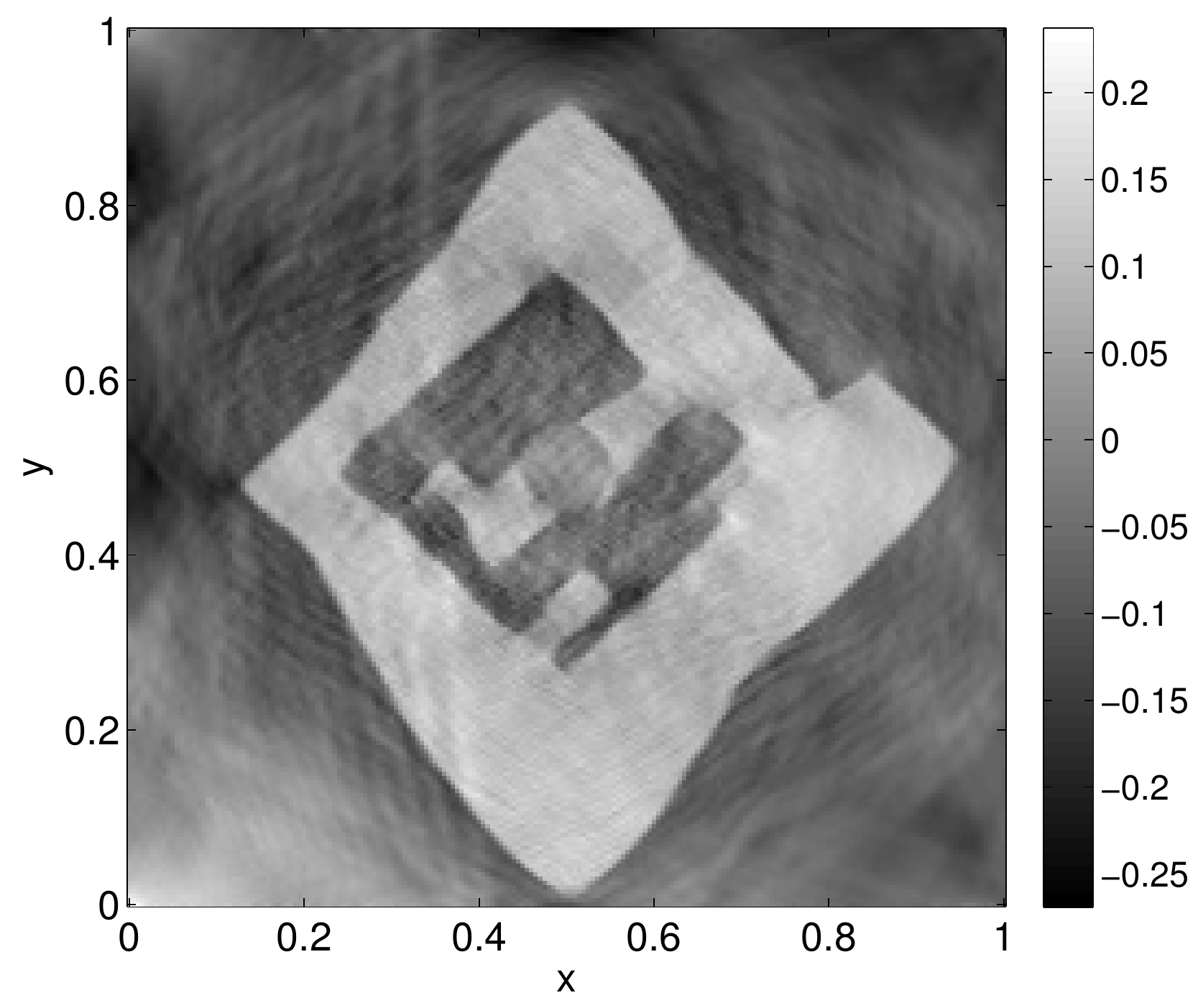}
			\\
			\includegraphics[width = 75mm]{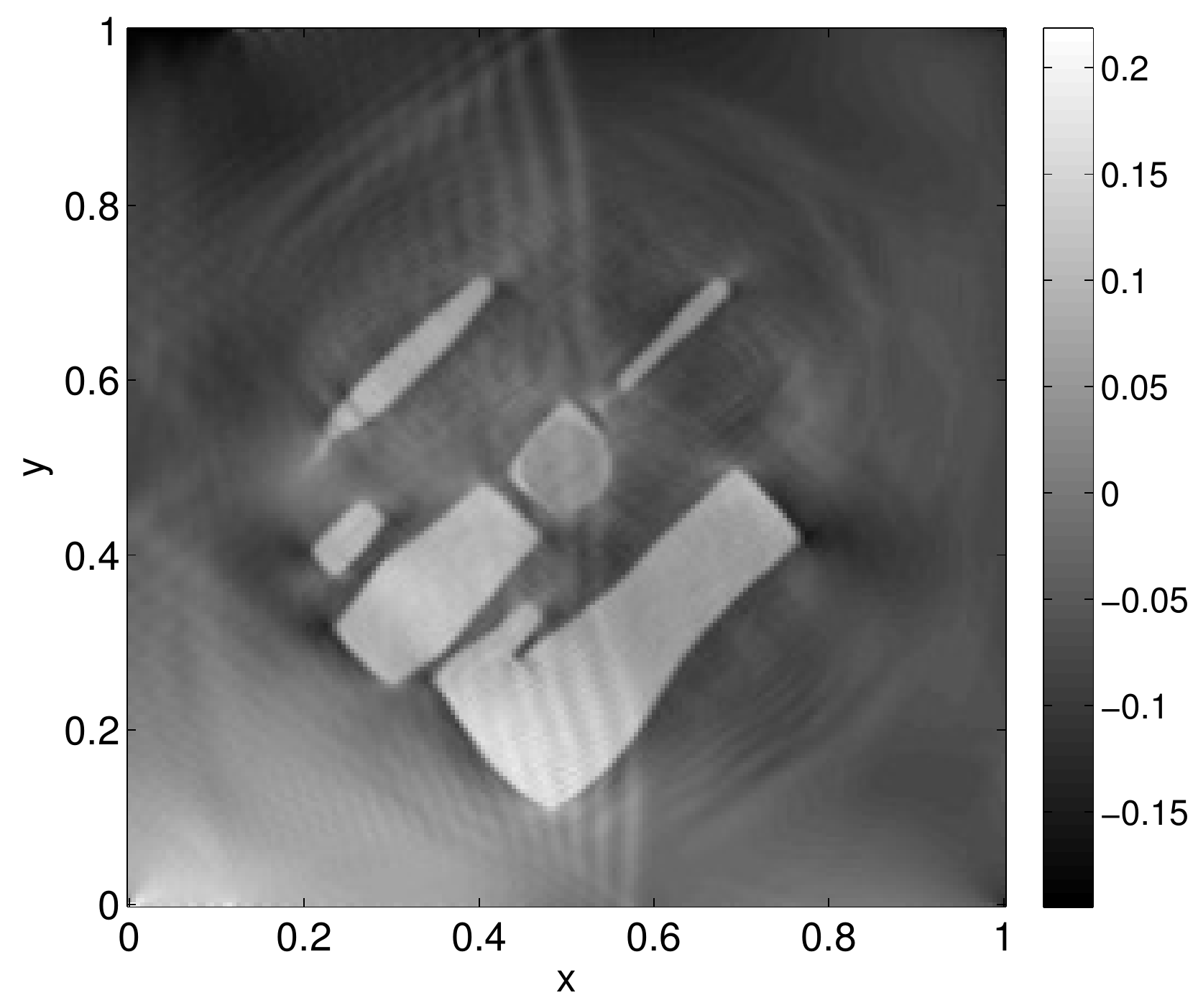}
			\includegraphics[width = 75mm]{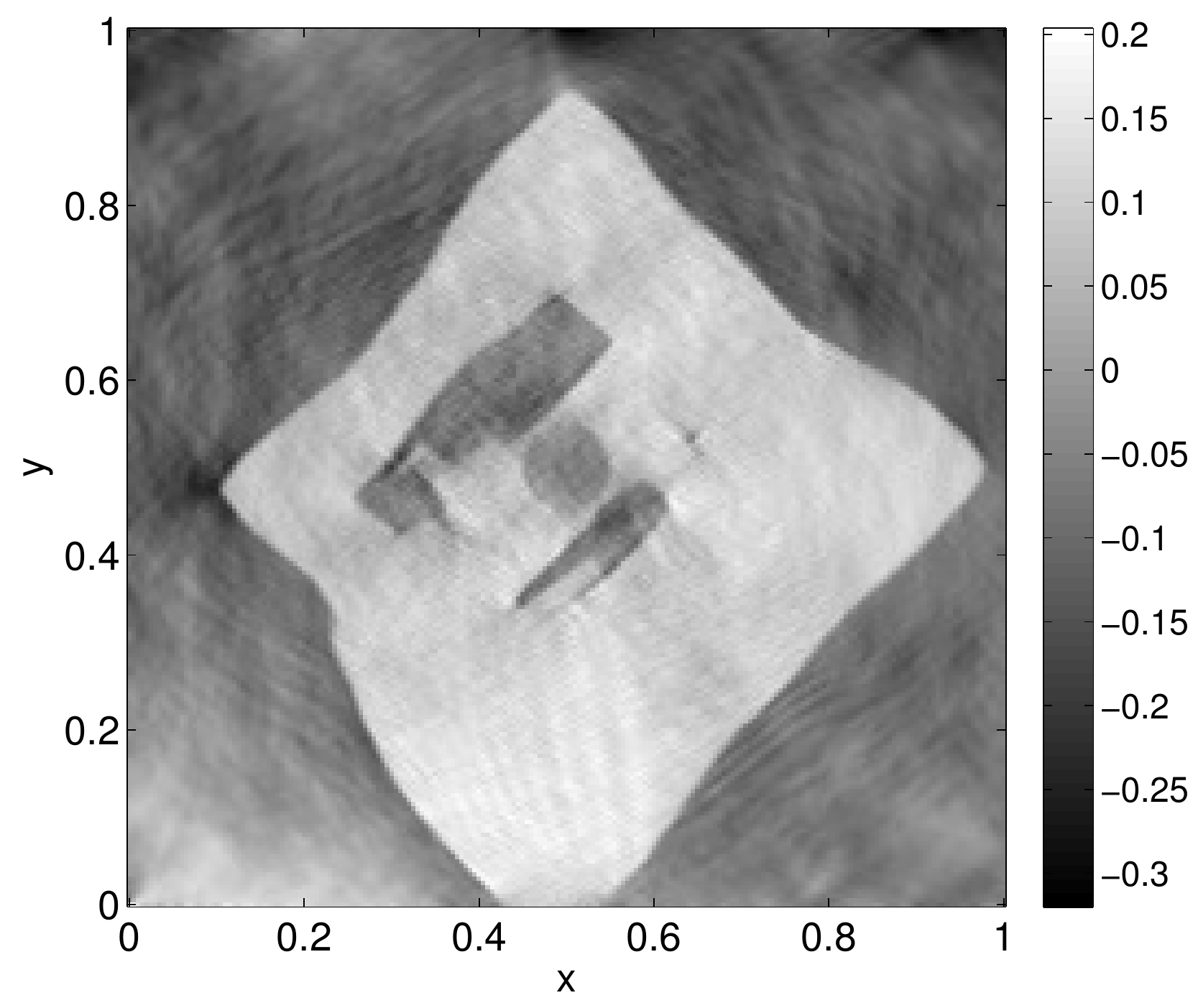}
			\caption{The left column is the evolution with large nucleation rate ($A_{0\to 1} = 2$), and the right column is the evolution with small nucleation rate ($A_{0\to 1} = 20$).  The plots show snapshots of the $F_{11} - 1$ field taken at the same time for both processes.}
			\label{fig:kin-vs-nucleation}
		\end{center}
	\end{minipage}
	}
\end{center}
\end{figure}

The results of the calculations are shown in Fig. \ref{fig:kin-vs-nucleation}.
A key qualitative difference is that when the nucleation rate is smaller, we find that the microstructure is finer as compared to the larger nucleation rate.
Heuristically, it appears that the high nucleation rate enables the formation of larger nuclei that quickly coalesce to form relatively coarse-microstructure.
For lower nucleation rates, the microstructure takes longer to develop and there are a number of smaller nuclei.
It is important to note that the total area of the transformed regions is smaller when the nucleation rate is lower, as is reasonable.

%%%%%%%%%%%%%%%%%%%%%
%%%%%%%%%%%%%%%%%%%%%
%%%%%%%%%%%%%%%%%%%%%
%%%%%%%%%%%%%%%%%%%%%

\section{Competition Between Thermodynamics and Momentum Balance}
\label{sec:thermo-vs-momentum}

Phase interfaces in continuum mechanics provide an interesting demonstration of the competition between momentum balance and thermodynamics.
Consider elastic shocks or their analog in gasdynamics: the behavior of these interfaces is almost completely constrained by momentum balance.
Thermodynamics -- in the form of positive dissipation -- typically serves only to select one of two possible solutions permitted by momentum balance.
Typical phase interfaces are quite different from elastic shocks.
Momentum balance provides only a weak constraint on the solutions, and there is a massive non-uniqueness that is left open.
Thermodynamics -- in the form of a kinetic relation -- selects the unique solution.

Regularized models of elastic shocks and phase interfaces also display this character.
The addition of viscosity and gradient terms serves to regularize elastic shocks, but does not significantly change the kinetics if these regularizing mechanisms are sufficiently small.
On the other hand, viscosity and gradient terms completely determine the kinetics of phase interfaces regardless of how small they may be.
If one thinks of viscosity and these higher-order terms as related to thermodynamics, we see again the contrast between elastic shocks and phase interfaces.

In our model, this interplay may be observed in 2 ways.
First, we recover the classical continuum driving force only in the quasistatic limit without inertia\footnote{Further assumptions are necessary, but not relevant to this discussion}. 
In addition, when we compare the prescribed kinetic response with the observed relation between classical driving force and interface velocity, we find that the disagreement becomes larger as we approach the sonic velocity.
Therefore, it is reasonable to consider the Mach number $M$ as a measure of the relative dominance of momentum balance and thermodynamics; $M=0$ corresponds to thermodynamic-dominance, and $M=1$ corresponds to momentum balance dominance.

\begin{figure}[htb!]
\begin{center}
	\fbox{
	\begin{minipage}{125mm}
		\begin{center}
			\includegraphics[width = 75mm]{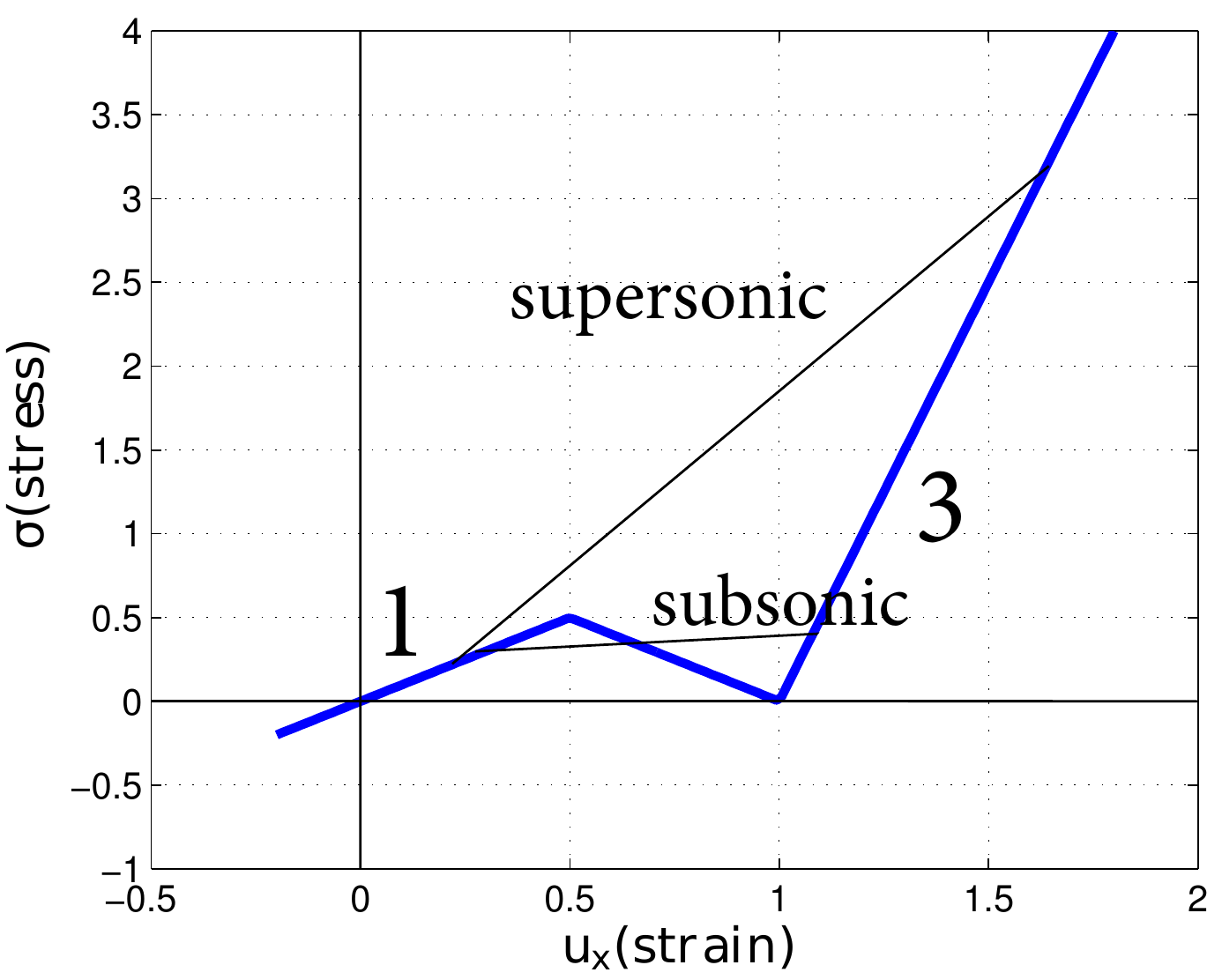}
			\caption{The stress-strain curve showing phases $1, 2, 3$.  The chords link the states on either side of the phase interface.  The upper chord is a supersonic interface, and the lower chord is subsonic. The slope of a chord gives the interface velocity, and the slope of a stress-strain branch gives the sonic speed for that branch.}
			\label{fig:supersonic-kinetics}
		\end{center}
	\end{minipage}
	}
\end{center}
\end{figure}

This brings us to an interesting example studied by \cite{abeyaratne1991kinetic,rosakis-straingrad}.
Consider a 1D problem with the material model shown in Fig. \ref{fig:supersonic-kinetics}.
Let the elastic modulus of phase $1$ be less than the modulus of phase $3$, but let them have the same mass density.
Therefore, the sonic speeds in these phases satisfy $c_1 < c_3$.
For an interface moving at velocity $v$, define $M_1 := v/c_1$ and $M_3 := v/c_3$.
Note that $M_3 < M_1$ for all $v$.

\cite{abeyaratne1991kinetic,rosakis-straingrad} consider, among various topics, a phase interface that bridges the phases $1$ and $3$.
To briefly summarize their findings regarding this problem, they find that momentum balance permits this phase interface to have $M_1 > 1$ if it is propagating into phase $3$.
Further, when $M_1 < 1$, then the interface requires a kinetic relation for unique evolution, but when $M_1 > 1$ the interface {\em evolution is fully determined by momentum balance}.
Note that $M_1 < 1$ is required when the interface propagates into phase $1$, and $M_3 <1$ always.

Therefore, an important challenge for the model that we have proposed is whether it can naturally capture this transition from thermodynamic-dominated evolution to momentum-balance-dominated evolution.
In other words, suppose we perform a dynamic calculation in which the velocity of the interface starts off subsonic, but at some point becomes supersonic.
Will the model ``automatically'' know that the interface should not be governed by the kinetic response once it transitions to supersonic?

We examine this question using a combination of traveling wave analysis and dynamic calculations.
We  work with the following model:
\begin{align}
	\Wcirc(u_x,\phi) &= \Big(1-H_a(\phi-0.5)\Big)\half C_1 u_x^2 + H_a(\phi-0.5) \half C_3 (u_x - \eps_3)^2 \\
	\sigma &= \parderiv{\Wcirc}{(u_x)} = \Big(1-H_a(\phi-0.5)\Big) C_1 u_x + H_a(\phi-0.5) C_3 (u_x - \eps_3)
\end{align}
where the stress-free strains are $0$ and $\epsilon_3 > 0$.
The elastic moduli are $C_1, C_3$ with $C_1 < C_3$.

We use the traveling wave ansatz $u(x,t) = U(x - Vt), \phi(x,t) = \Phi(x-Vt), \sigma(x,t) = \Sigma(x-Vt)$ and similarly for other fields, with $V>0$.
Integrating $\rho \ddot{u} = \deriv{\sigma}{x} \Rightarrow \rho V^2 U'' = \Sigma'$ once, we get:
\begin{equation}
	\rho V^2 U' = \Big(1-H_a(\Phi-0.5)\Big) C_1 U' + H_a(\Phi-0.5) C_3 \cdot (U' - \eps_3) + \const
\end{equation}
Using $M_1 = \frac{V}{(C_1 / \rho)^\half}$ and $M_3 = \frac{V}{(C_3 / \rho)^\half}$ and collecting the terms multiplying $U'$, we get:
\begin{equation}
\label{eqn:uprime-unequal}
	U' = \frac{C - \epsilon_3 M_1^2 H_a(\Phi-0.5)}{M_3^2 (M_1^2 - 1) - (M_1^2 - M_3^2) H_a(\Phi-0.5)}
\end{equation}
We can get 3 useful results from \eqref{eqn:uprime-unequal}.

First, consider that $\Phi|_{-\infty} = 0, \Phi|_{+\infty} = 1$, and define $\epsilon^- \equiv U'|_{-\infty}, \epsilon^+ \equiv U'|_{+\infty}$.
Evaluating \eqref{eqn:uprime-unequal} at $\pm\infty$ and subtracting gives $\epsilon^+ - \epsilon^- = \epsilon_3 \frac{M_1^2}{M_3^2} \frac{1}{1-M_1^2}$.
The jump in $\epsilon$ is positive if $M_1 < 1$ and negative if $| M_1 | > 1$.
From the fact that we have phase $1$ on the left, the jump in $\epsilon$ is expected to be positive.

Second, consider precisely $M_1 = 1$.
We have $U' = \frac{C - \epsilon_3 H_a(\Phi-0.5)}{- (1 - M_3^2) H_a(\Phi-0.5)}$.
Since $\Phi|_{-\infty} = 0$, we require that $C=0$ for this case to have $U'$ bounded.
Therefore, $U' = \frac{- \epsilon_3 H_a(\Phi-0.5)}{- (1 - M_3^2) H_a(\Phi-0.5)}$, which implies that at a given spatial location, either (i) $U'=\frac{\epsilon_3}{1 - M_3^2}$, or (ii) $H_a(\Phi-0.5) = 0$.
These conditions imply that the strain is constant in the vicinity of an interface in $\Phi$, and the strain can transition from one phase to another only {\em away} from an interface in $\Phi$.
Our dynamic calculations, described below, show this feature that the interfaces in the $U$ field and the $\Phi$ field are at different spatial locations as $M_1 \to 1$.
It appears that the system responds to over-constraining by this mechanism of separating the evolution of $\phi$ from the evolution of $u$.

Third, recall that $M_3 < M_1$, and examine the denominator in \eqref{eqn:uprime-unequal} for $M_1 < 1$ and $M_1 \geq 1$.
For all values of $M_1 < 1$, the denominator is positive and therefore $U'$ is bounded everywhere.
On the other hand, when $M_1 \geq 1$, the denominator goes to $0$ when $H_a(\Phi-0.5) = \frac{M_3^2 (M_1^2 - 1)}{M_1^2 - M_3^2} = \frac{M_1^2 - 1}{M_1^2 / M_3^2 - 1}$.
Using that $M_3 < 1$, we have that $0 < \frac{M_1^2 - 1}{M_1^2 / M_3^2 - 1} < 1$.
Noting that $H_a$ takes values between $0$ and $1$, it follows that the condition $H_a(\Phi-0.5) = \frac{M_1^2 - 1}{M_1^2 / M_3^2 - 1}$ is satisfied at some spatial location(s) for every $M_1 \geq 1$.
Therefore, our model will display unbounded strain at some point in the domain for interfaces that move faster than $M_1 > 1$.

We now examine this question through direct dynamic calculations.
While the traveling wave framework has already ruled out existence of supersonic interfaces in our model, it is based on the assumption of steadily-propagating interfaces.
Dynamic simulations enable us to probe the transient behavior as interfaces accelerate towards the sonic speed.
Essentially, we expect that the dynamics will not tend to a steady traveling wave state because such a state has been shown above to not exist.

\begin{figure}[htb!]
\begin{center}
	\fbox{
	\begin{minipage}{175mm}
		\begin{center}
			\includegraphics[width = 85mm]{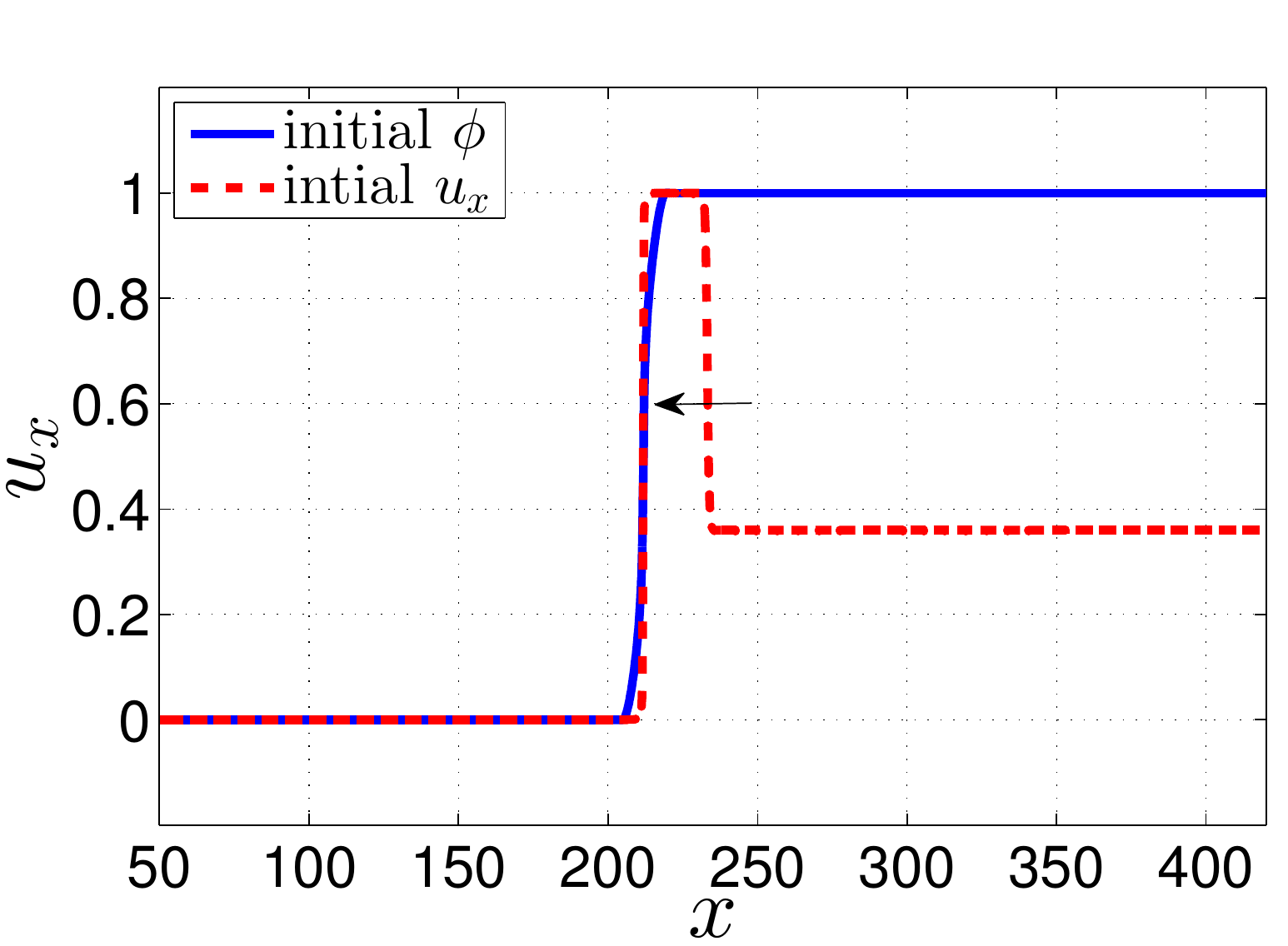}
			\\
			\includegraphics[width = 85mm]{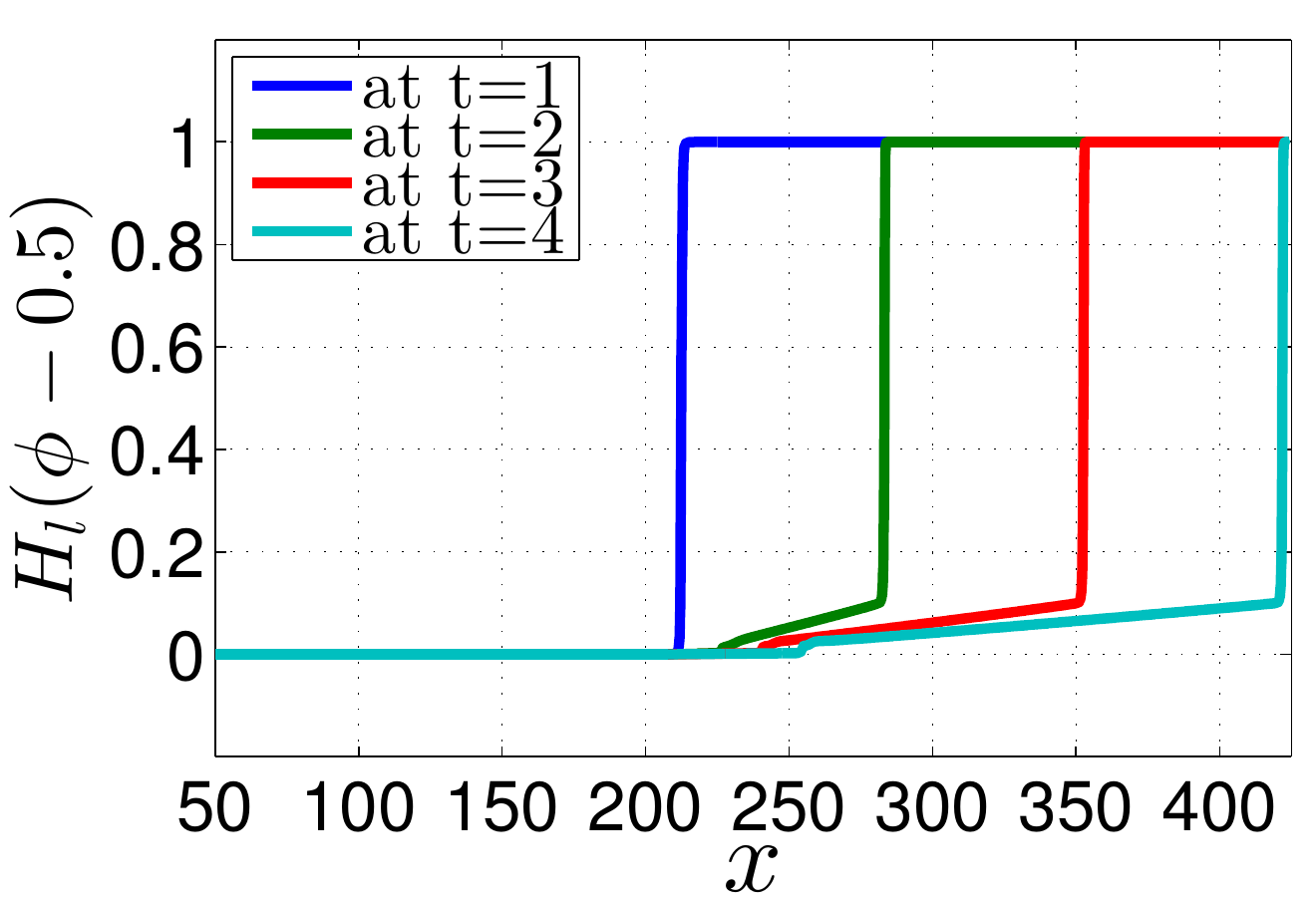}
			\includegraphics[width = 85mm]{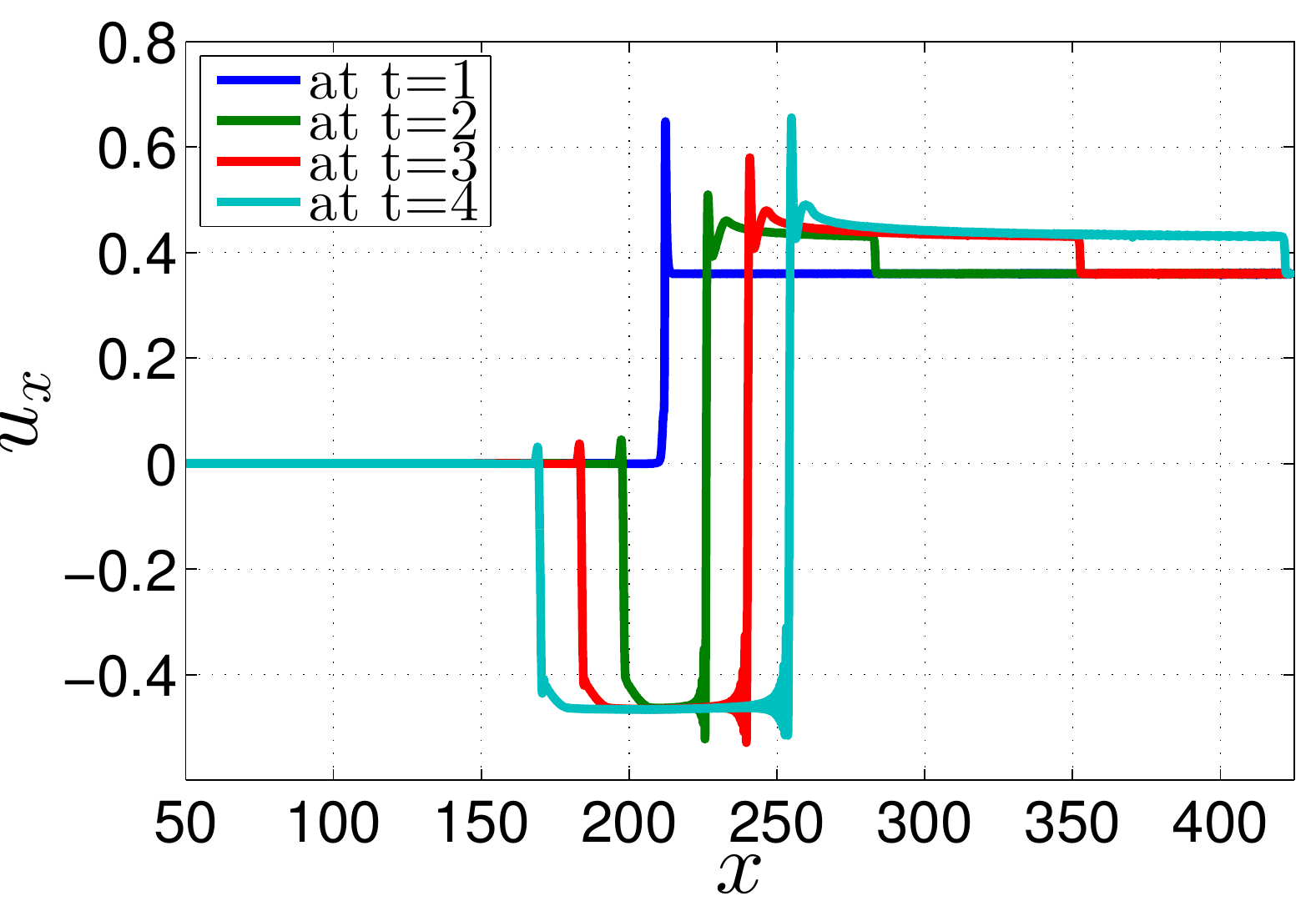}
			\caption{Top: A stationary interface as a compressive elastic wave approaches from the right. Left: The evolution of $H_l(\phi-0.5)$ shows the phase interface in $\phi$.  Right: The evolution of $u_x$ shows the phase interface in $u_x$.  The key point is that the evolution of phase interfaces in $\phi$ and $u_x$ are decoupled.
}
			\label{fig:supersonic-kinetics-dynamics}
		\end{center}
	\end{minipage}
	}
\end{center}
\end{figure}

We consider a material model with $C_3/C_1 = 2.25 \Rightarrow M_3 = M_1/1.5$. 
Fig. \ref{fig:supersonic-kinetics-dynamics} (top) shows the initial state with a stationary interface and a compressive shock approaching from the right.
The other plots in Fig. \ref{fig:supersonic-kinetics-dynamics} show the evolution of $\phi$ and $u_x$.
We find that (1) the interface in $\phi$ moves extremely rapidly, and is in fact significantly above sonic with respect to all wave speeds in the problem!, but (2) the interface in $u_x$ moves completely independently of the $\phi$-interface, and is subsonic with respect to both phases.
The $\phi$-interface carries a small elastic wave with it, but this can be considered as the response to a supersonic moving external load rather than as a supersonic wave.

In conclusion, it appears that our model does not handle the transition of phase interfaces from regimes where a kinetic relation is required, to regimes where a kinetic relation would overconstrain the evolution.
Our model simply does not allow supersonic transitions of the type studied in \cite{abeyaratne1991kinetic,rosakis-straingrad}.
However, it is encouraging that the model does not produce spurious seemingly-realistic solutions, but provides some warning.
That is, if we observe that the interfaces in $\phi$ and $u$ do not have the same spatial location, or if there are no steady traveling wave-like solutions, it is likely that the evolution is over-constrained and has transitioned from thermodynamics-dominated to momentum-balance-dominated.

We show in Appendix \ref{sec:phase-field-supersonic} that the standard phase-field model suffers from this same deficiency.

Finally, we briefly mention another example of the tension between momentum balance and thermodynamics: in 
\cite{purohit2003dynamics}, they find that certain phase interfaces in continuum string models also do not require an additional kinetic relation for unique evolution.

%%%%%%%%%%%%%%%%%%%%%
%%%%%%%%%%%%%%%%%%%%%
%%%%%%%%%%%%%%%%%%%%%
%%%%%%%%%%%%%%%%%%%%%

\section{Boundary Kinetics}
\label{sec:boundary-kinetics}

\begin{figure}[htb!]
\begin{center}
	\fbox{
	\begin{minipage}{155mm}
		\begin{center}
			\includegraphics[width = 150mm]{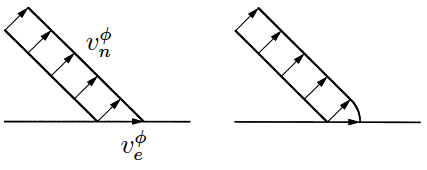}
			\caption{The same bulk interface velocity field $v_n^\phi$ can lead to very different overall evolution depending on the kinetic response $v_e^\phi$ of the interface-boundary junction line.  Adapted with permission from \cite{simha-bhatta-1}.}
			\label{fig:bdy-kinetics-schematic}
		\end{center}
	\end{minipage}
	}
\end{center}
\end{figure}

The most widely-used boundary condition (BC) on $\phi$ in standard phase-field modeling imposes $\nabla \phi \cdot \bfN = 0$ on $\partial\Omega$.
Among other things, it forces the interface to be normal to the boundary in the vicinity of the boundary, and it does not associate a kinetics to the evolution on the boundary beyond that which is driven by the evolution in the bulk.
However, as \cite{simha-bhatta-1,simha-bhatta-2,simha-bhatta-3} examined in the sharp-interface setting, it is possible to associate a separate kinetics to the junction line (or junction point in 2D) of the interface with the boundary.
Along the junction line, we require not only the normal interface velocity field $v_n^\phi$, but also the interface velocity component that is tangent to the boundary.
As illustrated in Fig. \ref{fig:bdy-kinetics-schematic}, the boundary velocity can have a profound effect on the evolution on the interior.
It also controls the surface evolution of $\phi$, and ignoring the effect of boundary kinetics can potentially lead to errors in analyzing experiments that are based only on surface measurements.

Another perspective is that the standard phase-field has roughly the character of a diffusion equation if one considers the highest derivatives: $\dot{\phi} = F(\phi) + \divergence \nabla\phi$ where $F(\phi)$ is the nonconvex term.
Standard phase-field uses Neumann BCs, and our use of boundary kinetics is roughly equivalent to time-dependent Dirichlet BCs, with the time-dependence being related to the fields in a complex and nonlinear fashion.

In this section, we use the interface balance principle to deduce a kinetic law for the junction lines, and apply the formulation to some illustrative calculations.
%%%%%%%%%%%%%%%%%%%%%
%%%%%%%%%%%%%%%%%%%%%

\subsection{Formulation of Boundary Kinetics}

As in the balance principle posed in Section \ref{sec:evolution-law}, we consider the flux of interfaces that thread a curve, with the only difference being that in this case the curve is restricted to lie on the boundary.
Hence, the kinetic law is the same as the bulk, $\dot{\phi} = |\nabla\phi| v_e^\phi + G_e$, with the gradient operator here interpreted as restricted to the boundary.
As with $v_n^\phi$, the edge velocity $v_e^\phi$ can be an arbitrary function of any arguments as long as it satisfies positive dissipation.
To ensure positive dissipation, as well as identify the thermodynamic conjugate driving force, we re-examine the dissipation statement, but with an augmentation to the energy to account for the boundary evolution.

We start by rewriting the energy to include an ``elastic energy on the boundary'':
\begin{equation}
\label{eq:bdy_energy}
	\int\limits_{\Omega_0}\Wcirc(\bfF,\phi) + \int\limits_{\Omega_0}\half \epsilon |\nabla\phi|^2 \bfalpha  
		+ t \int\limits_{\partial\Omega_0} \Wcirc(\bfF,\phi)+ \int\limits_{\partial\Omega_0}\half\epsilon_S |\nabla_S\phi|^2
\end{equation}
$\nabla_S$ is the gradient on the surface.

The first surface integral is the ``elastic energy on the boundary''.
The constant $t$ that multiplies it has dimensions of length, and is required for dimensional consistency.
Heuristically, it can be considered a measure of the distance that the surface effects penetrate into the bulk.

The term containing $\epsilon_S$ regularizes the interface-boundary junction.
In principle, it is possible for the elastic energy on the boundary to drive the interface to become singularly sharp at the junction, and this term prevents that.
Physically, it allows for the junction to have a width that can differ from the width of interfaces in the bulk.
We use $\epsilon_S$ to define $t \sim \epsilon_S/ \epsilon$ which has dimensions of length.

For notational convenience, we write the energy in \eqref{eq:bdy_energy} as $V[\bfu,\phi] + S[\bfu,\phi]$, where $V$ consists of the volume integrals and $S$ consists of the surface integrals.
We write the kinetic energy also as a sum of a volume integral and surface integral:
\begin{equation}
	K = K_v + K_s = \int\limits_{\Omega_0}\half \rho \bfV.\bfV + t \int\limits_{\partial\Omega_0}\half \rho \bfV.\bfV
\end{equation}
We now examine dissipation to obtain the thermodynamic conjugate driving forces for the the bulk and the edge interface velocities.
Recall that $\mathcal{D}$ is the difference between rate of external working and the rate of change of stored energy. 
Computing $\mathcal{D}$ gives us precisely the same result as the calculations in Section \ref{sec:dissipation}, except that we have the additional contribution from the surface terms in the stored energy.
This additional contribution is $-\deriv{}{t}\left(S[\bfu,\phi] + K_s\right)$, and can be simplified as follows:
\begin{equation}
\label{eq:surf_part}
	\deriv{}{t}(S+K_s) 
		= t\int\limits_{\partial\Omega_0}\parderiv{\Wcirc}{\bfF}:\dot{\bfF} + \rho\bfV\cdot\dot{\bfV} + t\int\limits_{\partial\Omega_0}\parderiv{\Wcirc}{\phi}\dot{\phi} -  \int\limits_{\partial\Omega_0}\epsilon_S \nabla_S \dot{\phi} \cdot \nabla_S\phi 
\end{equation}

The first term above vanishes as follows.
Following precisely the calculations in Section \ref{sec:dissipation}, we can write $\dot{F}$ as the material velocity gradient.
 Using integration-by-parts and the divergence theorem, we obtain balance of linear momentum which consequently vanishes, and another term that is a total derivative integrated over the closed region $\partial\Omega_0$ and therefore is set to $0$.

The last term above can be written as $\int\limits_{\partial\Omega_0}\epsilon_S  \dot{\phi} \divergence_S \nabla_S\phi $ and another term that is again a total derivative integrated over the closed region $\partial\Omega_0$ and therefore is set to $0$.

We can now write the dissipation $\mathcal{D}$.  It consists of the same expressions as in Section \ref{sec:dissipation}, and with the additional remaining nonzero terms from \eqref{eq:surf_part}:
\begin{equation}
	\mathcal{D}  
	= 
		\underbrace{-\int\limits_{\Omega_0} \left[ \parderiv{\Wcirc}{\phi} + \epsilon (\divergence \nabla \phi)\right] \dot{\phi}}_{\text{Bulk contributions}}
		\underbrace{- \int\limits_{\partial \Omega_0} \left[ \epsilon\nabla \phi \cdot \bfN - t \parderiv{\Wcirc}{\phi} +  \epsilon_S \divergence_S \nabla_S\phi \right] \dot{\phi}}_{\text{Surface contributions}}
\end{equation}
We use the bulk and boundary balance laws to replace $\dot{\phi}$ by $|\nabla\phi| v_n^\phi$ and $|\nabla\phi| v_e^\phi$ to obtain the thermodynamic conjugates to $v_n^\phi$ and $v_e^\phi$ as the bulk and edge driving forces:
\begin{equation}
	f_{bulk} := -\parderiv{\Wcirc}{\phi} + \epsilon \divergence\nabla\phi, \quad f_{edge} := -t\parderiv{\Wcirc}{\phi} + \epsilon_S \divergence_S \nabla_S\phi - \epsilon\nabla\phi\cdot\bfN
\end{equation}
The expression for $f_{edge}$ has some similarities to the sharp-interface version \cite{simha-bhatta-1}, in much the same way as $f_{bulk}$ has similarities to the classical driving force on an interface.

We have ignored the source $G^e$ above, but it provides a natural avenue to control nucleation of new interfaces on the boundary.
	
%%%%%%%%%%%%%%%%%%%%%
%%%%%%%%%%%%%%%%%%%%%

\subsection{Faceting of a Flat Stressed Interface}

We consider the energy in Section \ref{sec:2D-energy} using the unrotated specimen.
As noted there, the stress-free compatible twin interfaces are oriented at $\pm\pi/4$.
Here, we begin with an interface oriented vertically, and examine the evolution driven by the stress field.
As expected, the interface facets to locally align along the stress-free compatible directions.
However, we examine the effect of two contrasting choices of the boundary kinetics: (1) linear boundary kinetics that matches the linear bulk kinetics, and (2) pinned on the boundary with linear bulk kinetics.
We fix the left face of the specimen, and leave the other faces traction-free.

\begin{figure}[htb!]
\begin{center}
	\fbox{
	\begin{minipage}{175mm}
		\begin{center}
			\includegraphics[width = 85mm]{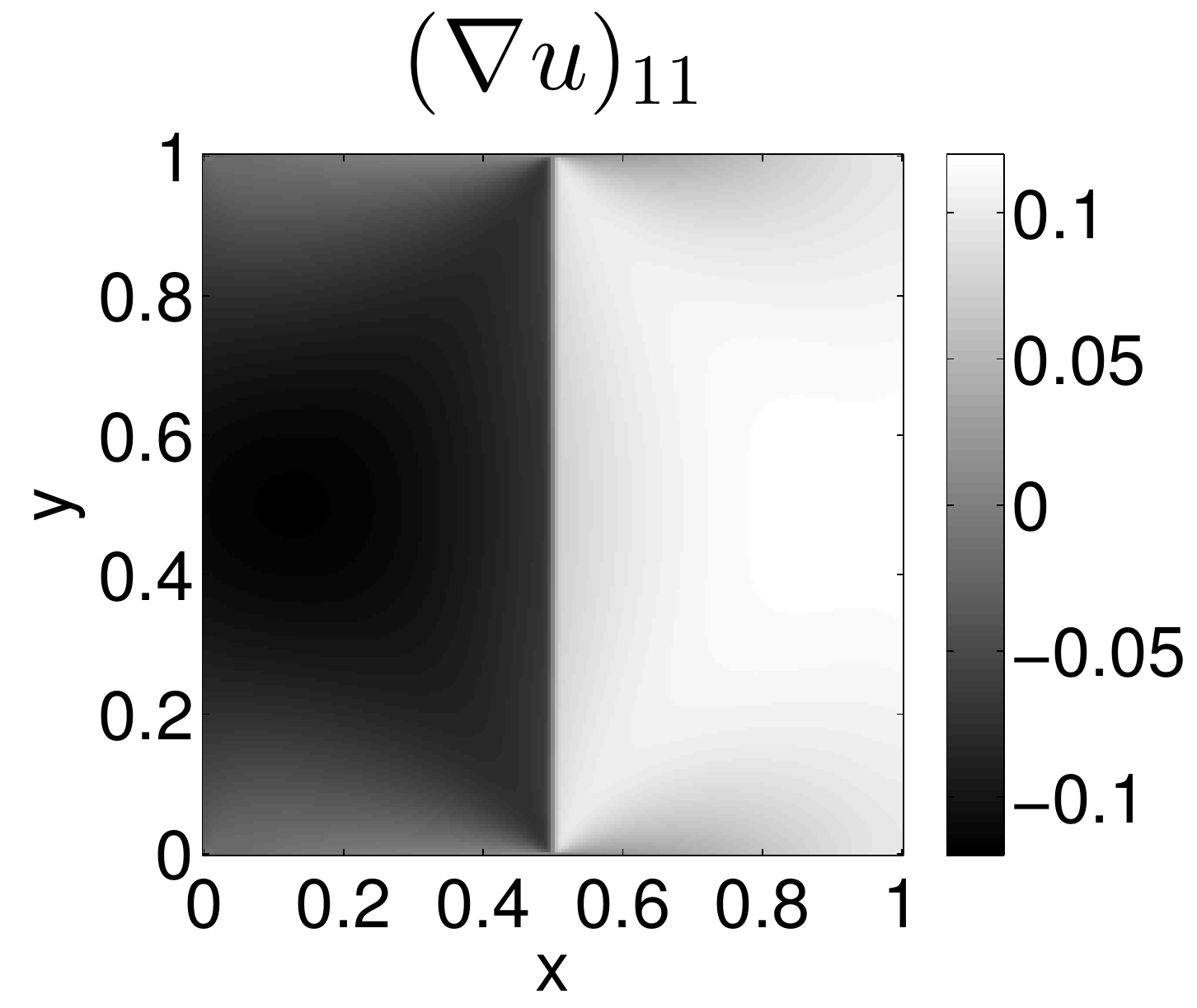}
			\\
			\includegraphics[width = 85mm]{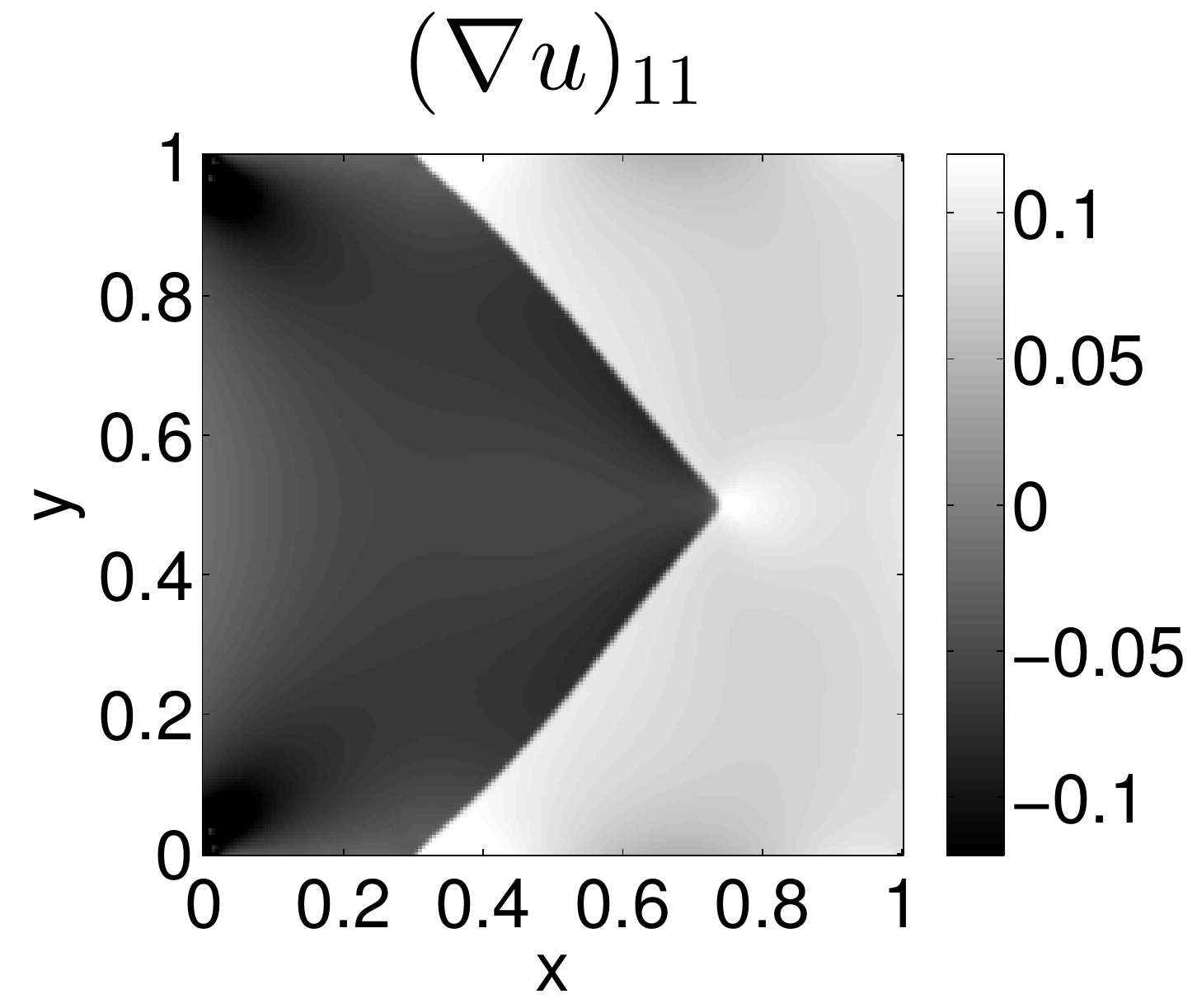}
			\includegraphics[width = 85mm]{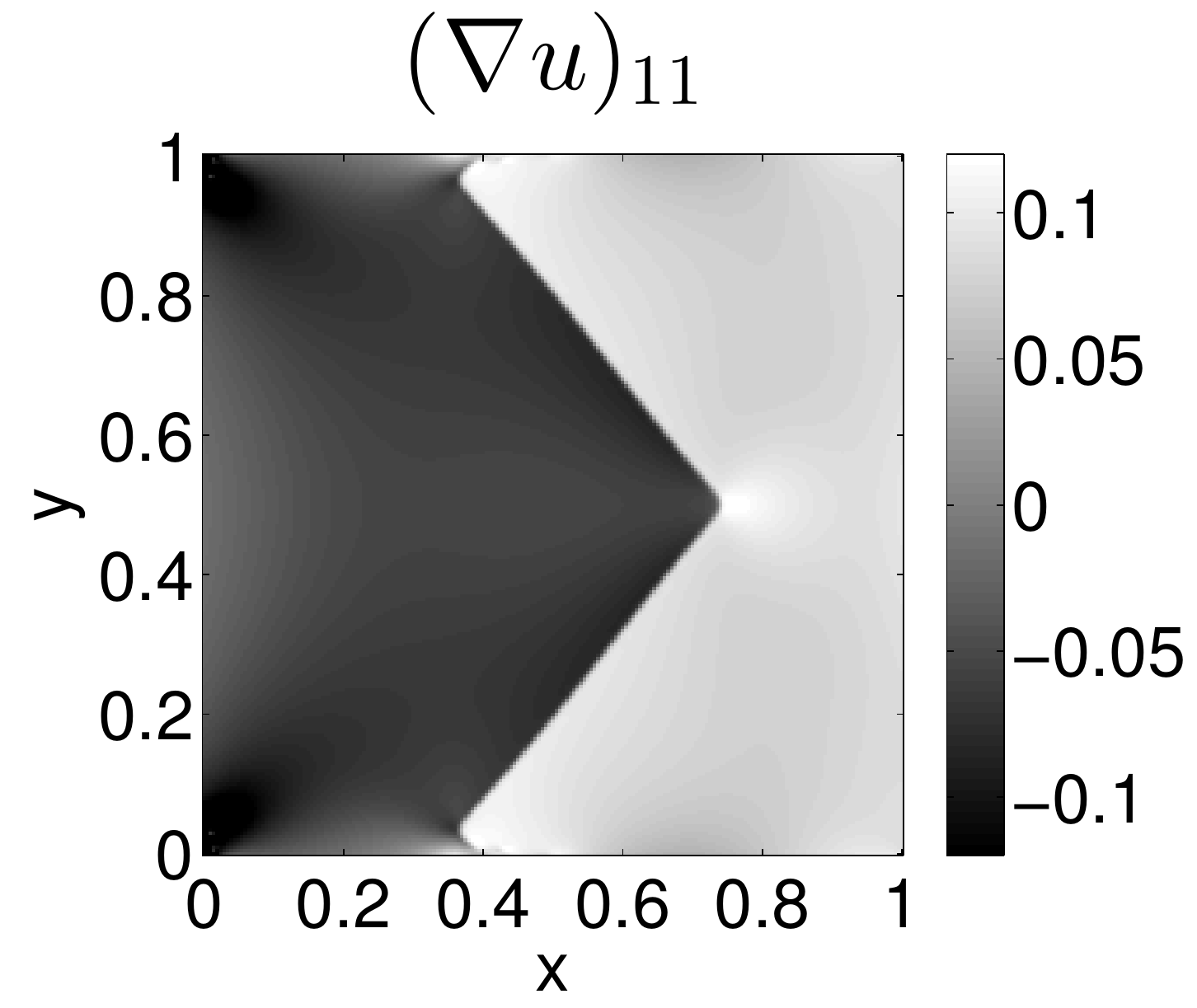}
			\\
			\includegraphics[width = 85mm]{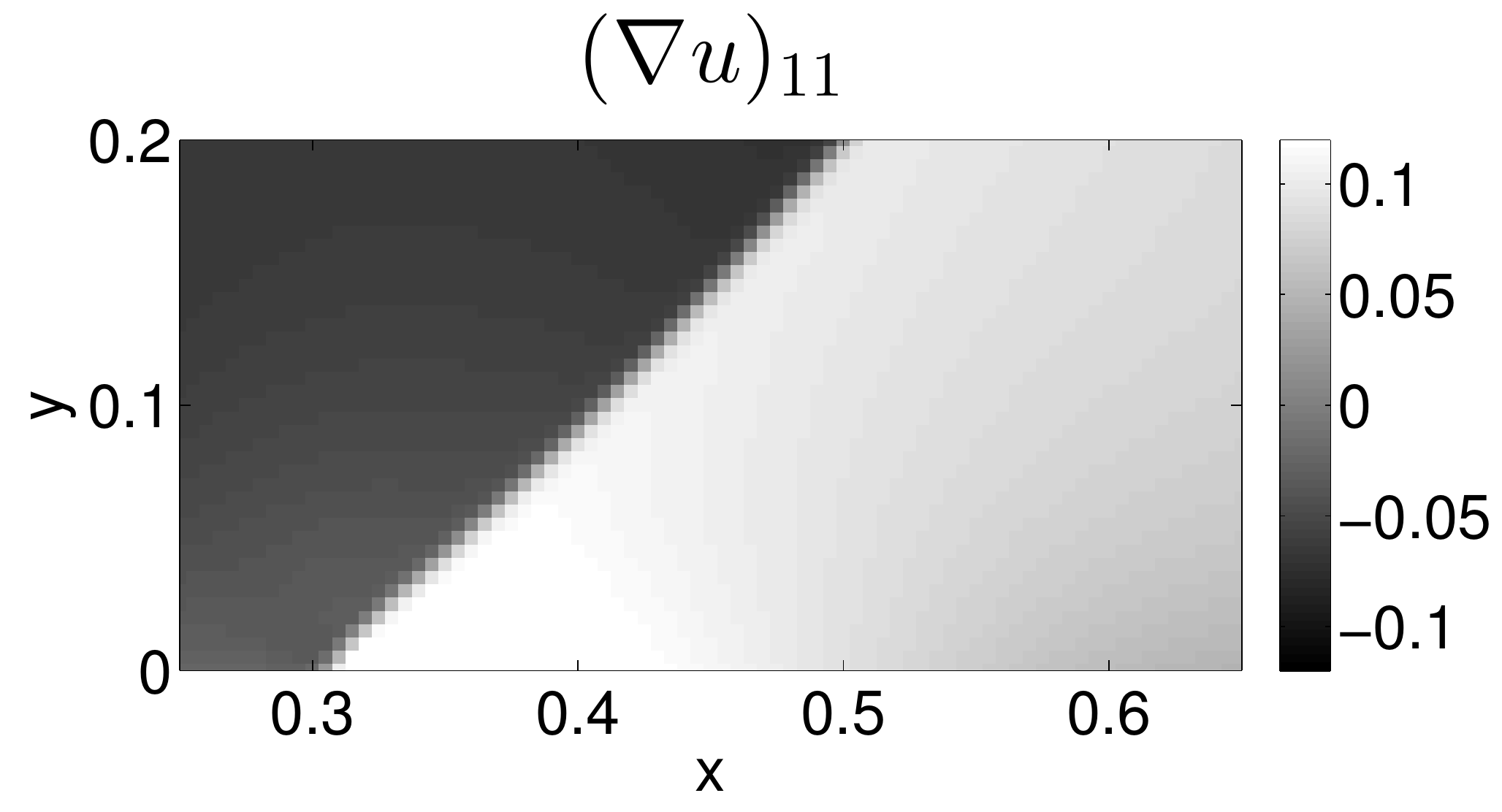}
			\includegraphics[width = 85mm]{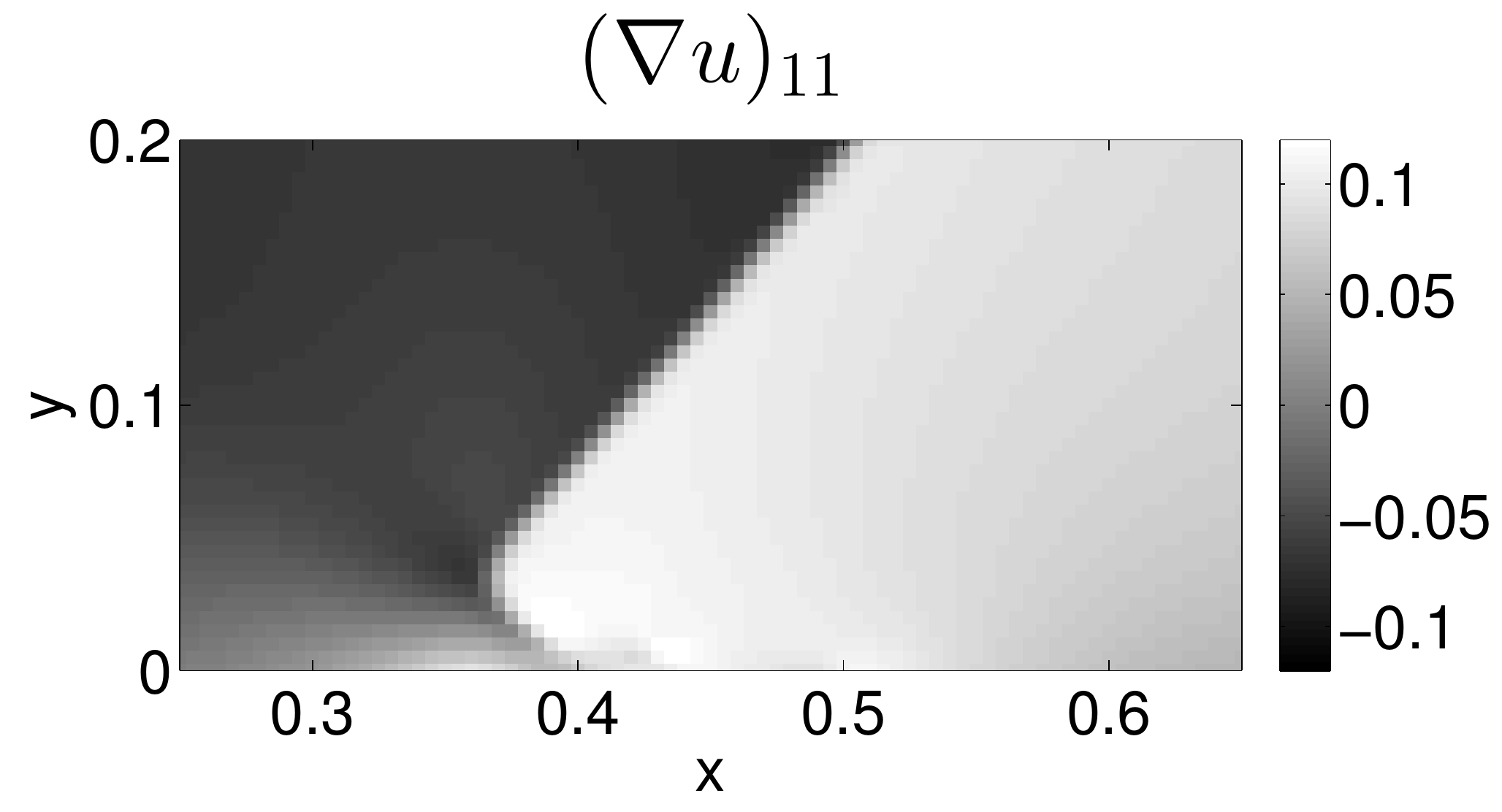}
			\caption{The top figure shows the initial condition.  The left column shows the final state with linear kinetics on the boundary (the entire sample and zoomed-in at the boundary).  The right column is the same with the interface pinned on the boundary.  All plots are of the $F_{11} - 1$ field.} 
			\label{fig:corrug-full}
		\end{center}
	\end{minipage}
	}
\end{center}
\end{figure}

Fig. \ref{fig:corrug-full} shows the initial condition for both cases, and final equilibrium state.
An interesting result when the interface is pinned on the boundary is that the orientation of the interface at the boundary is exactly the opposite of the interface orientation in the bulk.
Experimental probes of microstructure at the surface could completely mis-estimate the bulk microstructure.

%%%%%%%%%%%%%%%%%%%%%
%%%%%%%%%%%%%%%%%%%%%

\subsection{Competition Between Bulk and Boundary Kinetics}

We compare the evolution of a stress-free compatible interface for 3 different cases: when the boundary kinetics is (i) slower, (ii) the same, and (iii) faster, than the bulk kinetics.
For both bulk and boundary kinetics, we use a linear dependence on driving force, and change only the leading coefficient multiplying them.
The ratios of the coefficients are $0.4$ (boundary slower than bulk), $1.0$ (boundary and bulk are the same), and $1.6$ (boundary faster than bulk).

\begin{figure}[htb!]
\begin{center}
	\fbox{
	\begin{minipage}{175mm}
		\begin{center}
			\includegraphics[width = 85mm]{BK-F-initial}
			\includegraphics[width = 85mm]{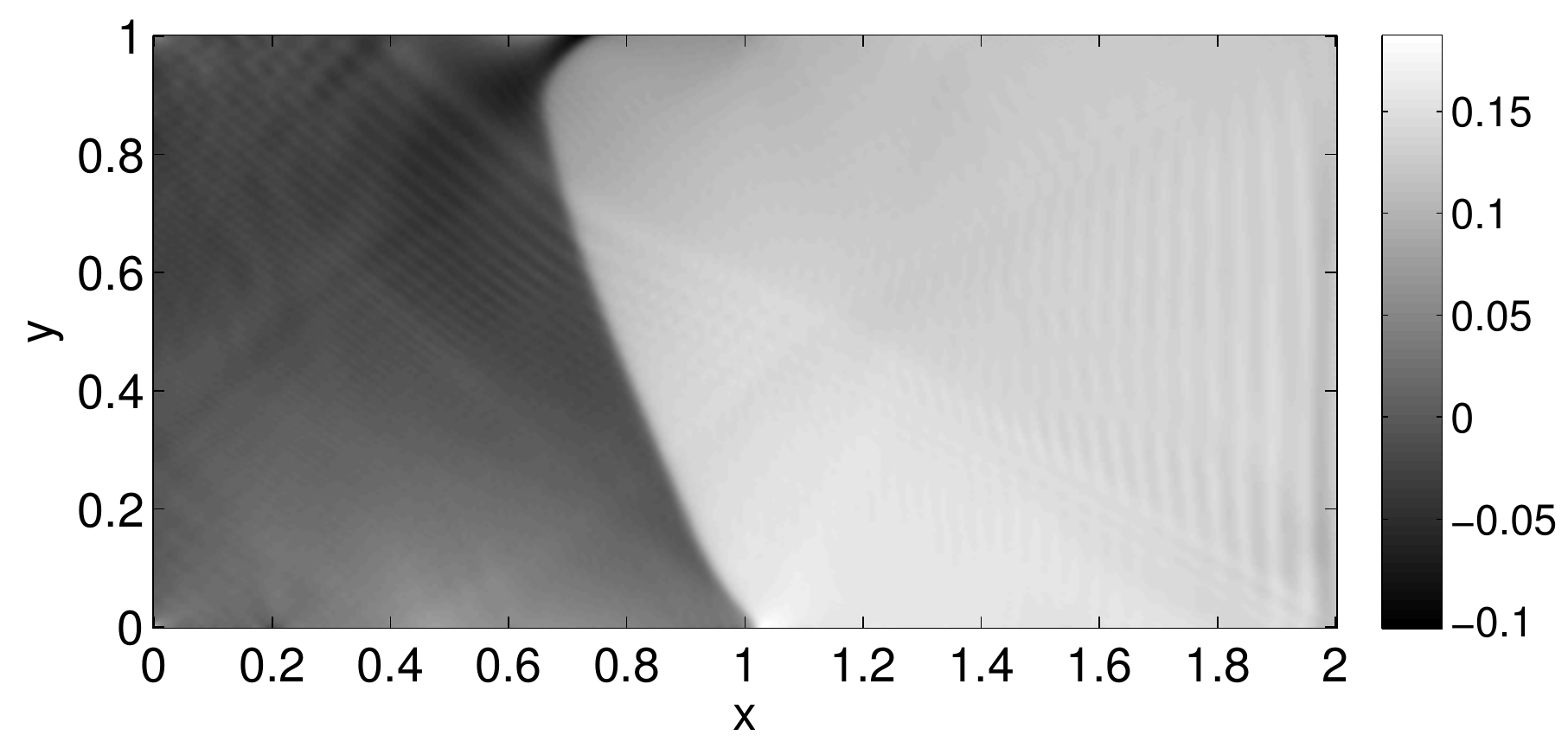}
			\\
			\includegraphics[width = 85mm]{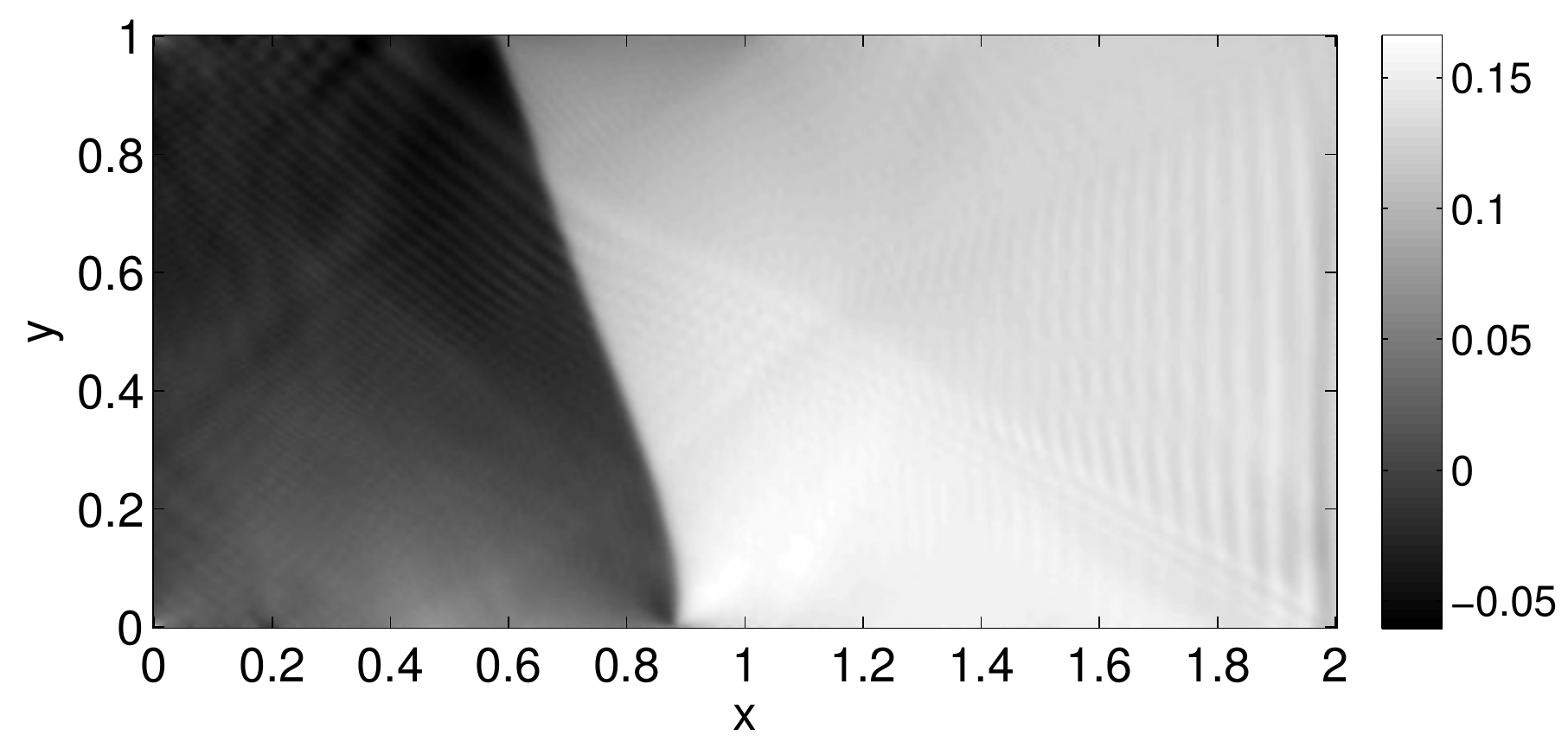}
			\includegraphics[width = 85mm]{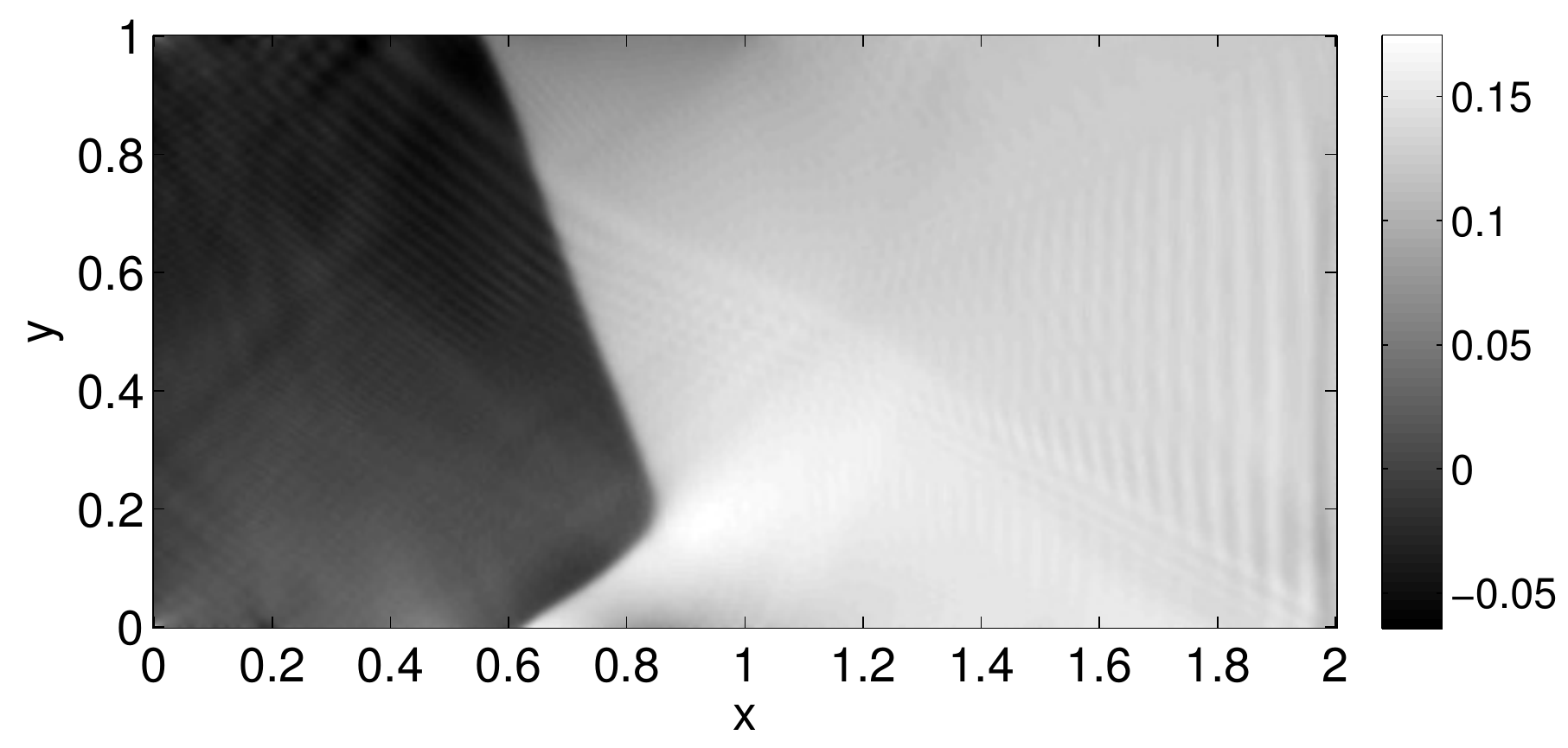}
			\caption{The first figure shows the initial configuration with a stress-free compatible interface.  The other figures show the evolution under applied load after some time for (i) slower, (ii) equal, and (iii) faster, boundary kinetics compared to bulk kinetics.  All plots are of the $F_{11} - 1$ field.}
			\label{fig:bdy-v-bulk}
		\end{center}
	\end{minipage}
	}
\end{center}
\end{figure}

We use the material described in Section \ref{sec:2D-energy} with the rotated specimen.
The initial configuration with the stress-free compatible interface for all 3 cases is the same and shown in Fig. \ref{fig:bdy-v-bulk}.
A constant tensile load is applied at the right face of the domain, and this causes the interface to begin to move.
The top and bottom faces are traction-free, and the left end is clamped.
The results from the 3 cases are shown in Fig. \ref{fig:bdy-v-bulk}.

%%%%%%%%%%%%%%%%%%%%%
%%%%%%%%%%%%%%%%%%%%%

\subsection{A Singularity In Boundary Kinetics}

The simplest boundary kinetic relation is to assume that the edge velocity field $v_e^\phi$ is a function of only the driving force $f_{edge}$.
However, we discuss here a simple but realistic example where this can lead to unexpected and likely-unphysical results.

Consider the energy in Section \ref{sec:2D-energy} for twinning, but with the specimen oriented such that the stress-free twin interface is vertical.
The interface is normal to the boundary on the top and bottom faces which are traction-free.
We apply a shear force on the right face while the left face is held fixed.

These phases in this problem are related by a shear, and hence the driving force for transformation is directly proportional to the shear traction on the plane that is parallel to the interface.
Therefore, a simple choice of linear kinetics -- in the bulk as well as on the boundary -- will set the interface velocity roughly proportional to the local value of the shear traction on the interface.

\begin{figure}[htb!]
\begin{center}
	\fbox{
	\begin{minipage}{175mm}
		\begin{center}
			\includegraphics[width = 85mm]{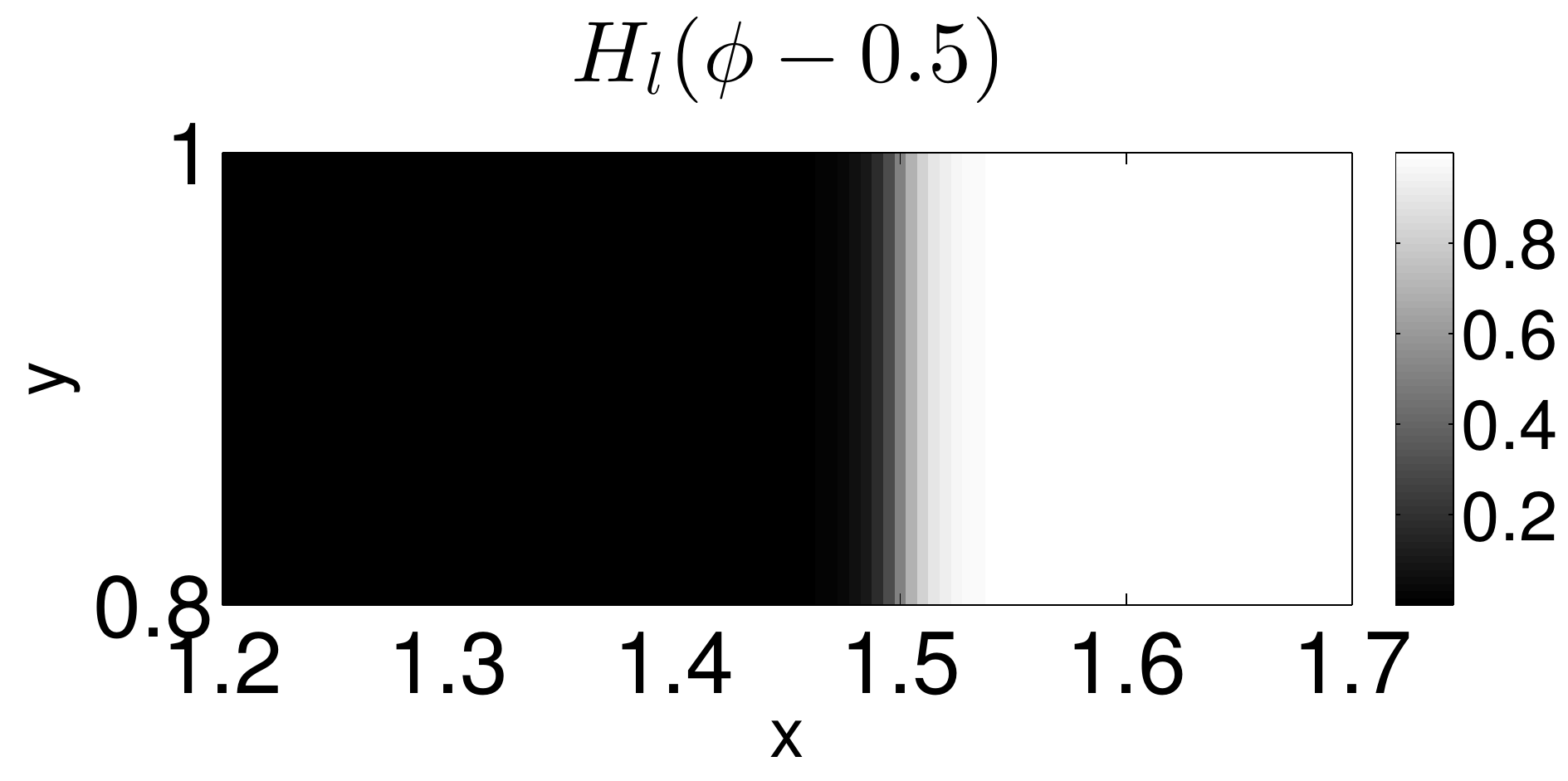}
			\includegraphics[width = 85mm]{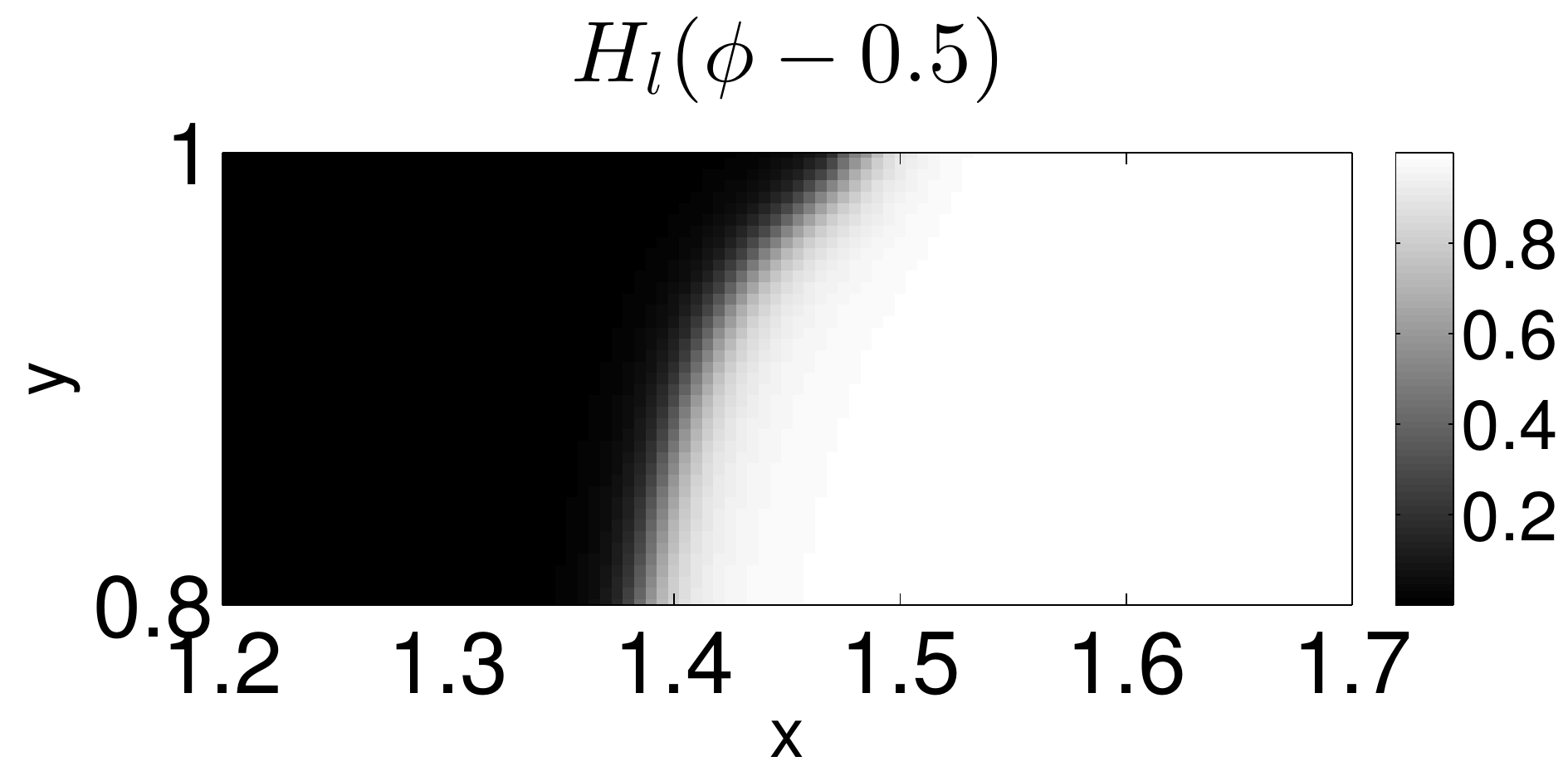}
			\caption{Left: the initial configuration with an interface normal to a traction-free boundary.  Right: the evolution of the interface.  There is no shear traction on the plane parallel to the interface at the surface, thereby not providing a driving force for the interface-boundary junction to evolve.}
			\label{fig:singular-bdy}
		\end{center}
	\end{minipage}
	}
\end{center}
\end{figure}

However, the top and bottom faces of the domain are traction-free; therefore, the shear stresses along the interface at the top and bottom faces are $0$.
The boundary driving force then largely vanishes, except for a small contribution due to the regularization parameters $\epsilon$ and $\epsilon_S$.
Fig. \ref{fig:singular-bdy} shows the time evolution of the interface in the vicinity of the boundary.
We find the unexpected development of an extremely curved interface as the interface in the interior moves forward while the interface-boundary junction barely moves.
The presence of $\epsilon$ and $\epsilon_S$ prevent this extremely curved interface from becoming completely singular.
With this regularization, the interface moves slowly along the boundary as the large curvature leads to contributions from the higher-order derivatives.

While this regularization keeps the problem from becoming singular, it does not seem physically attractive for the evolution to be dominated by ad-hoc regularization terms.
An approach that may perhaps provide more physically-meaningful evolution is  to set the boundary kinetics based on $\nabla\phi\cdot\bfN$; recall that the typical boundary condition in standard phase-field methods is $\nabla\phi\cdot\bfN = 0$.
However, as pointed out previously, this will drive the interface to be normal to the boundary in the vicinity of the boundary.

%%%%%%%%%%%%%%%%%%%%%
%%%%%%%%%%%%%%%%%%%%%
%%%%%%%%%%%%%%%%%%%%%
%%%%%%%%%%%%%%%%%%%%%

\section{Discussion}
\label{sec:discussion}

We have presented the formulation and characterization of a phase-field model that has non-singular interfaces yet allows for transparent prescription of kinetics and nucleation of interfaces.
The key elements are a re-parametrization of the energy and an evolution law that enables us to separate nucleation from kinetics.
In standard phase-field approaches, these are mixed together in an extremely opaque manner.
For instance, a uniform phase can nucleate a new phase through the kinetic equation, and there is no separate nucleation equation.
This mixing between kinetics and nucleation makes calibration extremely challenging.
In addition, standard phase-field models do not have a direct connection between evolution of $\phi$ and the kinetics of interfaces, and nucleation is completely opaque.
In our formulation, the calibration of nucleation and kinetics is simple and transparent.

We find two positive indications regarding our model:
(i) a formal limit of the kinetic driving force recovers the classical continuum sharp-interface driving force of Abeyaratne and Knowles;
(ii) examining interfaces as traveling waves in our model, we find that the driving force is constant in space, unlike the variational derivative of standard phase-field models.
The latter result implies that kinetic relations -- i.e., a relation between velocity and driving force --  is a well-defined notion in our model, and provides confidence in both the re-parametrized energy  and the evolution statement.
In addition, our re-parametrization of the energy modifies barriers in a drastic way, but preserves the structure of the energy away from the barriers.
However, we recall that the driving force in the classical setting \cite{abeyaratne2006evolution} depends only on the end-states on either side of the interface and not on the details of the barrier.
The information about the barrier is retained in our model through the kinetics and nucleation that we prescribe.

Our formulation allows us to prescribe both kinetics and nucleation independently.
From various numerical calculations, we find that microstructure patterns and mechanical response are not very sensitive to the precise kinetics, excepting the case of stick-slip kinetics.
Being able to model stick-slip kinetics is important for many physical situations of interest.
Standard phase-field does not handle stick-slip well when it is imposed by simply stopping the evolution when the driving force is below some critical value \cite{levitas2010interface}, but it is unclear if this difficulty is due to the energy structure or the evolution law.

We find that nucleation plays an extremely significant role in the development of microstructure patterns and mechanical response.
We have presented a number of examples to show the transparent and easy prescription of complex nucleation rules in our formulation.
These nucleation rules can involve various components of stress and strain, their rates, nonlocal quantities, and so on.
We emphasize that the nucleation criteria that we have used are toy models to enable the demonstration that we can readily and transparently incorporate an extremely broad range of nucleation criteria. 
In addition to enabling calibration to experiment or atomic calculations, this capability enables us to systematically probe the roles of different physical mechanisms for nucleation.
Recently, \cite{beyerlein2010probabilistic,wang2010atomic} presented a first-principles-motivated statistical model of twin nucleation in HCP materials.
While we have only examined deterministic nucleation in this work, it is straightforward to replace the deterministic source term in our evolution law by a statistical object that incorporates the insights from  \cite{beyerlein2010probabilistic,wang2010atomic}.

An important work with an alternative approach to the goal of numerical simulations of interface kinetics and nucleation is \cite{Rosakis-levelset}.
They use a level-set like approach but with a careful approach to regularization that aims to preserve the sharp-interface kinetics.
While there are many similarities between our work and theirs, there are also some key differences.
In our approach, we aim to formulate a model with certain key features such as regularized interfaces and the ability to prescribe nucleation and kinetics.
Their approach, however, is to develop a careful and sophisticated numerical method that reproduces the sharp-interface model.
A further difference appears to be the relative ease of numerical implementation of our model compared to that approach, but this could change.

An important open problem for our formulation is the inability to handle supersonic phase interfaces that do not require kinetics to obtain a unique evolution; in fact, prescription of kinetics overconstrains the evolution in the classical setting.
We find that our model cannot describe such supersonic interfaces in a traveling-wave setting corresponding to steady propagation, and dynamic calculations show strange behavior in these settings.
We further find that standard phase-field models also have this deficiency.
Regularized models that are based on adding viscosity and strain-gradients to the stress-response can correctly describe this interface \cite{rosakis-straingrad}.
But strain-gradient models have an important disadvantage compared to phase-field models.
Strain-gradient models do not use a phase-field at all but work exclusively with the displacement field.
The strain-gradients then impose severe smoothness requirements on the displacement field, making it challenging for finite elements in higher dimensions.  
In contrast, phase-field models -- both the existing approaches as well as our formulation -- require only the usual levels of smoothness because the gradients are on $\phi$ and only of second-order, and finite element implementations are commonplace.

Phase-field methods have largely been restricted  to linear elastic models.
This has the advantage that the elastic problem has nice properties such as existence, uniqueness etc.
While we have used a nonlinear deformation model in some of our calculations, it is convex (and quadratic) in the nonlinear strain for a fixed value of $\phi$.
Other recent phase-field models with large-deformation elasticity include \cite{lookman-porta, knap-clayton,levitas2014phase}.

Ongoing extensions of this work are as follows:
\begin{enumerate}
	\item The extension to $3$ or more phases is straightforward following the ideas presented here.  This is an important capability for realistic problems, and is a focus of our ongoing work.
	\item Our model extends the domain of stable phases, and allows them to exist in regions of strain space that should correspond to unstable phases.  This is not an ideal situation, and our ongoing work aims to remedy this by appropriately using nucleation criteria.
	\item Phase-field models for fracture are gaining in popularity.  Our ongoing work aims to extend the work presented here to that setting to enable dynamic fracture calculations.  In addition, we find that traveling wave analyses of the type presented in Appendix \ref{sec:phase-field-supersonic} are providing much insight into the behavior of supersonic cracks in the existing phase-field models of fracture, e.g. \cite{borden2012phase}.
	\item We have used finite elements for the elasticity and finite differences for the evolution of $\phi$.  However, the evolution of $\phi$ comes from a conservation principle, which can provide important advantages for robust numerical implementations such as a natural weak form \cite{trangenstein-book}.  The development of such a scheme may be a useful future extension. 
\end{enumerate}

%%%%%%%%%%%%%%%%%%%%%
%%%%%%%%%%%%%%%%%%%%%
%%%%%%%%%%%%%%%%%%%%%
%%%%%%%%%%%%%%%%%%%%%

\section*{Acknowledgments}

We thank ARO (W911NF-10-1-0140) for financial support. 
Vaibhav Agrawal also thanks the Carnegie Mellon University College of Engineering for financial support through the Bertucci Graduate Fellowship.
Kaushik Dayal also thanks the Carnegie Mellon University College of Engineering for financial support through the Early Career Fellowship.
This research was also supported in part by the National Science Foundation through TeraGrid resources provided by Pittsburgh Supercomputing Center. 
We thank Phoebus Rosakis for many useful discussions, encouragement, and pointers to the literature.

%%%%%%%%%%%%%%%%%%%%%
%%%%%%%%%%%%%%%%%%%%%
%%%%%%%%%%%%%%%%%%%%%
%%%%%%%%%%%%%%%%%%%%%

\appendix

\section{Connection to Noether's Theorem}
\label{sec:noether}

A seminal observation by Noether was that conservation principles in physics often have their roots in continuous symmetries.
For instance, conservation principles for linear momentum, angular momentum, and energy can be shown to arise from the facts that the energy of an isolated body is independent of translations in space, rotations in space, and translations in time respectively.
A concrete and extremely useful manifestation of Noether's principle is in \cite{knowles-sternberg-noether}, where they show that the essential path-independent properties of the $J$-integral of fracture mechanics are related to the application of Noether's theorem.
The idea can be roughly summarized as the procedure: (i) find the Lagrangian or action corresponding to the equation, and then (ii) take the variation with respect to the translations mentioned above.

We briefly examine here if a similar procedure provides any insight into the the conservation principle for interfaces.
For simplicity, we consider the one-dimensional setting with the interface velocity field being a constant that we set to $1$, i.e. our evolution equation is simply $\phi_t = \phi_x$.
Write the action $A = \int \int \half \left( \phi_t^2 \phi_x - \phi_x^2 \phi_t \right) \ dx dt$.
We have constructed $A$ by trial-and-error, using \cite{whitham1965general} as a starting point.
Subscripts $x$ and $t$ refer to space and time derivatives respectively.
We take the variation of $A$ by using the perturbation $\phi \to \phi + \epsilon \eta$, and evaluating $\left. \deriv{A}{\epsilon} \right|_{\epsilon=0} = 0$ provides:
\begin{equation}
	\phi_t \phi_{xt} - \phi_{xx} \phi_t -\phi_x\phi_{xt} + \phi_{tt}\phi_x + \phi_t \phi_{xt} - \phi_x \phi_{xt} = 0
\end{equation}
We have assumed that $\phi$ is sufficiently smooth for all derivatives to exist.

The Euler-Lagrange equation is a very nonlinear PDE and possibly has non-unique solutions.
However, we observe that $\phi_t = \phi_x$ is a solution, and also $\phi_t = \phi_x \Rightarrow \phi_{tt} = \phi_{xx} = \phi_{xt}$.
Further, we observe that $\phi_{tt} = \phi_{xx}$ by itself, without assuming $\phi_t=\phi_x$, does not appear to be a solution.
Therefore, this action is promising in that it provides the first-order wave equation but not the second-order equation.
The Hamiltonian is computed from the Legendre transform $\int \left( \phi_t \parderiv{L}{\phi_t} - L \right) \ dx$ to be $\half \phi_t^2 \phi_x$.

We examine then the effect of enforcing invariance under translations of the form $t \to t+\epsilon$ and $x \to x+\epsilon$.
Unfortunately, these do not provide any further insights.
This is not too surprising, because Noether's theorem typically does not provide new evolution laws, but rather can provide conserved quantities that are already implied by the original evolution laws; e.g., given the force-acceleration equation for an isolated system of particles, Noether's principle can show that the total momentum is conserved.

%%%%%%%%%%%%%%%%%%%%%
%%%%%%%%%%%%%%%%%%%%%
%%%%%%%%%%%%%%%%%%%%%
%%%%%%%%%%%%%%%%%%%%%

\section{Non-existence of Supersonic Interfaces in Standard Phase-field Models} 
\label{sec:phase-field-supersonic}

We consider a 1D phase-field model with energy as follows:
\begin{equation}
\label{eqn:standard-phase-field-1}
	\int_\Omega W(\phi) + \half C(\phi) \left(\epsilon - \epsilon_0(\phi)\right)^2 + \half \epsilon \left(\deriv{\phi}{x}\right)^2
\end{equation}
where $W$ is a nonconvex energy density with multiple wells.
Using the standard gradient-flow assumption, the evolution equations are:
\begin{equation}
\label{eqn:standard-phase-field-2}
	\rho \ddot{u} = \deriv{}{x}\left[ C(\phi) \left(\epsilon - \epsilon_0(\phi)\right) \right], \quad \dot{\phi} = \deriv{W}{\phi} + \epsilon \deriv{^2\phi}{x^2}
\end{equation}

The context of the calculation in this section is based on the discussion in Section \ref{sec:thermo-vs-momentum}.
Analysis of classical continuum models \cite{abeyaratne1990driving} and strain-gradient models \cite{rosakis-straingrad} show the possibility of interfaces that are supersonic with respect to the softer phase.
But our model does not admit such interfaces, and requires that interfaces are subsonic with respect to both phases.
We examine this question in the context of the standard phase-field energy \eqref{eqn:standard-phase-field-1}, \eqref{eqn:standard-phase-field-2}.
We note that by setting the elastic modulus to be an explicit function of the phase $C(\phi)$, we have enabled the moduli of the phases to be different.

We now analyze just the momentum balance equation from \eqref{eqn:standard-phase-field-2}.
Assuming a traveling wave framework, we write $u(x,t) = U(x-Vt), \phi(x,t) = \Phi(x-Vt)$.
This gives us $\rho V^2 U'' = \left[ C(\Phi) \left(U' - \epsilon_0(\Phi)\right) \right]'$.
Integrating once, we get $\rho V^2 U' = C(\Phi) \left(U' - \epsilon_0(\phi)\right) + D$.
We rearrange to write:
\begin{equation}
	U' = \frac{D - \epsilon_0(\Phi)}{1 - M^{-2}(\Phi)}
\end{equation}
where $M(\Phi) := \frac{V}{(C(\Phi)/\rho)^\half}$ is the local Mach number.

If $C(\phi)$ is a continuous function of its argument\footnote{Assuming that $C(\phi)$ is not continuous would require us to explicitly track the discontinuity in numerical calculations, thereby destroying the essential advantage of phase-field approaches.}, it follows that $M(\phi)$ is a continuous function of its argument.
Therefore, if $M(\Phi) > 1$ at any point in space, then $M = 1$ somewhere in the domain using continuity of $\Phi$ in its argument, and hence $U'$ is unbounded at the point where $M=1$.
Hence, phase-field models cannot support steadily-propagating phase interfaces that are supersonic with respect to the softer phase.

%%%%%%%%%%%%%%%%%%%%%
%%%%%%%%%%%%%%%%%%%%%
%%%%%%%%%%%%%%%%%%%%%
%%%%%%%%%%%%%%%%%%%%%

\bibliographystyle{amsalpha}
\bibliography{references}

\end{document}